\documentclass[authoryear,12pt, a4]{elsarticle}
\usepackage[latin1]{inputenc}
\usepackage{graphics, graphicx}
\usepackage{color}
\usepackage{psboxit,pstcol}
\usepackage{rotating}
\usepackage{amssymb}
\usepackage{lineno}
\usepackage{hyperref}

\biboptions{longnamesfirst}

\journal{Icarus}
\begin{document}
\begin{frontmatter}

\title{Saturn's icy satellites investigated by Cassini - VIMS. \\
IV. Daytime temperature maps}

\author[label1]{Gianrico Filacchione\corref{cor1}} 
\author[label1]{Emiliano D'Aversa}
\author[label1]{Fabrizio Capaccioni}
\author[label2]{Roger N. Clark}
\author[label3]{Dale P. Cruikshank}
\author[label1]{Mauro Ciarniello}
\author[label1]{Priscilla Cerroni}
\author[label1]{Giancarlo Bellucci}
\author[label4]{Robert H. Brown}
\author[label5]{Bonnie J. Buratti}
\author[label6]{Phillip D. Nicholson}
\author[label7]{Ralf Jaumann}
\author[label8]{Thomas B. McCord}
\author[label5]{Christophe Sotin}
\author[label7]{Katrin Stephan}
\author[label3]{Cristina M. Dalle Ore}

\address[label1]{INAF-IAPS, Istituto di Astrofisica e Planetologia Spaziali, Area di Ricerca di Tor Vergata, via del Fosso del Cavaliere, 100, 00133, Rome, Italy}
\address[label2]{PSI Planetary Science Institute, Tucson, AZ, USA}
\address[label3]{NASA Ames Research Center, Moffett Field, CA 94035-1000, USA}
\address[label4]{Lunar and Planetary Laboratory and Steward Observatory, University of Arizona, Tucson, AZ 85721, USA}
\address[label5]{Jet Propulsion Laboratory, California Institute of Technology, 4800 Oak Groove Drive, Pasadena, CA 91109, USA}
\address[label6]{Cornell University, Astronomy Department, 418 Space Sciences Building, Ithaca, NY 14853, USA}
\address[label7]{Institute for Planetary Exploration, DLR, Rutherfordsta{\ss}e 2, 12489, Berlin, Germany}
\address[label8]{The Bear Fight Institute, Winthrop, WA 98862, USA}
\cortext[cor1]{Corresponding author, email gianrico.filacchione@iaps.inaf.it}

\begin{abstract}
The spectral position of the 3.6 $\mu m$ continuum peak measured on Cassini-VIMS I/F spectra is used as a marker to infer the temperature of the regolith particles covering the surfaces of Saturn's icy satellites. This feature is characterizing the crystalline water ice spectrum which is the dominant compositional endmember of the satellites' surfaces.
Laboratory measurements indicate that the position of the 3.6 $\mu m$ peak of pure water ice is temperature-dependent, shifting towards shorter wavelengths when the sample is cooled, from about 3.65 $\mu m$ at T=123 K to about 3.55 $\mu m$ at T=88 K.
A similar method was already applied to VIMS Saturn's rings mosaics to retrieve ring particles temperature \citep{Filacchione2014}. We report here about the daytime temperature variations observed on the icy satellites as derived from three different VIMS observation types: a) a sample of 240 disk-integrated I/F observations of Saturn's regular satellites collected by VIMS during years 2004-2011 with solar phase in the 20$^\circ$-40$^\circ$ range, corresponding to late morning-early afternoon local times. This dataset is suitable to exploit the temperature variations at hemispherical scale, resulting in average temperature  T$<$88 K for Mimas, T$<<$88 K for Enceladus, T$<$88 K for Tethys, T=98-118 K for Dione, T=108-128 K for Rhea, T=118-128 K for Hyperion, T=128-148 K and T$>$168 K for Iapetus' trailing and leading hemispheres, respectively. A typical $\pm$5 K uncertainty is associated to the temperature retrieval. On Tethys and Dione, for which observations on both leading and trailing hemispheres are available, in average daytime temperatures higher of about 10 K on the trailing than on the leading hemisphere are inferred. b) Satellites disk-resolved observations taken at 20-40 km/pixel resolution are suitable to map daytime temperature variations across surfaces' features, such as Enceladus' tiger stripes and Tethys' equatorial dark lens. These datasets allow to disentangle solar illumination conditions from temperature distribution when observing surface's features with strong thermal contrast. c) Daytime average maps covering large regions of the surfaces are used to compare the inferred temperature with geomorphological features (impact craters, chasmatae, equatorial radiation lenses and active areas) and albedo variations. Temperature maps are built by mining the complete VIMS dataset collected in years 2004-2009 (pre-equinox) and in 2009-2012 (post equinox) by selecting pixels with max 150 km/pixel resolution. VIMS-derived temperature maps allow to identify thermal anomalies across the equatorial lens of Mimas and Tethys. A temperature T$>$115K is measured above Enceladus' Damascus and Alexandria sulci in the south pole region. VIMS has the sensitivity to follow seasonal temperature changes: on Tethys, Dione and Rhea higher temperature are measured above the south hemisphere during pre-equinox and above the north hemisphere during post-equinox epochs. The measured temperature distribution appears correlated with surface albedo features: in fact temperature increases on low albedo units located on Tethys, Dione and Rhea trailing hemispheres. The thermal anomaly region on Rhea's Inktomi crater detected by CIRS \citep{Howett2014} is confirmed by VIMS: this area appears colder with respect to surrounding terrains when observed at the same local solar time. 

\end{abstract}


\end{frontmatter}

\section{Introduction}
With the aim to retrieve Saturn's icy satellites surface temperature, a method based on the spectral properties at 3.6 $\mu m$ of water ice at cryogenic temperatures has been developed and applied to the Cassini-VIMS, the Visual and Infrared Mapping Spectrometer, dataset. Considered as a mineral by many authors \citep{Fletcher1970, Hobbs1974} water ice is capable of many rotovibrational modes: solid state lattice's rotational modes are driven by interactions with nearby molecules while vibrational modes appear at characteristic frequencies given by the molecular structure. Since all these modes are temperature-dependent, in principle is possible to derive the temperature of a water ice sample by analyzing the properties of its spectral reflectance. 

Water ice lattice undergoes temperature dependent phase-transitions: in vacuum the structure is amorphous for $T<137$ K, cubic for $137\le T \le 197$ K and hexagonal for $T>197$ K \citep{Hobbs1974}. The non-amorphous state is in general referred as crystalline, showing a diagnostic spectral feature at 1.65 $\mu m$ not present in the amorphous state. Another temperature-induced effect is the variation of the absorption coefficient which decreases when the sample is cooled. As a consequence, cold ice absorbs more photons than warm ice. 

At the end of the fifties, the first positive infrared identification of water ice on the surfaces of Europa and Ganymede was reported \citep{Kuiper1957}. With the progress of infrared imaging spectroscopy techniques, advanced spaceborne instrumentation has been developed to study the icy surfaces in the outer solar system: the Near Infrared Mapping Spectrometer (NIMS) on Galileo mission achieved the first compositional maps of the galilean icy satellites \citep{Carlson1996} while a decade later VIMS, the Visual and Infrared Mapping Spectrometer \citep{Brown2004} on Cassini has observed in detail the entire population of the icy satellites orbiting around Saturn, giving the first identification of crystalline water ice on the south polar region of the active moon Enceladus \citep{Brown2006, Jaumann2008} and the first compositional maps \citep{Jaumann2006}. The analysis of VIMS data indicates that other species, like  $CO_2$ and organics correlated with C-H and possibly C-N or adsorbed H$_2$ \citep{Clark2012}, appear mixed with water ice on the surfaces of Hyperion \citep{Cruikshank2007, Cruikshank2010}, Iapetus \citep{Cruikshank2008, Clark2012} and Phoebe \citep{Clark2005, Buratti2008, Coradini2008}. Exogenic dark material dispersed on Dione's trailing hemisphere \citep{Clark2008, Stephan2010, Scipioni2013}, across Rhea \citep{Stephan2012, Scipioni2014} and on Iapetus leading hemisphere \citep{Pinilla2011} was also observed and mapped in detail. Extensive studies have been published about the analysis of disk-integrated I/F spectra of Saturn's satellites and resulting comparative properties, including hemispheric asymmetries observed on the surfaces of the regular satellites \citep{Filacchione2007, Filacchione2010}, the retrival of phase and light curves with corresponding photometric parameters \citep{Pitman2010, Ciarniello2011} and the radial distribution of water ice and chromophores across Saturn's system of satellites and rings \citep{Filacchione2012, Filacchione2013}. 

All these remotely-sensed observations have boosted an intensive laboratory activity devoted to the characterization of water ice spectral properties. Relevant to the retrieval of water ice temperature was the investigation of the temperature-dependent 1.65 $\mu m$ absorption feature seen in the reflectance of crystalline water ice \citep{FinkLarson1975}. Further analyses, focused on water ice optical constants, have allowed to infer real and imaginary parts of the index of refraction in the VIS and NIR up to 2.8 $\mu m$ at -7$^\circ$C \citep{Warren1984}, to characterize the water ice absorptions and FWHM as a function of temperature between 20-270 K \citep{GrundySchmitt1998} and to use the 1.65 $\mu m$ band as temperature proxy \citep{Grundy1999, Leto2005}. \cite{Taffin2012} have acquired and characterized water ice sample in the 1-1.8 $\mu m$ spectral range by varying grains in the 80-700 $\mu m$ size range and temperature from 80 to 140 K.

Optical constants of amorphous and crystalline water ice for 20 $\le$ T$\le$ 150 K in the 1.1-2.6 $\mu m$ spectral range have been inferred 
by \cite{Mastrapa2008}.
A further spectral extension towards longer wavelengths (2.5-22 $\mu m$) for amorphous and crystalline forms \citep{Mastrapa2009} has shown that the reflectance around 3.6 $\mu m$ is influenced by temperature. This spectral range was never used before to infer the temperature of the ice sample. 
Located in between the intense absorption at 3.0 $\mu m$ due to the O-H-O symmetric stretch vibration ($\nu_1$) and the weaker 4.6 $\mu m$ feature generated by a combination of ($\nu_2$) and a weak libration mode, the shape of the continuum around 3.6 $\mu m$ is strongly influenced by the thermal behavior of those two features. The resulting spectral variation caused by temperature changes is described later in section \ref{sct:a}. 

In this work we are aiming to infer surface temperatures of Saturn's icy satellites from infrared observations at different spatial scales (hemispheric, regional and local). While the FP1 thermopile detector radiometer channel of Cassini/CIRS \citep{Flasar2014} is specifically designed to measure satellites' temperatures through fitting the thermal emission peak in the a 17-1000 $\mu m$ spectral range, VIMS is a 0.35-5.0 $\mu m$ imaging spectrometer devoted to infer surface composition \citep{Brown2004}. Despite being limited in spectral range, VIMS offers the advantage of having much better imaging capabilities than CIRS-FP1: with a nominal FOV of 1.83$^\circ$ and an IFOV of 500 $\mu rad$, VIMS is able to acquire hyperspectral cubes up to 64$\times$64 spatial pixels while CIRS-FP1 channel relies on a single pixel with a circular FOV of 3.9 mrad. 
Moreover, CIRS mid-infrared Michelson interferometer channel uses two HgCdTe detectors (FP3, FP4), with a 0.273 mrad IFOV, designed to measure icy satellites daytime temperatures. Conversely, since VIMS operates only in reflectance conditions, e.g. on dayside hemispheres, only daytime temperatures can be measured, therefore preventing the possibility to infer nocturnal temperatures and derive accurate thermal inertia data. This task can be achieved only by CIRS. 
For a given composition, thermal inertia depends upon surface roughness and porosity: as a general rule a grainy water ice pack has high bond albedo and low thermal inertia. On contrary, packed grains show high bond albedo and high thermal inertia. By analyzing CIRS data \cite{Howett2010} have derived an average thermal inertia lower than 20 $J \ m^{-2} \ K^{-1} \ s^{-1/2}$ across the surfaces of the Saturn's satellites which increases up to 89 $J \ m^{-2} \ K^{-1} \ s^{-1/2}$ above thermal anomaly regions of Mimas and Tethys, as discussed later.

The properties of the water ice optical constants and resulting I/F spectra as function of the temperature are the topics of the next section \ref{sct:a}. In section \ref{sct:b} we discuss how this particular spectral behavior of the ice allows us to use the VIMS instrument as a thermal mapper and to infer the icy satellites daytime temperature. This method is applied in subsection \ref{sct:c} to disk-integrated I/F observations suitable to explore the daytime temperature variability of the regular satellites at hemispherical scale and to characterize the differences between their leading and trailing hemispheres. The detection of temperature variations across Enceladus south hemisphere and in Tethys equatorial radiation lens are discussed in section \ref{sct:d}. Global temperature maps for Mimas, Enceladus, Tethys, Dione, Rhea, Hyperion and Iapetus during the pre Saturn's equinox (2004-2009) and post equinox (2009-2012) periods are discussed in section \ref{sct:e}.

\section{Water ice optical constants and temperature-dependent effects}\label{sct:a}

\begin{figure}[h!]
	\centering
		\includegraphics[width=14cm]{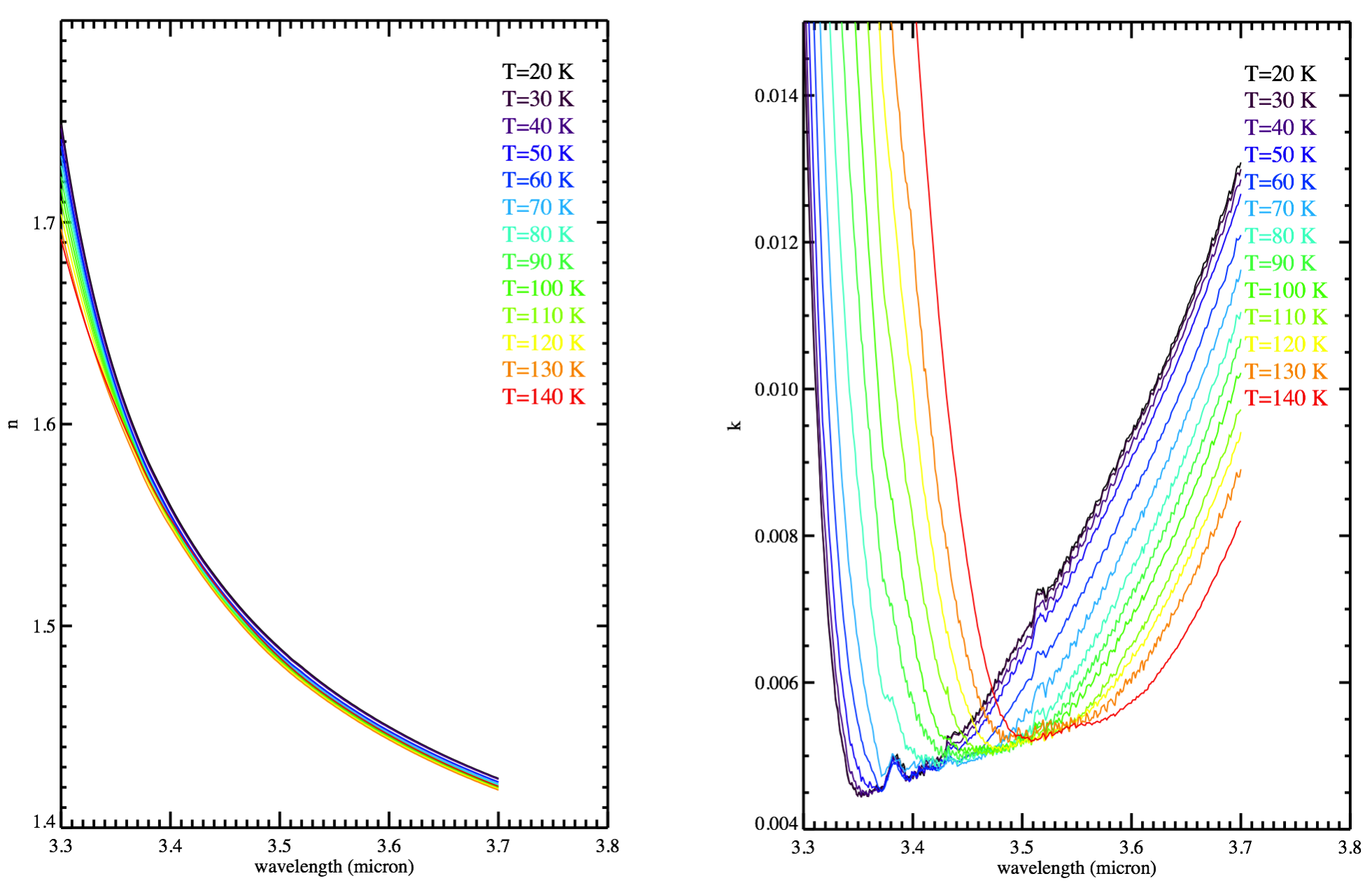}
		\caption{Water ice optical constants at temperature between T=20 to 140 K by \cite{Mastrapa2009}. Left panel: real index n. Right panel: imaginary index k. Note how the k index minimum shifts towards shorter wavelengths with temperature decrease.}
	\label{fig:figure1}
\end{figure}

Crystalline water ice reflectance is characterized by a broad peak at about 3.6 $\mu m$ that has temperature-dependent properties. Optical constants measured in transmittance by \cite{Mastrapa2009} in the range of temperature between 20 and 140 K show that in the 3.2-3.6 $\mu m$ spectral range the real n index is almost constant while the imaginary k index minimum position is temperature-dependent, moving to shorter wavelengths when the temperature decreases (Fig. \ref{fig:figure1}, right panel). 

\begin{figure}[h!]
	\centering
		\includegraphics[width=14cm]{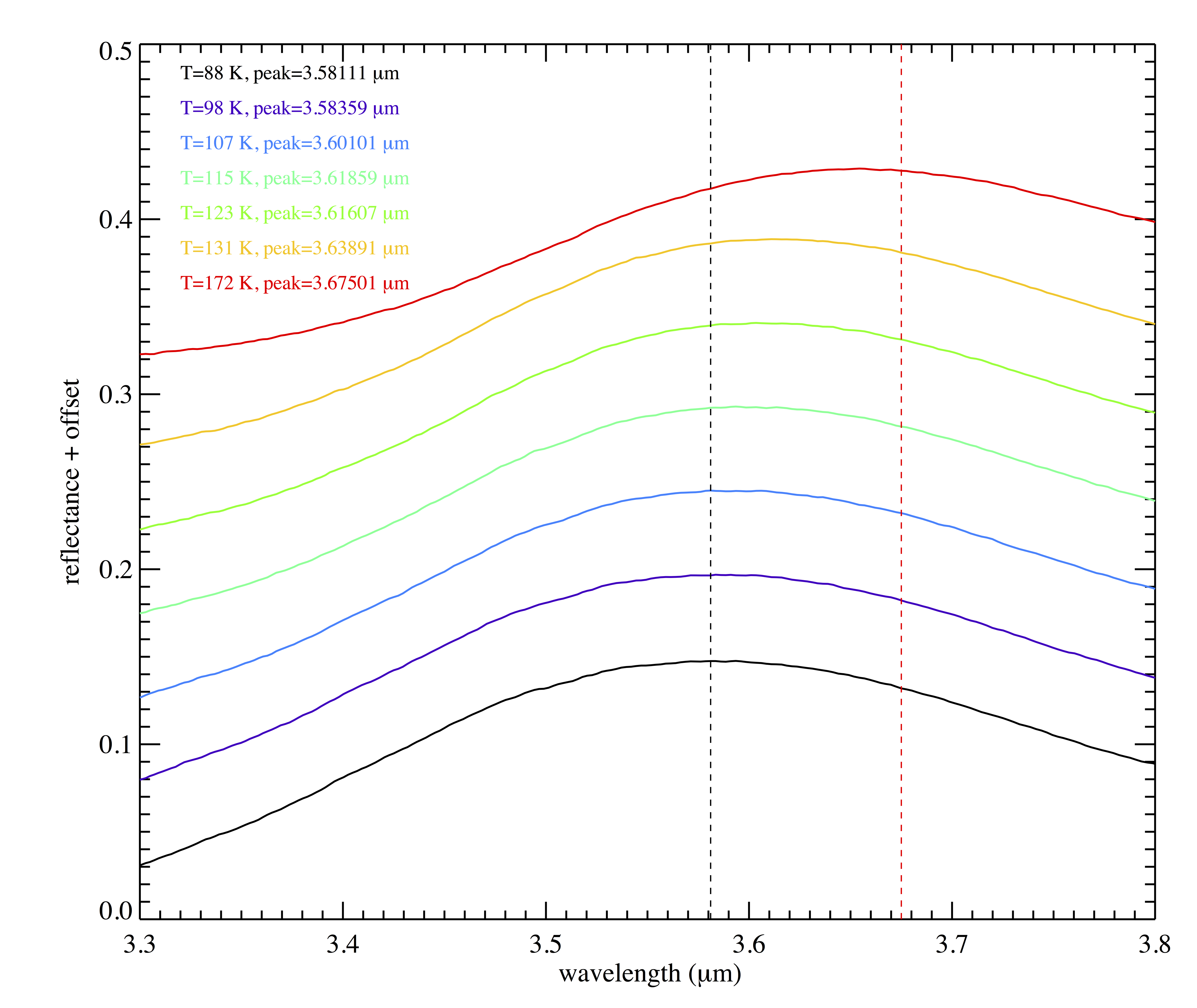}
		\caption{Water ice reflectance measured at temperature 88 $\le T \le$ 172 K. The best-fit wavelength of the 3.6 $\mu m$ peak is reported for each temperature. The peak position at T=88 K and T 172 K is indicated by vertical black and red dashed lines, respectively.} 
	\label{fig:figure2}
\end{figure}

The same trend is observed on reflectance measurements performed by \cite{Clark2012} on small grains of pure water ice illuminated at standard conditions (phase=30$^\circ$) for temperatures of the sample varying between T=88 K and T=172 K (Fig. \ref{fig:figure2}). The analysis of these data demonstrates that the 3.6 $\mu m$ reflectance peak shifts towards shorter wavelengths when the ice sample is cooled, moving from 3.675 $\mu m$ at T=172 K to 3.58 $\mu m$ at T=88 K. This means a shift of about 90 nm corresponding to about 6 VIMS spectral bands.

\section{VIMS as a thermal mapper}\label{sct:b}
\begin{figure}[h!]
	\centering
		\includegraphics[width=18cm]{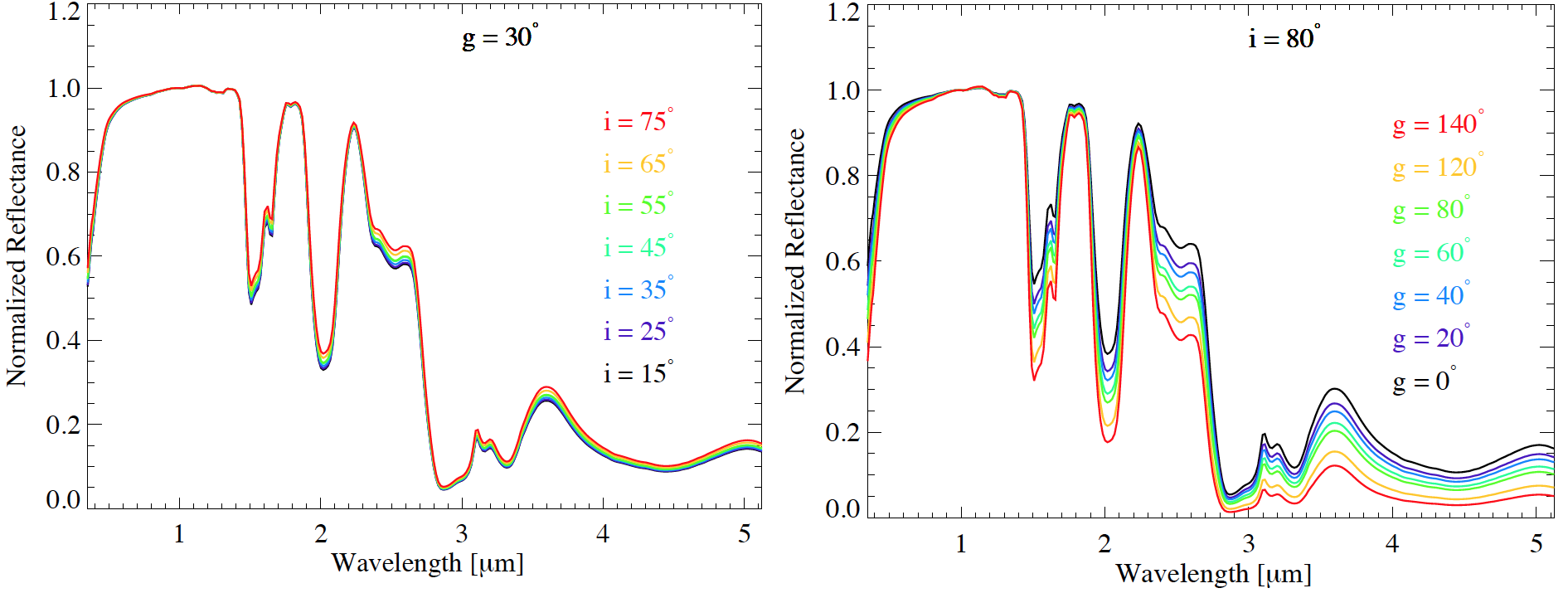}
		\caption{Water ice synthetic normalized reflectances of pure water ice grains of 30 $\mu m$ diameter at fixed phase angle g=30$^\circ$ and incidence angles i=15$^\circ$, 25$^\circ$, 35$^\circ$, 45$^\circ$, 55$^\circ$, 65$^\circ$, 75$^\circ$ (left panel) and at fixed incidence angle i=80$^\circ$ at phase g=0$^\circ$, 20$^\circ$, 40$^\circ$, 60$^\circ$, 80$^\circ$, 120$^\circ$, 140$^\circ$ (right panel). Simulation performed with the Monte Carlo ray-tracing formalism discussed in \cite{Ciarniello2014} by using water ice optical constants at T=90 K \citep{Warren1984, Mastrapa2009, Clark2012}.} 
	\label{fig:figure3}
\end{figure}

Starting from the experimental evidences discussed in the previous section, a 4th-degree polynomial fit in the 3.2-3.8 $\mu m$ range has been applied to laboratory measurements shown in Fig. \ref{fig:figure2} to determine the wavelength where the peak occurs with the view toward using it as a marker to retrieve similar temperatures on the regular icy satellites of Saturn. While band depths and I/F level are both grain size-dependent, the position of the 3.6 $\mu m$ peak is insensitive to grain size and illumination geometry: this effect is well shown on synthetic spectra of pure water ice grains of 10-100 $\mu m$ diameters discussed in \cite{Filacchione2012}, Fig. 5. Moreover, the 3.6 $\mu m$ peak position is not affected by illumination geometry as shown on synthetic VIS-IR reflectance spectra of pure water ice grains of 30 $\mu m$ diameter simulated at fixed phase angle g=30$^\circ$ and incidence angles i=15$^\circ$, 25$^\circ$, 35$^\circ$, 45$^\circ$, 55$^\circ$, 65$^\circ$, 75$^\circ$ (Fig. \ref{fig:figure3}, left panel) and at fixed incidence angle i=80$^\circ$ at phase g=0$^\circ$, 20$^\circ$, 40$^\circ$, 60$^\circ$, 80$^\circ$, 120$^\circ$, 140$^\circ$ (Fig. \ref{fig:figure3}, right panel). In both cases the synthetic spectra are simulated following the Monte Carlo ray-tracing formalism discussed in \cite{Ciarniello2014} by using water ice optical constants at T=90 K measured by \cite{Warren1984, Mastrapa2009, Clark2012}. At fixed phase, changes of the incidence angle have small effects on the continuum level and on band depths. Conversely, fixing the incidence angle and varying the phase, large variations are induced: in fact when the phase angle increases the 1.5, 2.0 and 2.9 $\mu m$ absorptions become larger while the continuum level decreases. In both cases, however, the position of the 3.6 $\mu m$ is not changing being it only dependent from temperature through the imaginary index of the optical constant, as previously shown in Fig. \ref{fig:figure1}.

The same fitting technique used for laboratory data has been applied to VIMS I/F data. The retrieval of the temperature is therefore performed through the comparison of the 3.6 $\mu m$ peak wavelength between VIMS and laboratory data.  
The position of the peak has been derived by applying a 4th degree polynomial fit to VIMS I/F data in the 3.2-3.8 $\mu m$ range and then determining the position of the peak on the fitted curve. A sampling equal to 1/3 of VIMS band width has been used to determine the peak wavelength, corresponding to a temperature's error of about $\pm$5 K to be associated to all results reported hereafter. Data filtering, as discussed later in section 6, allow us to apply this method on pixels having a typical SNR$>$25 at 3.6 $\mu m$. The error resulting from the presence of other compositional endmembers is not included. 
As a general rule, the spectral position and shape of the 3.6 $\mu m$ peak can be altered only in case of water ice mixed with materials having absorption features in this spectral range. As an example, tholins \citep{Imanaka2012} and organic materials are characterized by a broad absorption feature in the 3.2-3.6 $\mu m$ range which could affect the peak properties and therefore the retrieval of the temperature. Conversely, amorphous carbon \citep{Zubko1996} is featureless at these wavelengths: while small amount of carbon mixed within water ice reduces significantly the reflectance level, being spectrally neutral it is not able to alter the peak position.

The same temperature retrieval method has been already applied to several VIMS mosaics of Saturn's ring and the results were compared with CIRS observations occurring at the same time \citep{Filacchione2014}. In general the temperatures measured by VIMS with this method are higher than corresponding ones reported by CIRS. This is a consequence of the shallow layer (few microns) to which VIMS is sensitive compared to CIRS which measures temperature at greater depth, from a few millimeters to a centimeter. Since in this work data taken around noon time have been selected, the shallower skin depth sampled by VIMS could be the reason of the higher temperatures inferred. 

Moreover, both instruments are influenced by surface roughness and sub-pixel shadows but the discussion of these effects goes well beyond the scopes of this work. Also grain size and amount of contaminants embedded in water ice play a role in the 3.6 $\mu m$ peak's properties and these effects have yet to be investigated. Certainly, combining VIMS and CIRS measurements together will allow us to better understand the regolith's physical properties and heat transport mechanisms. The outer layer in fact is warmer than the material  below it when measured around noon as the surface receive the maximum of energy that the subsurface will later get by conduction. It is noteworthy that the method based on the 3.6 $\mu m$ peak fit greatly differs with respect to the derivation of the temperature from thermal infrared radiance fitting with a blackbody law. In the first case water ice is used as a proxy and spectral properties measured in laboratory conditions are used to infer the temperature of the satellites' surfaces. The major advantages of this method are the wide excursion of the peak as function of temperature, which allows to derive its spectral position with great precision and the good signal to noise ratio in the 3.6 $\mu m$ range in VIMS spectra. The drawback is the effect of contaminants in changing the shape and position of the water ice feature which is difficult to characterize without specific laboratory measurements. In the second case the temperature is derived by fitting radiance in the thermal range with a blackbody function. While this second method is in principle more robust and working well for warm objects orbiting close to the Sun \citep{Coradini2011}, it can be scarcely used to derive icy satellites temperature from VIMS data because the thermal emission is lower than the solar contribution in the 3-5 $\mu m$ range and signal to noise is poor. Up to the present time, only \cite{Goguen2013} have successfully used the blackbody fit to derive Baghdad Sulcus color temperature $T_c$=197$\pm$20 K from VIMS high spatial resolution data acquired above Enceladus' south pole region during the 14$^{th}$ April 2012 South Pole Cassini's flyby.

In the next sections we discuss about the results obtained by applying the 3.6 $\mu m$ fit method to three different VIMS datasets:
\begin{enumerate}
\item a collection of 240 disk-integrated observations of Saturn's regular satellites collected by VIMS between 2004 and 2011 with solar phase in the 20$^\circ$-40$^\circ$ range, corresponding to late morning-early afternoon local times is described in section \ref{sct:c}. This dataset is suitable to analyze average surface temperature at global scale and to retrieve temperature variations across the leading and trailing hemispheres. 
\item Disk-resolved observations (20-40 km/pixel) of Enceladus and Tethys, discussed in section \ref{sct:d}, which are suitable to map temperature variations across geomorphological regional features, such as Enceladus' tiger stripes or Tethys equatorial dark lens. 
\item Global temperature maps are derived for Mimas, Enceladus, Tethys, Dione, Rhea, Hyperion and Iapetus by using the entire VIMS geo-referenced dataset. These results are given in section \ref{sct:e}.  

\end{enumerate}

\section{Icy satellites disk-integrated temperatures.}\label{sct:c}
\begin{figure}[h!]
	\centering
		\includegraphics[width=14cm]{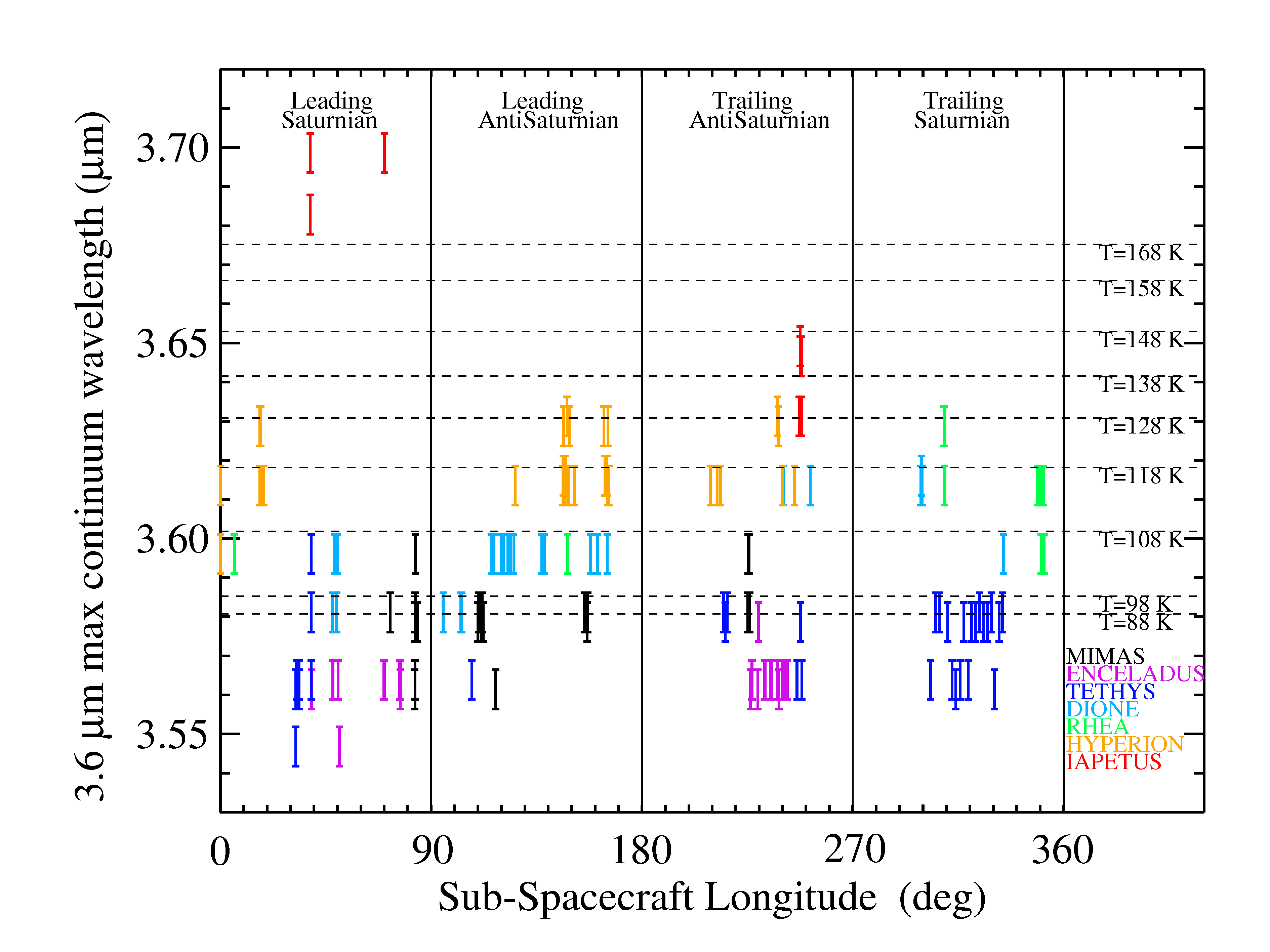}
		\caption{Wavelength of the 3.6 $\mu m$ peak measured on the icy satellites as function of the sub-spacecraft longitude. On the left vertical axis is reported the water ice temperature scale.} 
	\label{fig:figure4}
\end{figure}

A first temperature retrieval for all regular satellites from Mimas to Iapetus is carried out on the dataset listed in Tables \ref{table:1}-\ref{table:7} which includes observations taken with solar phase in the 20$^\circ$-40$^\circ$ range, e.g., at late morning-early afternoon conditions, selected from the dataset already discussed in \cite{Filacchione2012}. The filtering on phase is imposed in order to match the laboratory 30$^\circ$ phase illumination condition \citep{Clark2012} for which the 3.6 $\mu m$ peak position is calibrated. By fitting the position of the 3.6 $\mu$m peak on those disk-integrated I/F spectra we have measured average daytime temperatures of T=88 K for Mimas, T$<<$88 K for Enceladus, T$<$88 K for Tethys, T=100 K for Dione, T=108 K for Rhea, T=113 K for Hyperion, T=138K and T$>$168K for Iapetus trailing and leading hemispheres, respectively. The distribution of the temperature as a function of the subspacecraft longitude on the 240 observations processed is shown in Fig. \ref{fig:figure4}. Three different trends are visible in the scatterplot: 1) constant temperature is observed independently from the observed hemisphere on Mimas, Enceladus and Hyperion; 2) On Tethys, Dione and Rhea, for which observations taken on both leading and trailing hemispheres are available, temperatures are higher by about 10 K on the trailing hemisphere than on the leading hemisphere. In this case higher temperature is correlated with the prevalence of lower albedo terrain units on the trailing hemispheres of these moons; 3) On Iapetus the higher temperature is observed on the dark leading than on the bright trailing hemisphere. The lack of laboratory measurements for T$<88$ K and T$>168$ K allows us only to give upper limits to Mimas, Enceladus, Tethys temperatures and a lower limit to Iapetus leading hemisphere temperature, respectively.

\begin{table}
\centering  
\begin{tiny}   
\begin{tabular}{|c|c|c|c|c|c|c|c|c|c|c|}      
\hline\hline      
SEQUENCE & START-END & S & L & EXP & PHASE & SUBSPC  & SUBSPC & SSOL  & SSOL & SPC ALT \\
OBS ID   & TIME (UT)      &   &   & (s) &($^\circ$) & LON ($^\circ$) & LAT ($^\circ$) & LON ($^\circ$) & LAT ($^\circ$) & (km) \\   
\hline
S25-V1542757940 & 2006-324T23:19:59-23:20:14 & 12 & 12 & 80 & 38.7 &      225.1 &     -47.4 &      206.1 &     -13.1 &     152554. \\
S25-V1542757954 & 2006-324T23:20:14-23:20:28 & 12 & 12 & 80 & 38.7 &      225.1 &     -47.4 &      206.1 &     -13.1 &      152554. \\
S25-V1542757969 & 2006-324T23:20:29-23:20:43 & 12 & 12 & 80 & 38.7 &      225.1 &     -47.4 &      206.1 &     -13.1 &      152554. \\
S25-V1542757998 & 2006-324T23:20:58-23:21:12 & 12 & 12 & 80 & 38.4 &      225.3 &     -47.1 &      206.3 &     -13.1 &      152444. \\
S25-V1542758013 & 2006-324T23:21:12-23:21:27 & 12 & 12 & 80 & 38.4 &      225.3 &     -47.1 &      206.3 &     -13.1 &      152444. \\
S25-V1542758042 & 2006-324T23:21:42-23:21:56 & 12 & 12 & 80 & 38.4 &      225.3 &     -47.1 &      206.3 &     -13.1 &      152444. \\
S25-V1542758057 & 2006-324T23:21:56-23:22:11 & 12 & 12 & 80 & 38.1 &      225.5 &     -46.8 &      206.6 &     -13.1 &      152339. \\
S25-V1542758071 & 2006-324T23:22:11-23:22:25 & 12 & 12 & 80 & 38.1 &      225.5 &     -46.8 &      206.6 &     -13.1 &      152339. \\
S25-V1542758086 & 2006-324T23:22:26-23:22:40 & 12 & 12 & 80 & 38.1 &      225.5 &     -46.8 &      206.6 &     -13.1 &      152339. \\
S25-V1542758101 & 2006-324T23:22:40-23:22:55 & 12 & 12 & 80 & 38.1 &      225.5 &     -46.8 &      206.6 &     -13.1 &      152339. \\
S25-V1542758115 & 2006-324T23:22:55-23:23:09 & 12 & 12 & 80 & 37.8 &      225.7 &     -46.5 &      206.8 &     -13.1 &      152238. \\
S25-V1542758130 & 2006-324T23:23:09-23:23:24 & 12 & 12 & 80 & 37.8 &      225.7 &     -46.5 &      206.8 &     -13.1 &      152238. \\
S25-V1542758144 & 2006-324T23:23:24-23:23:38 & 12 & 12 & 80 & 37.8 &      225.7 &     -46.5 &      206.8 &     -13.1 &      152238. \\
S25-V1542758159 & 2006-324T23:23:39-23:23:53 & 12 & 12 & 80 & 37.8 &      225.7 &     -46.5 &      206.8 &     -13.1 &      152238. \\
S25-V1542758174 & 2006-324T23:23:53-23:24:08 & 12 & 12 & 80 & 37.5 &      225.9 &     -46.1 &      207.1 &     -13.2 &      152143. \\
S25-V1542758188 & 2006-324T23:24:08-23:24:22 & 12 & 12 & 80 & 37.5 &      225.9 &     -46.1 &      207.1 &     -13.2 &      152143. \\
S30-V1558932227 & 2007-147T04:09:45-04:09:59 & 12 & 12 & 80 & 24.4 &      72.5 &      -34.1 &      67.1 &     -10.9 &      207129. \\
S30-V1558941481 & 2007-147T06:43:58-06:44:24 & 24 & 12 & 80 & 26.0 &      109.9 &     -35.8 &      113.3 &     -10.9 &      224918. \\
S30-V1558941534 & 2007-147T06:44:51-06:45:16 & 24 & 12 & 80 & 26.0 &      110.2 &     -35.8 &      113.6 &     -10.9 &      224968. \\
S30-V1558941566 & 2007-147T06:45:23-06:46:02 & 24 & 12 & 120 &  26.0 &      110.6 &     -35.8 &      113.9 &     -10.9 &      225018. \\
S30-V1558941605 & 2007-147T06:46:03-06:46:41 & 24 & 12 & 120 &  26.0 &      110.9 &     -35.8 &      114.2 &     -10.9 &      225068. \\
S30-V1558941644 & 2007-147T06:46:42-06:47:20 & 24 & 12 & 120 &  26.0 &      111.2 &     -35.8 &      114.4 &     -10.9 &      225119. \\
S30-V1558941692 & 2007-147T06:47:29-06:48:08 & 24 & 12 & 120 &  26.0 &      111.5 &     -35.8 &      114.7 &     -10.9 &      225169. \\
S30-V1558941731 & 2007-147T06:48:09-06:48:47 & 24 & 12 & 120 &  26.0 &      111.8 &     -35.7 &      115.0 &     -10.9 &      225219. \\
S30-V1558941770 & 2007-147T06:48:48-06:49:26 & 24 & 12 & 120 &  26.0 &      112.2 &     -35.7 &      115.3 &     -10.8 &      225269. \\
S30-V1558949456 & 2007-147T08:56:53-08:57:19 & 24 & 12 & 80 &  25.0 &      155.5 &     -31.9 &      149.1 &     -10.1 &      238232. \\
S30-V1558949482 & 2007-147T08:57:20-08:57:45 & 24 & 12 & 80 &  25.0 &      155.8 &     -31.8 &      149.3 &     -10.1 &      238445. \\
S30-V1558949509 & 2007-147T08:57:46-08:58:11 & 24 & 12 & 80 &  25.0 &      156.2 &     -31.8 &      149.6 &     -10.1 &      238662. \\
S30-V1558949535 & 2007-147T08:58:12-08:58:38 & 24 & 12 & 80 &  25.0 &      156.5 &     -31.7 &      149.8 &     -10.1 &      238880. \\
S30-V1558949561 & 2007-147T08:58:38-08:59:04 & 24 & 12 & 80 &  25.0 &      156.9 &     -31.7 &      150.1 &     -10.1 &      239102. \\
S57-V1644780761 & 2010-044T18:48:38-18:49:58 & 30 & 30 & 80 & 27.1 &      117.5 &     -10.6 &      141.7 &      1.8 &      30537. \\
S66-V1675134165 & 2011-031T02:15:11-02:15:37 & 24 & 12 & 80 &  39.0 &      84.1 &     -2.5 &      46.2 &      7.0 &      139223. \\
S66-V1675134191 & 2011-031T02:15:38-02:16:03 & 24 & 12 & 80 &  38.7 &      84.0 &     -2.5 &      46.4 &      7.0 &      139266. \\
S66-V1675135681 & 2011-031T02:40:27-02:40:53 & 24 & 12 & 80 &  31.9 &      83.3 &     -2.6 &      52.8 &      7.0 &      140809. \\
S66-V1675135707 & 2011-031T02:40:54-02:41:19 & 24 & 12 & 80 &  31.6 &      83.3 &     -2.6 &      53.1 &      7.0 &      140892. \\
S66-V1675135741 & 2011-031T02:41:27-02:41:53 & 24 & 12 & 80 &  31.6 &      83.3 &     -2.6 &      53.1 &      7.0 &      140892. \\
S66-V1675135767 & 2011-031T02:41:54-02:42:19 & 24 & 12 & 80 &  31.4 &      83.2 &     -2.6 &      53.3 &      7.0 &      140977. \\
S66-V1675137246 & 2011-031T03:06:32-03:06:58 & 24 & 12 & 80 &  25.3 &      83.0 &     -2.8 &      59.7 &      7.0 &      143351. \\
S66-V1675137272 & 2011-031T03:06:59-03:07:24 & 24 & 12 & 80 &  25.0 &      83.0 &     -2.8 &      60.0 &      7.0 &      143462. \\
S66-V1675137332 & 2011-031T03:07:59-03:08:24 & 24 & 12 & 80 &  24.8 &      83.0 &     -2.8 &      60.2 &      7.0 &      143574. \\
S66-V1675138026 & 2011-031T03:19:32-03:19:58 & 24 & 12 & 80 &  22.3 &      83.1 &     -2.8 &      63.1 &      7.0 &      144851. \\
S66-V1675138052 & 2011-031T03:19:59-03:20:24 & 24 & 12 & 80 &  22.0 &      83.1 &     -2.8 &      63.4 &      7.0 &      144971. \\
\hline \hline
\end{tabular}
\end{tiny}
\caption{Mimas disk-integrated observations processed in this work. Columns S and L indicate the cube dimensions along sample and lines axis, respectively.}  
\label{table:1}                   
\end{table}

\begin{table}
\centering  
\begin{tiny}   
\begin{tabular}{|c|c|c|c|c|c|c|c|c|c|c|}      
\hline\hline      
SEQUENCE & START-END & S & L & EXP & PHASE & SUBSPC  & SUBSPC & SSOL  & SSOL & SPC ALT \\
OBS ID   & TIME (UT)     &   &   & (s) &($^\circ$) & LON ($^\circ$) & LAT ($^\circ$) & LON ($^\circ$) & LAT ($^\circ$) & (km) \\   
\hline
S17-V1516162569 & 2006-017T03:46:39-03:47:46 & 32 & 24 & 80 & 23.7  &    226.3 &  -0.3 &    241.7 &   -18.3 &    148814. \\
S17-V1516162727 & 2006-017T03:49:17-03:50:24 & 32 & 24 & 80 & 23.8  &    226.8 &  -0.3 &    242.4 &   -18.3 &    148805. \\
S17-V1516162803 & 2006-017T03:50:33-03:51:40 & 32 & 24 & 80 & 23.8  &    227.0 &  -0.3 &    242.6 &   -18.3 &    148805. \\
S17-V1516163887 & 2006-017T04:08:37-04:10:48 & 48 & 32 & 80 & 24.4  &    229.4 &  -0.3  &   246.1  &  -18.3   &  148929. \\
S17-V1516164030 & 2006-017T04:11:00-04:13:11 & 48 & 32 & 80 & 24.5  &    229.8 &  -0.3  &   246.7  &  -18.3   &  148973. \\ 
S17-V1516165299 & 2006-017T04:32:09-04:34:20 & 48 & 32 & 80 & 25.4  &    232.4 &  -0.2  &   250.7  &  -18.4   &  149454. \\
S17-V1516165442 & 2006-017T04:34:32-04:36:43 & 48 & 32 & 80 & 25.5  &     232.6 &  -0.2 &    251.1 &   -18.4  &   149515. \\
S17-V1516166711 & 2006-017T04:55:40-04:57:52 & 48 & 32 & 80 & 26.4  &    234.8  & -0.2  &   254.8  &  -18.4   &  150235. \\
S17-V1516166854 & 2006-017T04:58:03-05:00:15 & 48 & 32 & 80 & 26.8  &    235.4  & -0.1  &   255.9  &  -18.5   &  150512. \\
S17-V1516168270 & 2006-017T05:21:39-05:23:51 & 48 & 32 & 80 & 28.1  &    237.7  & -0.1  &   260.2  &  -18.5   &  151732. \\
S17-V1516168413 & 2006-017T05:24:02-05:26:14 & 48 & 32 & 80 & 28.3  &    237.9  & -0.1  &   260.8  &  -18.5   &  151923. \\
S17-V1516168802 & 2006-017T05:30:31-05:31:59 & 36 & 28 & 80 & 28.7  &    238.5  & -0.1  &   262.0  &  -18.5   &  152323. \\
S17-V1516169654 & 2006-017T05:44:43-05:46:11 & 36 & 28 & 80 & 29.8  &    239.8  & -0.1  &   264.7  &  -18.5   &  153352. \\
S17-V1516170506 & 2006-017T05:58:55-06:00:23 & 36 & 28 & 80 & 30.9  &    240.9  & -0.0  &   267.5  &  -18.5   &  154515. \\
S17-V1516171358 & 2006-017T06:13:07-06:14:35 & 36 & 28 & 80 & 32.2  &    242.1  & -0.0  &   270.4  &  -18.5   &  155913. \\
S55-V1637469704 & 2009-325T03:58:33-03:59:21 & 24 & 22 & 80 & 35.3  &     38.9  & -1.9  &    74.6  &    1.6   &  51731. \\
S60-V1652861972 & 2010-138T07:34:32-07:35:31 & 30 & 30 & 60 & 28.0  &     48.0  & -1.0  &    75.5  &    4.2   &  35699. \\
S60-V1652871322 & 2010-138T10:10:22-10:11:01 & 30 & 15 & 80 & 34.4  &     69.7  & -0.5  &   103.8  &    4.2   &  94938. \\
S60-V1652871418 & 2010-138T10:11:58-10:12:37 & 30 & 15 & 80 & 34.5  &     70.0  & -0.5  &   104.2  &    4.2   &  95611. \\
S60-V1652874331 & 2010-138T11:00:31-11:01:10 & 30 & 15 & 80 & 36.7  &     76.5  & -0.4  &   112.9  &    4.2   &  110604. \\
S60-V1652874379 & 2010-138T11:01:19-11:01:58 & 30 & 15 & 80 & 36.7  &     76.6  & -0.4  &   113.1  &    4.2   &  110891. \\
S60-V1652874427 & 2010-138T11:02:07-11:02:46 & 30 & 15 & 80 & 36.8  &     76.7  & -0.4  &   113.3  &    4.2   &  111177. \\
S60-V1652874475 & 2010-138T11:02:55-11:03:34 & 30 & 15 & 80 & 36.8  &     76.9  & -0.4  &   113.5  &    4.2   &  111463. \\
S70-V1696181352 & 2011-274T16:39:25-16:40:22 & 36 & 18 & 80 & 24.6  &     50.3  & -0.5  &    28.6  &   11.3   &  77778. \\
S70-V1696181623 & 2011-274T16:43:56-16:44:53 & 36 & 18 & 80 & 24.4  &     50.8  & -0.5  &    29.3  &   11.3   &  79700. \\
\hline \hline
\end{tabular}
\end{tiny}
\caption{Enceladus disk-integrated observations processed in this work. Columns S and L indicate the cube dimensions along sample and lines axis, respectively.}  
\label{table:2}                   
\end{table}

\begin{table}
\centering  
\begin{tiny}   
\begin{tabular}{|c|c|c|c|c|c|c|c|c|c|c|}      
\hline\hline      
SEQUENCE & START-END & S & L & EXP & PHASE & SUBSPC  & SUBSPC & SSOL  & SSOL & SPC ALT \\
OBS ID   & TIME (UT) &   &   & (s) &($^\circ$) & LON ($^\circ$) & LAT ($^\circ$) & LON ($^\circ$) & LAT ($^\circ$) & (km) \\   
\hline
S09-V1489020423 & 2005-068T00:20:27-00:22:55 & 48 & 36 & 80 & 23.5  &   214.9  &     1.3 &  208.6 & -20.7  &   214413. \\
S09-V1489020590 & 2005-068T00:23:14-00:25:42 & 48 & 36 & 80 & 23.6  &   215.5  &     1.3 &  209.0 & -20.7  &   213687. \\
S09-V1489020757 & 2005-068T00:26:01-00:28:29 & 48 & 36 & 80 & 23.7  &   216.3  &     1.3 &  209.5 & -20.7  &   212736. \\
S22-V1532362574 & 2006-204T15:45:00-15:45:52 & 24 & 24 & 80 & 24.3  &   107.4  &   -37.5 &   93.6 & -16.4  &   130244. \\
S30-V1558913033 & 2007-146T22:49:51-22:50:30 & 24 & 24 & 60 & 39.1  &   38.8   &     2.2 &    3.3 & -12.8  &   127606. \\
S30-V1558913501 & 2007-146T22:57:39-22:58:56 & 24 & 24 & 120 & 37.3  &   38.7   &     0.4 &    4.3 & -12.8  &   131768. \\
S30-V1558913605 & 2007-146T22:59:23-22:59:55 & 24 & 24 & 50 & 36.9  &   38.7   &     0.0 &    4.6 & -12.8  &   132841. \\
S36-V1577490717 & 2007-361T23:15:52-23:18:52 & 32 & 32 & 160 & 35.8  &   303.2  &    21.3 &  320.4 &  -9.8  &   2307240. \\
S36-V1577491678 & 2007-361T23:31:53-23:34:53 & 32 & 32 & 160 & 35.9  &   305.4  &    21.2 &  322.7 &  -9.8  &   2316120. \\ 
S36-V1577492448 & 2007-361T23:44:43-23:47:43 & 32 & 32 & 160 & 35.9  &   307.0  &    21.1 &  324.4 &  -9.8  &   2322380. \\
S36-V1577494179 & 2007-362T00:13:34-00:16:34 & 32 & 32 & 160 & 36.0  &   310.5  &    21.0 &  328.2 &  -9.7  &   2335900. \\
S36-V1577495140 & 2007-362T00:29:35-00:32:35 & 32 & 32 & 160 & 36.0  &   312.4  &    20.9 &  330.3 &  -9.7  &   2343070. \\
S36-V1577495910 & 2007-362T00:42:25-00:45:25 & 32 & 32 & 160 & 36.1  &   313.9  &    20.8 &  332.0 &  -9.7  &   2348740. \\
S36-V1577496871 & 2007-362T00:58:26-01:01:26 & 32 & 32 & 160 & 36.2  &   315.8  &    20.8 &  334.1 &  -9.7  &   2355520. \\
S36-V1577497641 & 2007-362T01:11:16-01:14:16 & 32 & 32 & 160 & 36.2  &   317.3  &    20.7 &  335.8 &  -9.7  &   2360860. \\
S36-V1577498602 & 2007-362T01:27:17-01:30:17 & 32 & 32 & 160 & 36.3  &   319.2  &    20.6 &  337.9 &  -9.7  &   2367210. \\
S36-V1577499372 & 2007-362T01:40:07-01:43:07 & 32 & 32 & 160 & 36.4  &   320.7  &    20.6 &  339.6 &  -9.7  &   2372200. \\
S36-V1577500333 & 2007-362T01:56:08-01:59:08 & 32 & 32 & 160 & 36.5  &   322.5  &    20.5 &  341.5 &  -9.7  &   2377760. \\
S36-V1577501103 & 2007-362T02:08:58-02:11:58 & 32 & 32 & 160 & 36.5  &   324.1  &    20.5 &  343.3 &  -9.7  &   2382740. \\
S36-V1577502064 & 2007-362T02:24:59-02:27:59 & 32 & 32 & 160 & 36.6  &   325.7  &    20.5 &  345.1 &  -9.7  &   2387530. \\
S36-V1577502834 & 2007-362T02:37:49-02:40:49 & 32 & 32 & 160 & 36.7  &   327.4  &    20.4 &  347.1 &  -9.7  &   2392440. \\
S36-V1577503795 & 2007-362T02:53:50-02:56:50 & 32 & 32 & 160 & 36.8  &   329.1  &    20.4 &  349.0 &  -9.7  &   2397110. \\
S36-V1577504565 & 2007-362T03:06:40-03:09:40 & 32 & 32 & 160 & 36.9  &   330.5  &    20.3 &  350.6 &  -9.7  &   2406700. \\
S36-V1577505526 & 2007-362T03:22:41-03:25:41 & 32 & 32 & 160 & 37.1  &   332.4  &    20.3 &  352.7 &  -9.7  &   2405450. \\
S36-V1577506296 & 2007-362T03:35:31-03:38:31 & 32 & 32 & 160 & 37.2  &   333.9  &    20.3 &  354.4 &  -9.7  &   2408900. \\
S40-V1587465454 & 2008-112T10:00:19-10:00:31 & 24 & 12 & 40 & 36.4  &   32.1   &  -38.2  &   11.3 &  -7.7  &    572815. \\
S40-V1587465468 & 2008-112T10:00:32-10:00:45 & 24 & 12 & 40 & 36.4  &   32.2   &  -38.2  &   11.4 &  -7.7  &    573143. \\
S40-V1587465844 & 2008-112T10:06:49-10:07:01 & 24 & 12 & 40 & 36.1  &   32.5   &  -38.1  &   12.0 &  -7.7  &    574770. \\
S40-V1587465858 & 2008-112T10:07:02-10:07:15 & 24 & 12 & 40 & 36.1  &   32.5   &  -38.1  &   12.2 &  -7.7  &    575093. \\
S40-V1587466234 & 2008-112T10:13:19-10:13:31 & 24 & 12 & 40 & 35.8  &   32.8   &  -38.0  &   12.9 &  -7.7  &    577005. \\
S40-V1587466248 & 2008-112T10:13:32-10:13:45 & 24 & 12 & 40 & 35.8  &   32.9   &  -38.0  &   13.1 &  -7.7  &    577321. \\
S40-V1587466624 & 2008-112T10:19:49-10:20:01 & 24 & 12 & 40 & 35.5  &   33.2   &  -38.0  &   13.8 &  -7.7  &    579190. \\
S40-V1587466638 & 2008-112T10:20:02-10:20:15 & 24 & 12 & 40 & 35.5  &   33.3   &  -38.0  &   14.0 &  -7.7  &    579498. \\
S40-V1587467014 & 2008-112T10:26:19-10:26:31 & 24 & 12 & 40 & 35.3  &   33.6   &  -37.9  &   14.6 &  -7.7  &    581022. \\
S40-V1587467028 & 2008-112T10:26:32-10:26:45 & 24 & 12 & 40 & 35.2  &   33.6   &  -37.9  &   14.7 &  -7.7  &    581324. \\
S54-V1634216514 & 2009-287T12:19:07-12:19:28 & 42 & 22 & 20 & 27.6  &   248.2  &    2.5  &  220.2 &   2.1  &    123558. \\
S54-V1634216781 & 2009-287T12:23:34-12:23:55 & 42 & 22 & 20 & 26.5  &   247.7  &    2.4  &  220.8 &   2.1  &    125888. \\
S54-V1634217726 & 2009-287T12:39:19-12:39:40 & 42 & 22 & 20 & 22.9  &   246.1  &    2.3  &  222.9 &   2.1  &    134467. \\
\hline \hline
\end{tabular}
\end{tiny}
\caption{Tethys disk-integrated observations processed in this work. Columns S and L indicate the cube dimensions along sample and lines axis, respectively.}  
\label{table:3}                   
\end{table}

\begin{table}
\centering  
\begin{tiny}   
\begin{tabular}{|c|c|c|c|c|c|c|c|c|c|c|}      
\hline\hline      
SEQUENCE & START-END & S & L & EXP & PHASE & SUBSPC  & SUBSPC & SSOL  & SSOL & SPC ALT \\
OBS ID   & TIME (UT)  &   &   & (s) &($^\circ$) & LON ($^\circ$) & LAT ($^\circ$) & LON ($^\circ$) & LAT ($^\circ$) & (km) \\   
\hline
S13-V1501609705  & 2005-213T17:20:29-17:23:38 & 42 & 30 & 140 &  39.9  &    251.8   &  -41.1  &   210.9 &  -20.8   &   273999. \\
S13-V1501605232  & 2005-213T16:05:56-16:07:39 & 24 & 24 & 160 &  37.4  &    240.3   &  -43.0  &   204.0 &  -20.8   &   287995. \\
S15-V1507717998  & 2005-284T10:04:43-10:06:12 & 32 & 32 & 80 &  21.9  &    137.2   &   -0.5  &   147.3 &  -20.1   &   281504. \\
S15-V1507718096  & 2005-284T10:06:21-10:07:50 & 32 & 32 & 80 &  21.9  &    137.4   &   -0.5  &   147.5 &  -20.1   &   280126. \\
S15-V1507718735  & 2005-284T10:17:00-10:18:15 & 36 & 24 & 80 &  22.0  &    138.2   &   -0.5  &   148.5 &  -20.1   &   272575. \\
S15-V1507718819  & 2005-284T10:18:24-10:19:39 & 36 & 24 & 80 &  22.0  &    138.3   &   -0.5  &   148.6 &  -20.1   &   271890. \\
S15-V1507700309  & 2005-284T05:09:54-05:10:45 & 24 & 24 & 80 &  20.2  &    115.7   &   -0.3  &   120.4 &  -20.1   &   500654. \\
S15-V1507700483  & 2005-284T05:12:48-05:13:39 & 24 & 24 & 80 &  20.2  &    115.9   &   -0.3  &   120.7 &  -20.1   &   498285. \\
S15-V1507700542  & 2005-284T05:13:47-05:14:38 & 24 & 24 & 80 &  20.2  &    115.9   &   -0.3  &   120.8 &  -20.1   &   497496. \\
S15-V1507733247  & 2005-284T14:18:52-14:20:44 & 36 & 36 & 80 &  22.8  &    157.9   &   -0.8  &   170.5 &  -20.1   &   119191. \\
S15-V1507735292  & 2005-284T14:52:56-14:54:33 & 40 & 28 & 80 &  22.8  &    160.9   &   -0.9  &   173.6 &  -20.1   &   99308. \\
S15-V1507703774  & 2005-284T06:07:39-06:08:17 & 24 & 18 & 80 &  20.5  &    119.8   &   -0.4  &   125.7 &  -20.1   &   455298. \\
S15-V1507704605  & 2005-284T06:21:30-06:22:08 & 24 & 18 & 80 &  20.6  &    120.7   &   -0.4  &   127.0 &  -20.1   &   444499. \\
S15-V1507706267  & 2005-284T06:49:12-06:49:50 & 24 & 18 & 80 &  20.7  &    122.7   &   -0.4  &   129.4 &  -20.1   &   423845. \\
S15-V1507707228  & 2005-284T07:05:13-07:06:04 & 24 & 24 & 80 &  20.8  &    123.9   &   -0.4  &   131.0 &  -20.1   &   410961. \\
S15-V1507708306  & 2005-284T07:23:11-07:24:02 & 24 & 24 & 80 &  20.9  &    125.2   &   -0.4  &   132.6 &  -20.1   &   397423. \\
S15-V1507701040  & 2005-284T05:22:05-05:22:43 & 24 & 18 & 80 &  20.3  &    116.5   &   -0.3  &   121.5 &  -20.1   &   491190. \\
S15-V1507738123  & 2005-284T15:40:07-15:40:41 & 32 & 32 & 30 &  22.7  &    165.2   &   -1.1  &   177.9 &  -20.1   &   72415. \\
S17-V1514076858  & 2005-358T00:25:02-00:26:26 & 36 & 18 & 120 &  21.2  &    95.1    &   -0.4  &   105.3 &  -19.2   &   597903. \\
S22-V1532405053  & 2006-205T03:32:59-03:33:25 & 24 & 12 & 80 &  30.1  &    299.5   &    6.9  &   280.4 &  -16.5   &   261515. \\
S22-V1532405086  & 2006-205T03:33:32-03:34:24 & 24 & 12 & 160 &  30.0  &    299.4   &    6.9  &   280.5 &  -16.5   &   261452. \\
S22-V1532405149  & 2006-205T03:34:35-03:35:01 & 24 & 12 & 80 &  29.9  &    299.3   &    7.0  &   280.6 &  -16.5   &   261390. \\
S22-V1532405609  & 2006-205T03:42:15-03:42:41 & 24 & 12 & 80 &  29.6  &    299.0   &    7.2  &   281.0 &  -16.5   &   261164. \\
S34-V1569831278  & 2007-273T07:39:27-07:42:44 & 48 & 48 & 80 &  32.1  &    334.3   &    4.7  &     2.7 &  -10.4   &   63758. \\ 
S34-V1569853207  & 2007-273T13:44:55-13:46:20 & 36 & 18 & 120 &  23.1  &    47.8    &    9.4  &    35.9 &  -10.4   &   183882. \\
S34-V1569853808  & 2007-273T13:54:56-13:56:21 & 36 & 18 & 120 &  23.0  &    48.7    &    9.3  &    36.8 &  -10.4   &   187928. \\
S34-V1569854409  & 2007-273T14:04:57-14:06:22 & 36 & 18 & 120 &  22.9  &    49.5    &    9.2  &    37.7 &  -10.4   &   191945. \\
S34-V1569854506  & 2007-273T14:06:34-14:07:59 & 36 & 18 & 120 &  22.9  &    49.7    &    9.2  &    37.9 &  -10.4   &   192744. \\
S34-V1569854700  & 2007-273T14:09:48-14:11:13 & 36 & 18 & 120 &  22.8  &    50.0    &    9.2  &    38.3 &  -10.4   &   194339. \\
S70-V1696219767  & 2011-275T03:19:40-03:20:56 & 34 & 17 & 120 &  28.8  &    102.7   &    0.0  &   129.4 &   11.4   &   199816. \\
S70-V1696219855  & 2011-275T03:21:08-03:22:24 & 34 & 17 & 120 &  28.8  &    102.8   &    0.0  &   129.5 &   11.4   &   199916. \\
S70-V1696219943  & 2011-275T03:22:36-03:23:52 & 34 & 17 & 120 &  28.7  &    103.0   &    0.0  &   129.6 &   11.4   &   200017. \\
\hline \hline
\end{tabular}
\end{tiny}
\caption{Dione disk-integrated observations processed in this work. Columns S and L indicate the cube dimensions along sample and lines axis, respectively.}  
\label{table:4}                   
\end{table}

\begin{table}
\centering  
\begin{tiny}   
\begin{tabular}{|c|c|c|c|c|c|c|c|c|c|c|}      
\hline\hline      
SEQUENCE & START-END & S & L & EXP & PHASE & SUBSPC  & SUBSPC & SSOL  & SSOL & SPC ALT \\
OBS ID   & TIME (UT) &   &   & (s) &($^\circ$) & LON ($^\circ$) & LAT ($^\circ$) & LON ($^\circ$) & LAT ($^\circ$) & (km) \\   
\hline
S17-V1516202693 & 2006-017T14:55:22-14:56:28 & 42 & 36 & 40 &   34.7   &   309.1   &   0.3   &  338.7   &  -18.5  &    226163. \\
S17-V1516202765 & 2006-017T14:56:34-14:57:40 & 42 & 36 & 40 &   34.8   &   309.0   &   0.3   &  338.8   &  -18.5  &    225845. \\
S43-V1597403216 & 2008-227T10:28:30-10:30:00 & 32 & 16 & 160 &  27.7   &   348.9   &  -4.1   &  321.2   &  -5.3  & 1638460. \\
S43-V1597403307 & 2008-227T10:30:01-10:31:30 & 32 & 16 & 160 &   27.7   &   348.9   &  -4.1   &  321.3   &  -5.3  & 1638800. \\
S43-V1597404177 & 2008-227T10:44:31-10:46:01 & 32 & 16 & 160 &   27.4   &   349.4   &  -3.9   &  322.1   &  -5.3  & 1642060. \\
S43-V1597405905 & 2008-227T11:13:19-11:14:49 & 32 & 16 & 160 &   26.9   &   350.4   &  -3.7   &  323.7   &  -5.3  & 1648300. \\ 
S43-V1597405996 & 2008-227T11:14:50-11:16:19 & 32 & 16 & 160 &   26.9   &   350.5   &  -3.7   &  323.8   &  -5.3  & 1648620. \\
S43-V1597406099 & 2008-227T11:16:33-11:18:03 & 32 & 16 & 160 &   26.8   &   350.5   &  -3.7   &  323.9   &  -5.3  & 1648990. \\ 
S43-V1597407060 & 2008-227T11:32:34-11:34:04 & 32 & 16 & 160 &   26.5   &   351.1   &  -3.6   &  324.8   &  -5.3  & 1652310. \\
S43-V1597407151 & 2008-227T11:34:05-11:35:34 & 32 & 16 & 160 &   26.5   &   351.2   &  -3.6   &  324.8   &  -5.3  & 1652610. \\
S43-V1597407254 & 2008-227T11:35:48-11:37:18 & 32 & 16 & 160 &   26.5   &   351.2   &  -3.6   &  324.9   &  -5.3  & 1652970. \\
S43-V1597407345 & 2008-227T11:37:19-11:38:48 & 32 & 16 & 160 &   26.4   &   351.3   &  -3.6   &  325.0   &  -5.3  & 1653270. \\
S43-V1597407448 & 2008-227T11:39:02-11:40:32 & 32 & 16 & 160 &   26.4   &   351.3   &  -3.6   &  325.1   &  -5.3  & 1653620. \\
S43-V1597407539 & 2008-227T11:40:33-11:42:02 & 32 & 16 & 160 &   26.4   &   351.4   &  -3.6   &  325.2   &  -5.3  & 1653920. \\
S43-V1597407642 & 2008-227T11:42:16-11:43:01 & 32 & 16 & 80 &   26.3   &   351.4   &  -3.6   &  325.3   &  -5.3  & 1654200. \\
S43-V1597407688 & 2008-227T11:43:02-11:43:46 & 32 & 16 & 80 &   26.3   &   351.5   &  -3.5   &  325.3   &  -5.3  & 1654350. \\
S43-V1597407740 & 2008-227T11:43:54-11:44:39 & 32 & 16 & 80 &   26.3   &   351.5   &  -3.5   &  325.4   &  -5.3  & 1654530. \\
S43-V1597407786 & 2008-227T11:44:40-11:45:24 & 32 & 16 & 80 &   26.3   &   351.5   &  -3.5   &  325.4   &  -5.3  & 1654680. \\
S43-V1597407838 & 2008-227T11:45:32-11:46:17 & 32 & 16 & 80 &   26.3   &   351.6   &  -3.5   &  325.5   &  -5.3  & 1654850. \\
S43-V1597407884 & 2008-227T11:46:18-11:47:02 & 32 & 16 & 80 &   26.3   &   351.6   &  -3.5   &  325.5   &  -5.3  & 1655000. \\
S43-V1597407936 & 2008-227T11:47:10-11:47:55 & 32 & 16 & 80 &   26.3   &   351.6   &  -3.5   &  325.5   &  -5.3  & 1655180. \\
S43-V1597407982 & 2008-227T11:47:56-11:48:40 & 32 & 16 & 80 &   26.2   &   351.6   &  -3.5   &  325.6   &  -5.3  & 1655330. \\
S43-V1597408034 & 2008-227T11:48:48-11:49:33 & 32 & 16 & 80 &   26.2   &   351.7   &  -3.5   &  325.6   &  -5.3  & 1655500. \\
S43-V1597833610 & 2008-232T10:01:41-10:02:20 & 24 & 12 & 120 &   26.5   &     5.9   &  -30.6  &  357.7   &  -5.2  &     427174. \\
S43-V1597833658 & 2008-232T10:02:29-10:03:08 & 24 & 12 & 120 &   26.6   &     5.9   &  -30.7  &  357.7   &  -5.2  &     427783. \\
S43-V1597833706 & 2008-232T10:03:17-10:03:56 & 24 & 12 & 120 &   26.6   &     5.9   &  -30.7  &  357.8   &  -5.2  &     428391. \\
S43-V1597833802 & 2008-232T10:04:53-10:05:32 & 24 & 12 & 120 &   26.7   &     6.0   &  -30.8  &  357.9   &  -5.2  &     429609. \\
S43-V1597833850 & 2008-232T10:05:41-10:06:20 & 24 & 12 & 120 &   26.7   &     6.0   &  -30.8  &  357.9   &  -5.2  &     430217. \\
S43-V1597833898 & 2008-232T10:06:29-10:07:08 & 24 & 12 & 120 &   26.7   &     6.0   &  -30.8  &  358.0   &  -5.2  &     430826. \\
S43-V1597833946 & 2008-232T10:07:17-10:07:56 & 24 & 12 & 120 &   26.7   &     6.0   &  -30.9  &  358.0   &  -5.2  &     431435. \\
S43-V1597833994 & 2008-232T10:08:05-10:08:44 & 24 & 12 & 120 &   26.8   &     6.0   &  -30.9  &  358.0   &  -5.2  &     432044. \\
S43-V1597834042 & 2008-232T10:08:53-10:09:32 & 24 & 12 & 120 &    26.8   &     6.1   &  -30.9  &  358.1   &  -5.2   &    432653. \\
S72-V1710091283 & 2012-070T16:30:08-16:32:13 & 54 & 54 & 40 &   32.1   &   148.2   &  -1.4   &  177.0   &   13.2  &     58652. \\
\hline \hline
\end{tabular}
\end{tiny}
\caption{Rhea disk-integrated observations processed in this work. Columns S and L indicate the cube dimensions along sample and lines axis, respectively.}  
\label{table:5}                   
\end{table}

\begin{table}
\centering  
\begin{tiny}   
\begin{tabular}{|c|c|c|c|c|c|c|c|c|c|c|}      
\hline\hline      
SEQUENCE & START-END & S & L & EXP & PHASE & SUBSPC  & SUBSPC & SSOL  & SSOL & SPC ALT \\
OBS ID   & TIME (UT) &   &   & (s) &($^\circ$) & LON ($^\circ$) & LAT ($^\circ$) & LON ($^\circ$) & LAT ($^\circ$) & (km) \\   
\hline
S11-V1497069822  &  2005-161T04:16:15-04:17:32 & 24 & 18 & 160 & 33.8  &    146.2  &   -5.3   &   171.0  &  -0.2  &    384855. \\
S11-V1497069911  &  2005-161T04:17:44-04:19:01 & 24 & 18 & 160 & 33.8  &    146.3  &   -5.3   &   171.0  &  -0.2  &    384584. \\
S11-V1497070000  &  2005-161T04:19:13-04:20:30 & 24 & 18 & 160 & 33.7  &    146.5  &   -5.3   &   171.1  &  -0.2  &    384040. \\
S11-V1497070437  &  2005-161T04:26:30-04:27:47 & 24 & 18 & 160 & 33.6  &    146.8  &   -5.2   &   171.3  &  -0.2  &    382140. \\
S11-V1497070874  &  2005-161T04:33:47-04:35:04 & 24 & 18 & 160 & 33.5  &    147.3  &   -5.1   &   171.5  &  -0.2  &    380244. \\
S11-V1497070963  &  2005-161T04:35:16-04:36:33 & 24 & 18 & 160 & 33.4  &    147.4  &   -5.1   &   171.6  &  -0.2  &    379703. \\
S11-V1497071546  &  2005-161T04:44:59-04:46:16 & 24 & 18 & 160 & 33.2  &    148.0  &   -5.0   &   171.9  &  -0.2  &    377270. \\ 
S11-V1497072041  &  2005-161T04:53:14-04:54:31 & 24 & 18 & 160 & 33.1  &    148.5  &   -4.9   &   172.2  &  -0.2  &    375112. \\
S11-V1497072478  &  2005-161T05:00:31-05:01:48 & 24 & 18 & 160 & 32.9  &    148.9  &   -4.9   &   172.4  &  -0.2  &    372958. \\
S11-V1497088451  &  2005-161T09:26:44-09:28:18 & 30 & 18 & 160 & 26.6  &    163.7  &   -2.3   &   180.8  &  -0.2  &    303736. \\
S11-V1497089002  &  2005-161T09:35:55-09:37:29 & 30 & 18 & 160 & 26.3  &    164.2  &   -2.2   &   181.1  &  -0.2  &    301505. \\
S11-V1497089933  &  2005-161T09:51:26-09:52:16 & 20 & 14 & 160 & 25.9  &    165.0  &   -2.1   &   181.6  &  -0.2  &    298050. \\
S11-V1497090342  &  2005-161T09:58:15-09:59:49 & 30 & 18 & 160 & 25.7  &    165.4  &   -2.0   &   181.8  &  -0.2  &    296085. \\
S11-V1497090451  &  2005-161T10:00:04-10:01:38 & 30 & 18 & 160 & 25.6  &    165.5  &   -2.0   &   181.9  &  -0.2  &    295594. \\
S11-V1497090986  &  2005-161T10:08:59-10:10:33 & 30 & 18 & 160 & 25.4  &    165.9  &   -1.9   &   182.1  &  -0.2  &    293637. \\
S11-V1497129440  &  2005-161T20:49:52-20:51:09 & 24 & 18 & 160 & 20.3  &    209.2  &    9.6   &   204.1  &  -0.2  &    174404. \\
S11-V1497130881  &  2005-161T21:13:53-21:15:10 & 24 & 18 & 160 & 22.1  &    212.1  &    10.5  &    205.1  &  -0.2 &     172413. \\
S11-V1497131666  &  2005-161T21:26:58-21:28:15 & 24 & 18 & 160 & 23.1  &    213.6  &    11.0  &    205.5  &  -0.2 &     171520. \\
S11-V1497140531  &  2005-161T23:54:43-23:56:00 & 24 & 18 & 160 & 36.6  &    237.8  &    18.3  &    211.9  &  -0.2 &     166551. \\
S11-V1497140620  &  2005-161T23:56:12-23:57:29 & 24 & 18 & 160 & 36.8  &    238.2  &    18.4  &    212.0  &  -0.2 &     166564. \\
S11-V1497141146  &  2005-162T00:04:58-00:06:15 & 24 & 18 & 160 & 37.6  &    239.9  &    18.8  &    212.3  &  -0.2 &     166639. \\
S11-V1497142550  &  2005-162T00:28:22-00:29:39 & 24 & 18 & 160 & 39.9  &    245.1  &    20.1  &    213.4  &  -0.2 &     167071. \\
S17-V1513877057  &  2005-355T16:55:02-16:55:53 & 24 & 24 & 160 & 22.0   &   125.8  &     7.9   &   139.6   &  -2.9 &    762971. \\
S17-V1513911329  &  2005-356T02:26:14-02:27:05 & 24 & 12 & 160 & 20.2  &  151.2  &    3.2   &   163.7   &  -2.4 &    560628. \\ 
S21-V1530174536  &  2006-179T07:57:52-07:59:35 & 24 & 24 & 160 & 37.3  & 16.7   &   13.0   &   14.6  &   -4.5  &    311371. \\
S21-V1530174748  &  2006-179T08:01:24-08:03:07 & 24 & 24 & 160 & 37.1  &    16.8   &   12.9   &   14.7  &   -4.5  &    310973. \\
S21-V1530176158  &  2006-179T08:24:54-08:26:37 & 24 & 24 & 160 & 35.9  &    17.0   &   12.3   &   15.5  &   -4.5  &    308026. \\
S21-V1530176823  &  2006-179T08:35:59-08:37:42 & 24 & 24 & 160 & 35.3  &    17.1   &   12.0   &   15.9  &   -4.5  &    306682. \\
S21-V1530178153  &  2006-179T08:58:09-08:59:52 & 24 & 24 & 160 & 34.0  &    17.3   &   11.4   &   16.7  &   -4.5  &    304013. \\
S21-V1530178818  &  2006-179T09:09:14-09:10:57 & 24 & 24 & 160 & 33.4  &    17.4   &   11.1   &   17.1  &   -4.5  &    302806. \\
S21-V1530179483  &  2006-179T09:20:19-09:22:02 & 24 & 24 & 160 & 32.8  &    17.5   &   10.8   &   17.5  &   -4.6  &    301644. \\
S21-V1530179598  &  2006-179T09:22:14-09:23:57 & 24 & 24 & 160 & 32.7  &    17.5   &   10.7   &   17.5  &   -4.6  &    301541. \\
S21-V1530179713  &  2006-179T09:24:09-09:25:52 & 24 & 24 & 160 & 32.6  &    17.5   &   10.6   &   17.6  &   -4.6  &    301233. \\
S21-V1530186404  &  2006-179T11:15:40-11:17:23 & 24 & 24 & 260 & 26.5  &    18.5   &   7.5    &   21.7  &   -4.7  &    292344. \\
S65-V1669622273  & 2010-332T07:10:54-07:11:48 & 34 & 12 & 80 & 32.8  & 152.2 & -0.1 & 280.7 & 7.1 &     96822. \\
S65-V1669622327  & 2010-332T07:11:48-07:12:42 & 34 & 18 & 80 & 32.7  & 152.7 & -0.1 & 281.3 & 7.1 &     97018. \\
S65-V1669622390  & 2010-332T07:12:51-07:13:45 & 34 & 18 & 80 & 32.6  & 152.7 & -0.1 & 281.3 & 7.1 &     97215. \\
S65-V1669622444  & 2010-332T07:13:45-07:14:39 & 34 & 18 & 80 & 32.5  & 152.8 & -0.1 & 281.9 & 7.1 &     97412. \\
S65-V1669622507  & 2010-332T07:14:48-07:16:35 & 34 & 18 & 160 & 32.3  & 154.9 & -0.1 & 283.0 & 7.1 &     97609. \\
S65-V1669622615  & 2010-332T07:16:36-07:18:22 & 34 & 18 & 160 & 32.1  & 155.0 & -0.1 & 283.6 & 7.1 &     98006. \\
S65-V1669622737  & 2010-332T07:18:38-07:20:25 & 34 & 18 & 160 & 31.9  & 156.6 & -0.1 & 283.6 & 7.1 &     98404. \\
S65-V1669622845  & 2010-332T07:20:26-07:22:12 & 34 & 18 & 160 & 31.6  & 157.8 & -0.1 & 286.4 & 7.1 &     98804. \\
S65-V1669626673  & 2010-332T08:24:14-08:25:08 & 34 & 18 & 80 & 25.4  & 193.3 & -0.1 & 322.4 & 7.1 &     112458. \\
S65-V1669626736  & 2010-332T08:25:17-08:26:11 & 34 & 18 & 80 & 25.4  & 193.8 & -0.1 & 323.7 & 7.1 &     112458. \\
S65-V1669626790  & 2010-332T08:26:11-08:27:04 & 34 & 18 & 80 & 25.3  & 194.4 & -0.1 & 323.7 & 7.1 &     112683. \\
S65-V1669626853  & 2010-332T08:27:14-08:29:01 & 34 & 18 & 160 &  25.1  & 195.5 & -0.1 & 324.7 & 7.1 &   113133. \\
S65-V1669626961  & 2010-332T08:29:02-08:30:48 & 34 & 18 & 160 & 24.9  & 196.1 & -0.1 & 325.3 & 7.1 &     113811. \\
S65-V1669627191  & 2010-332T08:32:52-08:34:38 & 34 & 18 & 160 & 24.7  & 198.3 & -0.1 & 327.5 & 7.1 &     114265. \\
S69-V1692997197  & 2011-237T20:10:31-20:12:49 & 40 & 20 & 160 & 39.5  & 14.6 & 0.1 & 26.1 & 10.9 &    67704. \\
S69-V1692997353  & 2011-237T20:13:07-20:15:25 & 40 & 20 & 160 & 39.3  & 16.8 & 0.1 & 28.4 & 10.9 &    68277. \\
S69-V1692997510  & 2011-237T20:15:44-20:16:53 & 40 & 20 & 80 & 39.2 & 17.9 & 0.1 & 29.5 & 10.9 &     68851. \\
S69-V1692997590  & 2011-237T20:17:04-20:18:13 & 40 & 20 & 80 & 39.0 & 18.5 & 0.1 & 30.1 & 10.9 &    69425. \\
\hline \hline
\end{tabular}
\end{tiny}
\caption{Hyperion disk-integrated observations processed in this work. Columns S and L indicate the cube dimensions along sample and lines axis, respectively.}  
\label{table:6}                   
\end{table}

\begin{table}
\centering  
\begin{tiny}   
\begin{tabular}{|c|c|c|c|c|c|c|c|c|c|c|}      
\hline\hline      
SEQUENCE & START-END & S & L & EXP & PHASE & SUBSPC  & SUBSPC & SSOL  & SSOL & SPC ALT \\
OBS ID   & TIME (UT) &   &   & (s) &($^\circ$) & LON ($^\circ$) & LAT ($^\circ$) & LON ($^\circ$) & LAT ($^\circ$) & (km) \\   
\hline
S07-V1483156810 &  2004-366T03:34:12-03:37:56 & 38 & 34 & 160 &  38.2  &    70.0  &   -23.7  &   106.1   &  -7.4  &    173986. \\
S15-V1510202713 &  2005-313T04:16:22-04:17:56 & 30 & 18 & 160 &  39.7  &    38.4  &    10.3  &    75.3   &  -4.7  &    719296. \\
S15-V1510202822 &  2005-313T04:18:11-04:24:28 & 30 & 18 & 640 &  39.7  &    38.4  &    10.3  &    75.3   &  -4.7  &    718907. \\
S15-V1510203548 &  2005-313T04:30:17-04:31:51 & 30 & 18 & 160 &  39.7  &    38.4  &    10.4  &    75.3   &  -4.7  &    717933. \\
S15-V1510203657 &  2005-313T04:32:06-04:38:23 & 30 & 18 & 640 &  39.7  &    38.4  &    10.4  &    75.3   &  -4.7  &    717544. \\
S15-V1510204383 &  2005-313T04:44:12-04:45:46 & 30 & 18 & 160 &  39.8  &    38.4  &    10.4  &    75.4   &  -4.7  &    716571. \\
S15-V1510204492 &  2005-313T04:46:01-04:52:18 & 30 & 18 & 640 &  39.8  &    38.4  &    10.4  &    75.4   &  -4.7  &    716183. \\
S15-V1510205218 &  2005-313T04:58:07-04:59:41 & 30 & 18 & 160 &  39.9  &    38.3  &    10.5  &    75.4   &  -4.7  &    715211. \\
S15-V1510205327 &  2005-313T04:59:56-05:06:13 & 30 & 18 & 640 &  39.9  &    38.3  &    10.5  &    75.4   &  -4.7  &    714823. \\
S15-V1510206053 &  2005-313T05:12:02-05:13:36 & 30 & 18 & 160 &  39.9  &    38.3  &    10.5  &    75.4   &  -4.7  &    713852. \\
S15-V1510206162 &  2005-313T05:13:51-05:20:08 & 30 & 18 & 640 &  39.9  &    38.3  &    10.5  &    75.4   &  -4.7  &    713561. \\
S33-V1568157352 &  2007-253T22:40:51-22:47:26 & 48 & 48 & 160 &   33.3  &    247.0  &   -10.7   &   215.8  &    1.5  &    85545. \\
S33-V1568162035 &  2007-253T23:58:54-00:00:47 & 36 & 36 & 80  & 33.4  &    247.4  &   -10.7   &   216.1  &    1.5  &    96436. \\
S33-V1568162861 &  2007-254T00:12:40-00:17:45 & 42 & 42 & 160 &   33.5  &    247.5  &   -10.7   &   216.1  &    1.5  &    98699. \\
S33-V1568163183 &  2007-254T00:18:02-00:23:07 & 42 & 42 & 160 &  33.5  &    247.5  &   -10.7   &   216.1  &    1.5  &    99407. \\
S33-V1568163600 &  2007-254T00:24:59-00:30:04 & 42 & 42 & 160 &  33.5  &    247.5  &   -10.7   &   216.2  &    1.5  &    100256. \\
S33-V1568171295 & 2007-254T02:33:14-02:35:07  & 36 & 36 & 80 &  33.6  &    248.1  &   -10.6   &   216.6  &    1.5  &    118217. \\
S33-V1568171804 & 2007-254T02:41:43-02:42:51 & 32 & 32 & 60 & 33.6  &    248.1  &   -10.6   &   216.6  &    1.5  &    119206. \\
\hline \hline
\end{tabular}
\end{tiny}
\caption{Iapetus disk-integrated observations processed in this work. Columns S and L indicate the cube dimensions along sample and lines axis, respectively.}  
\label{table:7}                   
\end{table}

\section{Disk-resolved temperature maps}\label{sct:d}
With the goal of deriving temperature maps across specific geomorphological units, we have selected two VIMS observations which offer best view of Enceladus' south pole region and Tethys' equatorial radiation lens.

\subsection{Enceladus}
Since the discovery of plumes \citep{Porco2006} ejected from the tiger stripes fissures located on the south polar region of Enceladus, many studies have investigated the mechanisms originating this activity \citep{Nimmo2007, Hedman2013, Porco2014}, the composition of the plumes particles and gaseous species \citep{Waite2006, Hansen2006, Postberg2011} and correlated them with the thermal properties of this region \citep{Spencer2006, Howett2011b}.
One of the best observations of Enceladus south pole region obtained by VIMS is cube V1500059894 which was acquired from 2005-195T18:50:28 to 18:52:57 (UT) using an IR channel integration time of 160 msec/line. In this observation VIMS resolution is equal to 16.9 km/pixel while the solar phase is $46^\circ$. A detailed discussion of the tiger stripes spectral features observed on this specific observation is reported in \cite{Brown2006}.

\begin{figure}[h!]
	\centering
		\includegraphics[width=15cm]{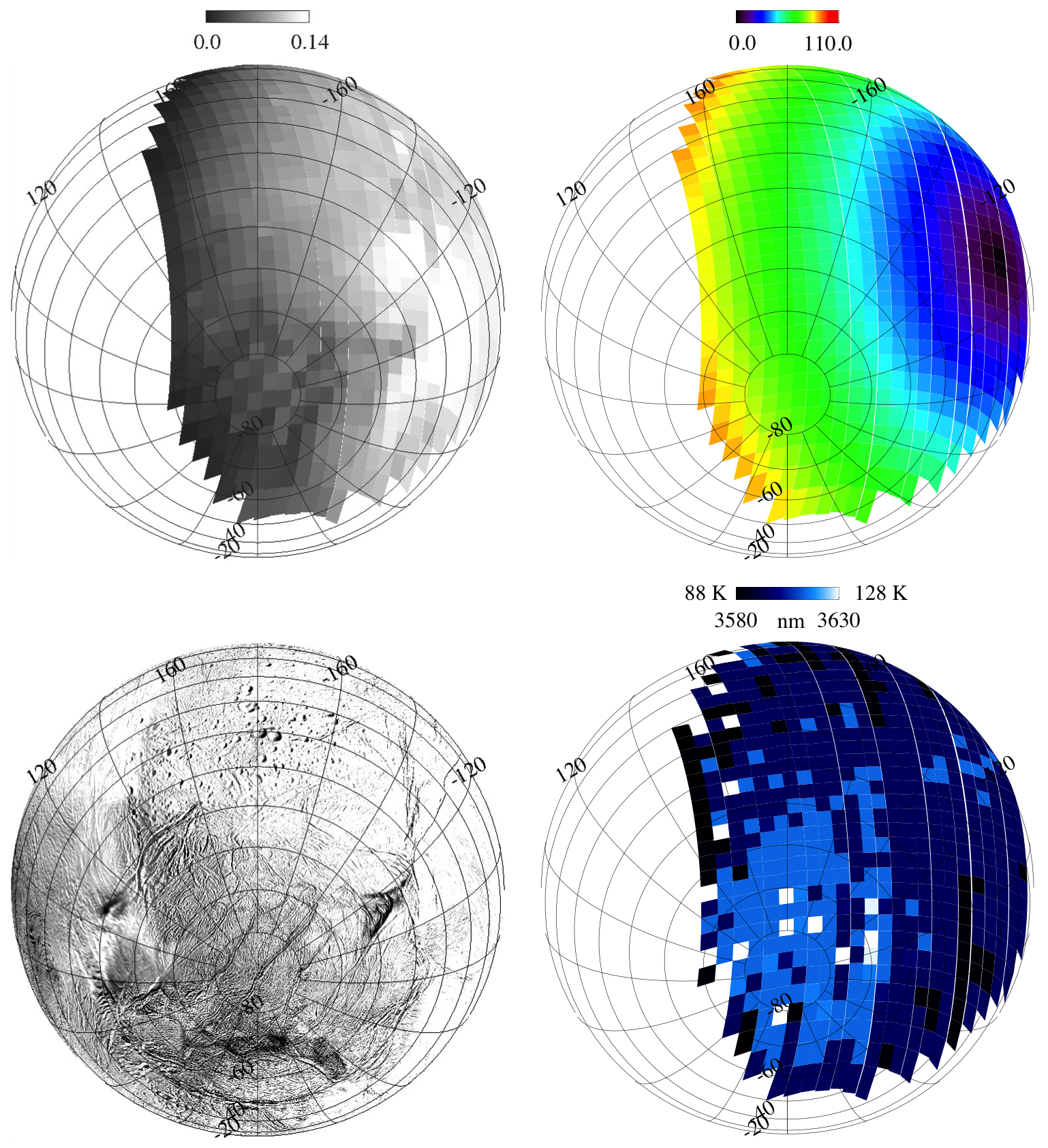}
		\caption{Enceladus disk-resolved observation V1500059894. Top left panel: I/F(3.6 $\mu$m) image mapped in orthographic projection. Top right panel: solar incidence angle (in deg) map projection. Bottom left panel: ISS basemap. Bottom right panel: VIMS-derived temperature projection.} 
	\label{fig:figure5}
\end{figure}

SPICE routines \citep{Acton1996} and reconstructed spacecraft's attitude and trajectory kernels have been used to calculate several geometry parameters (latitude, longitude, local solar time or LST, spacecraft's distance, phase, incidence, and emission angles) for each VIMS pixel (center and 4 corners). This method has been applied to all VIMS data discussed in the rest of this work.
In Figure \ref{fig:figure5} are shown the orthographic projection maps of I/F at 3.6 $\mu$m, the solar incidence angle, the ISS basemap and the temperature. Appearing dark on the I/F(3.6 $\mu$m) map, the four tiger stripes can be identified above the south pole. The area located at mid-latitudes towards the limb appears bright because photometric correction has been not applied to the VIMS data. 
The map of the solar incidence angle is shown in the top right panel of Fig. \ref{fig:figure5}. In this figure the sub-solar point (corresponding to an incidence angle of 0$^\circ$) is rendered in black, and is located near lon=120$^\circ$ W and lat=-30$^\circ$ (i.e. on the right side of the image). The distribution of the temperature, correlated with the wavelength of the 3.6 $\mu$m peak, is shown in the bottom right panel: in this map the light blue pixels located in the south polar region around tiger stripes have a temperature T$>$115K. A few scattered points, marked in white, correspond to maximum T=128 K. In average the rest of the surface has T=95 K (dark blue pixels), including regions close to the subsolar point at noon local time. This result demonstrates that the method is insensitive to illumination and photometric effects, being capable to measure a $\Delta T \approx 20 \ K$ between points having a $\Delta inc = 90^\circ$, e.g. between tiger stripes located at morning terminator and subsolar point.    
Conversely, the spatial resolution influences the temperature retrieval: as a general rule thermal contrast is strongly reduced by lower spatial resolution because high temperature features are mixed with cold surfaces within the same pixel. This effect is well-evident when comparing our low spatial resolution (16.9 km/pixel) results with \cite{Goguen2013} findings, which have analyzed VIMS high spatial resolution (38 $\times$ 214 m) data collected during a 74 km low altitude Cassini's flyby above Baghdad Sulcus and have measured a color temperature T$_c$=197$\pm$10 K by fitting VIMS radiance in the 3-5 $\mu m$ thermal range.   

\subsection{Tethys}
\begin{figure}[h!]
	\centering
		\includegraphics[width=15cm]{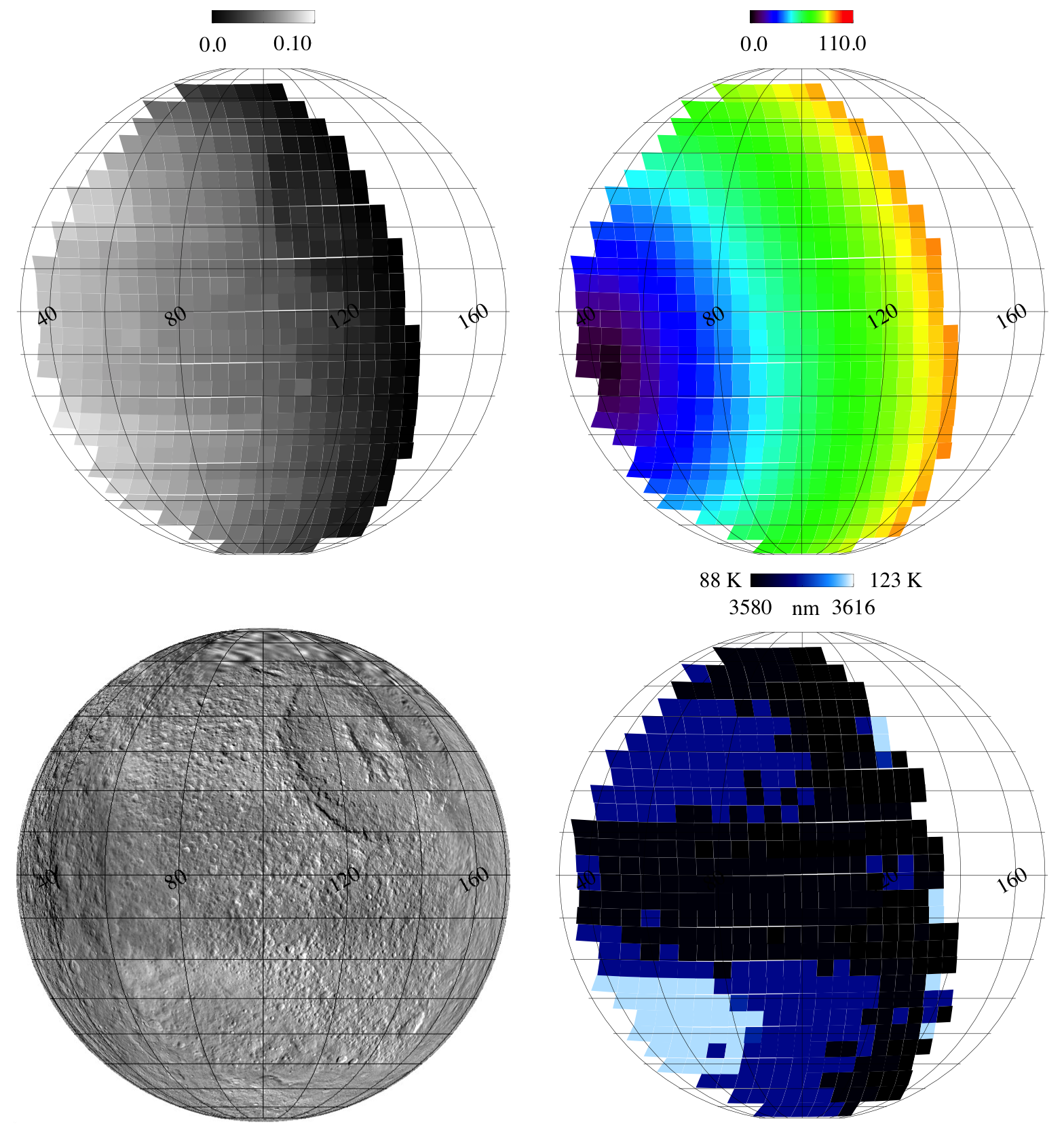}
		\caption{Tethys disk-resolved observation V1567089225. Top left panel: I/F(3.6 $\mu$m) image mapped in orthographic projection. Top right panel: solar incidence angle (in deg) map projection. Bottom left panel: ISS basemap. Bottom right panel: VIMS-derived temperature projection.} 
	\label{fig:figure6}
\end{figure}

Apart from active heating, as in the case of Enceladus' south polar region previously discussed, other processes can cause anomalous thermal behavior on the surfaces of the icy moons. On Tethys, as an example, significant temperature variation is observed across the equatorial region of the leading hemisphere \citep{Howett2012}. A similar effect has been reported also for Mimas \citep{Howett2011a}. These thermal anomalies are directly correlated with the irradiation of the surface by magnetospheric MeV electrons \citep{Paranicas2010a, Paranicas2010b}, which are focused on the leading hemisphere equatorial regions, resulting in the formation of a dark ``lens" visible in the IR/UV ratio ISS images \citep{Schenk2011}: as a consequence of the energy released by this space weathering, surface water ice grains are sintered together in a few centimeters-deep layer resulting in a local increase of thermal inertia inside the dark lens \citep{Howett2012}. 
Tethys equatorial lens appears well-resolved on VIMS observation V1567089225 shown in Fig. \ref{fig:figure6} taken from 2007-241T13:58:51 to 14:15:33 (UT) with an integration time of 640 msec/line. At the time of the observation VIMS spatial resolution is 35.1 km/pixel and solar phase 65$^\circ$.

The equatorial belt corresponds to the dark area visible in the center of the I/F(3.6 $\mu m$) image and ISS basemap mapped in orthographic projection shown in Figure \ref{fig:figure6}. On the same images is partially visible the 400 km-wide Odysseus impact basin centered along the north terminator at lon=130$^\circ$, lat=30$^\circ$. As shown in the top right panel, the subsolar point is located close to the limb at lat=-10$^\circ$. The temperature map, in the bottom right panel, reveals that south latitude regions close to the subsolar point have the highest temperature, T$>$115 K (light blue pixels). Black pixels identify cold areas at T$\le$88 K which are localized in the low-albedo equatorial lens and along the terminator, including the Odysseus crater floor. The remaining points at both north and south latitudes coded in blue are at T$\approx$100 K. It is remarkable that on this image a large thermal gradient ($\Delta T > 27 $ K) visible near the subsolar point (noon time) between the equatorial lens and the nearby southern regions. This effect could be the consequence of the large thermal inertia difference reported on these two regions by CIRS, corresponding to 5 and 25 $J \ m^{-2} \ K^{-1} \ s^{-1/2}$ outside and inside the equatorial lens, respectively \citep{Howett2012}.

\section{Averages temperature cylindrical maps}\label{sct:e}
By using data projection techniques the temperature distribution derived from VIMS observations has been mapped on the regular satellites' surfaces and correlated with geomorphological features, including albedo units, equatorial radiation lenses and impact craters. VIMS observations are filtered in two groups corresponding to the pre-equinox (years 2004-2009, heliocentric distance 9.0-9.4 AU) and post-equinox (years 2009-2012, heliocentric distance 9.4-9.7 AU) periods. 

This approach allows us to follow seasonal variations because the subsolar point moves from south latitudes during the pre-equinox to north latitudes during the post-equinox period. 
VIMS data are further filtered to retrieve all pixels of each satellite matching the following multiple conditions: 
\begin{enumerate}
\item acquired from distances lower than 300,000 km (corresponding to a VIMS spatial resolution better than 150 km/pixel) and 
\item with incidence and emission angles lower than $80^\circ$ (to avoid very oblique and distorted views of the surface) and
\item with 10:00 $\le$ LST $\le$ 14:00 hr, e.g., around noon time (to limit influence of local time) and
\item taken in the 2004-2009 (pre equinox) and 2009-2012 (post equinox) periods (to follow temperature seasonal changes between north and south hemispheres).
\end{enumerate} 
A summary of the number of cubes and pixels resulting from this selection is summarized in Table \ref{tbl:table1} for each satellite.
Temperature maps are rendered in cylindrical projection by applying a fixed $1^\circ \times 1^\circ$ grid in longitude and latitude. As satellites have different radii, the spatial resolution on the maps changes according to the values reported in Table \ref{tbl:table1}. Finally, for each $1^\circ \times 1^\circ$ bin, the median value of the temperature and the data redundancy (number of VIMS pixels covering a given bin) have been calculated and rendered on the cylindrical maps. A discussion of the results obtained is given in the next paragraphs for each satellite.

\begin{table}
\centering
\begin{tabular}{|c|c|c|c|}

\hline
Satellite & number of & number of & Resolution (km/bin) \\
          & observations & pixels & of a $1^\circ \times 1^\circ$ bin \\

\hline
Mimas & 703 & 10581 & 3.5 \\
Enceladus & 1917 & 104787 & 4.4 \\
Tethys & 802 & 60626 & 9.3 \\
Dione & 938 & 120838 & 9.8 \\
Rhea & 1070 & 144747 & 13.3 \\
Hyperion & 523 & 8087 & 2.3 \\
Iapetus & 159 & 46860 & 12.8 \\
\hline
\end{tabular}
\caption{Summary of Saturn's icy satellites observations processed in this work. }
\label{tbl:table1}
\end{table}

\subsection{Mimas}
The principal geological feature of the small Mimas is the presence of the 110 km-wide Herschel crater dominating the center of the leading hemiphere. Many other impact craters are spread across the surface which, in contrast with the outer moons, lack significant tectonic features. An equatorial lens on the leading hemisphere, caused by the bombardment of MeV energy magnetospheric electrons corresponds to the Thermal Anomaly Region (TAR) reported by \cite{Howett2011a}.
   
\begin{figure}[h!]
	\centering
		\includegraphics[width=14.5cm]{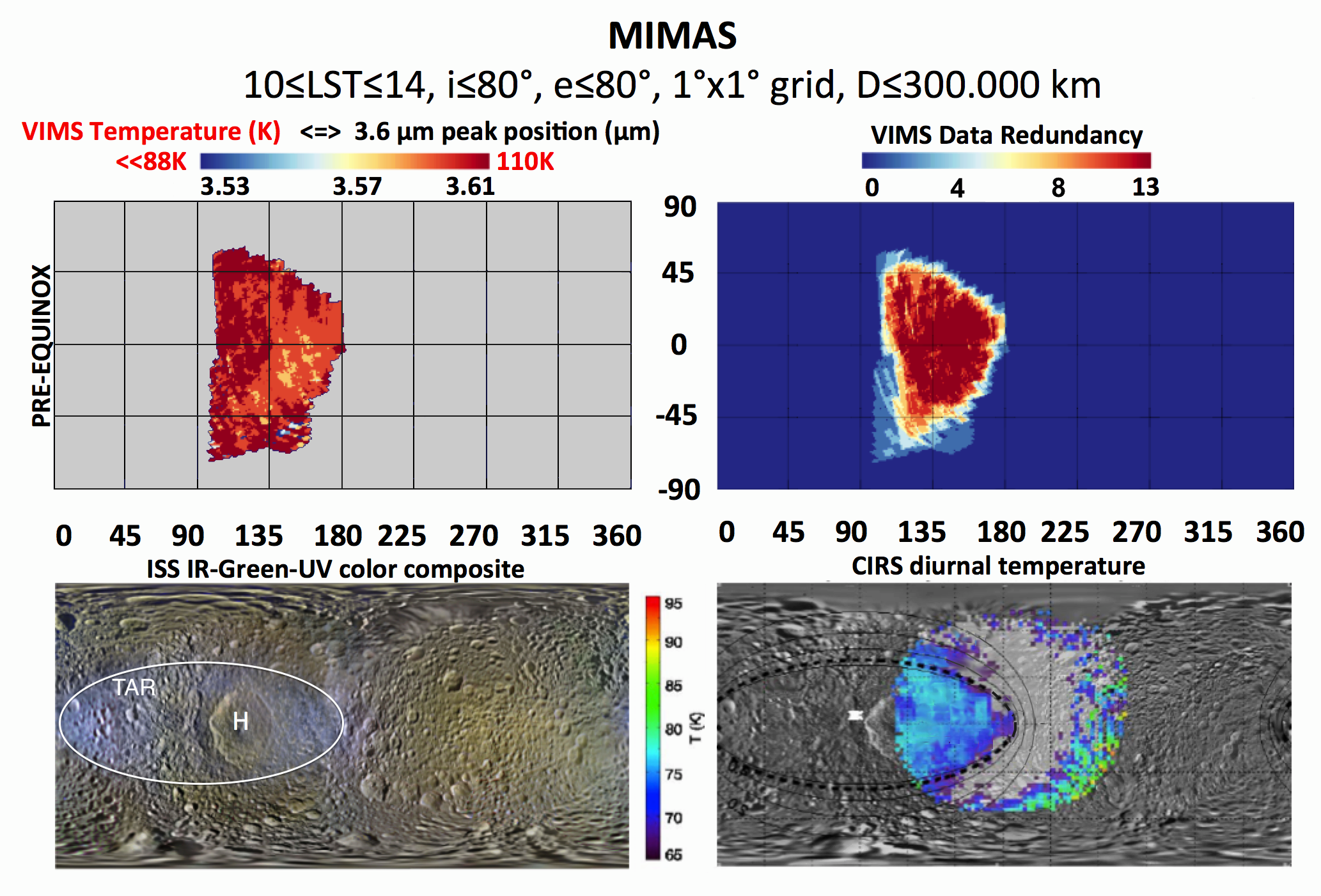}
		\caption{Mimas maps. Top left panel: Pre-equinox temperature map derived from the 3.6 $\mu m$ peak position. Top Right panel: VIMS data redundancy map (number of VIMS pixels/bin). Bottom left panel: context image map based on Cassini-ISS IR-Green-UV color composite (by P. Schenk, available from http://photojournal.jpl.nasa.gov/catalog/PIA18437). Thermal Anomaly Region (TAR) extension is indicated by the oval; Herschel crater is identified by the letter H. Bottom right panel: CIRS night-time temperature map \citep{Howett2011a}.}
	\label{fig:figure7}
\end{figure}

Following the data selection criteria previously exposed, only pre-equinox data are available for Mimas. The temperature map shown in Fig. \ref{fig:figure7}-top left panel covers the leading hemisphere between 95$^\circ$ $\le$ lon $\le$ 180$^\circ$, -60$^\circ$ $\le$ lat $\le$ +50$^\circ$. The VIMS dataset has up to 13 independent pixels covering a given $1^\circ \times \ 1^\circ$ bin (Fig. \ref{fig:figure7}-top right panel). On the equator of the leading hemisphere the TAR has a T=100 K, about 10 K colder than the external regions. The lens extends up to latitudes $\pm 45^\circ$ and is interrupted in a pattern corresponding to the Herschel crater located at the equator at about lon=$100^\circ$ where the temperature increases up to T=110 K (dark red pixels). VIMS temperature distribution across the TAR matches with the light blue regions visible on ISS IR-Green-UV color composite map (Fig. \ref{fig:figure7}-bottom left panel) and with CIRS night-time temperature map by \cite{Howett2011a} (Fig. \ref{fig:figure7}-bottom left panel).

\subsection{Enceladus}
Enceladus' surface is shaped by the active processes occurring in the warm ``tiger stripes" features located in the south pole region. These features consists of four main linear, almost parallel troughs (Alexandria, Cairo, Baghdad, and Damascus Sulci) about 500 m deep, 2 km wide and 130 km long \citep{Porco2006} from which plumes are ejected. According to \cite{Waite2006} the plumes are composed mainly by water vapor and carbon dioxide, carbon monoxide or molecular nitrogen, and methane. Moreover, the plume activity is not constant but shows a periodicity linked to orbital position \citep{Hedman2013} and tides \citep{Porco2014}. Wide fractured plains, resurfaced by tectonism, are located on both hemispheres while the north hemisphere shows more impact craters \citep{Jaumann2009}.
 
\begin{figure}[h!]
	\centering
		\includegraphics[width=14.5cm]{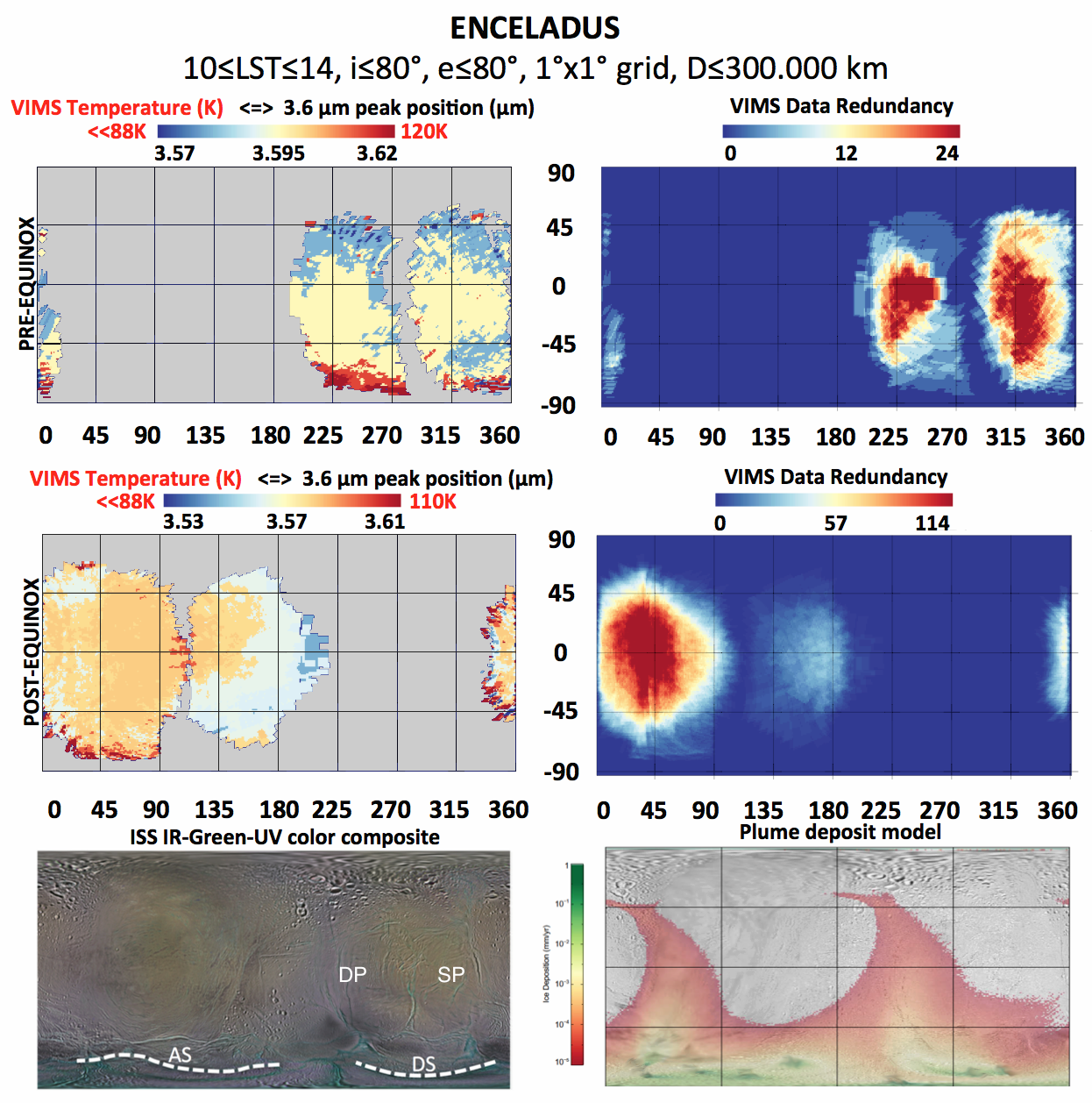}
		\caption{Enceladus maps. Top left panel: Pre-equinox temperature map. Top Right panel: VIMS pre-equinox data redundancy map. Center left panel: Post-equinox temperature map. Center Right panel: VIMS pre-equinox data redundancy map. Bottom left panel: context image map based on Cassini-ISS IR-Green-UV color composite (by P. Schenk, available from http://photojournal.jpl.nasa.gov/catalog/PIA18435). Locations of Diyar Planitia (DP), Sarandib Planitia (SP), Alexandria (AS) and Damascus Sulci (DS) are indicated. Bottom right panel: distribution of the plumes particles accumulation on the surface in mm/year by \cite{Kempf2010}.}
	\label{fig:figure8}
\end{figure}

For Enceladus both trailing and leading hemisphere coverage (Fig. \ref{fig:figure8} - top and center panels) is available during the pre-equinox and post-equinox periods. Pre-equinox temperature map includes the fractured and ridged plains of the trailing hemisphere spanning above Diyar and Sarandib Planitia (marked as DP and SP in Fig. \ref{fig:figure8} - bottom left panel), between 185$^\circ$ $\le$ lon $\le$ 360$^\circ$, -85$^\circ$ $\le$ lat $\le$ +50$^\circ$ with a narrow gap at about lon=270$^\circ$ (Fig. \ref{fig:figure8} -top left panel). A small coverage of the leading south hemisphere between 0$^\circ$ $\le$ lon $\le$ 15$^\circ$ is available. The maximum temperature measured is T=120 K above Damascus Sulcus (DS) rendered in red color on the temperature map. With the exclusion of the south pole active region, most of entire south hemisphere during the summer season has a T=105 K (yellow pixels) while the regions located northwards of lat=10$^\circ$, in winter season, most are colder at T$\le$ 95 K.

The post-equinox temperature map covers the heavy cratered terrains on the leading hemisphere between 0$^\circ$ $\le$ lon $\le$ 220$^\circ$, -85$^\circ$ $\le$ lat $\le$ +75$^\circ$ with a narrow gap at about lon=110$^\circ$ (Fig. \ref{fig:figure8} - center left panel). A partial coverage of the trailing equatorial hemisphere between 330$^\circ$ $\le$ lon $\le$ 360$^\circ$ is available. On this map the maximum temperature measured is T=110 K in the region corresponding to Alexandria Sulcus (rendered in red color, indicated as AS in Fig. \ref{fig:figure8} - bottom left panel). A wide region of the leading hemisphere, rendered in orange has T=100 K spanning on both north (summer) and south (winter) hemispheres. Towards the trailing hemisphere temperature drops below T$<$95 K (light blue pixels).

Temperature variations appear correlated with surface colors (see the ISS IR-Green-UV color composite map shown in Fig. \ref{fig:figure8} - bottom left panel). The two yellowish areas centered in the middle of the leading and trailing north-equatorial hemispheres on ISS map appear warmer in VIMS maps than the neutral gray areas on Diyar Planitia placed in the antisaturnian hemisphere. Conversely, the relationship with the distribution of the tiger stripes plumes deposits is less evident (see the deposit model by \cite{Kempf2010} shown in Fig. \ref{fig:figure8} - bottom right panel, valid for grains smaller than 5 $\mu m$). 
The downwelling fraction of the material released from tiger stripes deposits preferentially in the nearby regions (green areas in Fig. \ref{fig:figure8} - bottom right panel) and in the south hemisphere along two V-shaped regions centered on lon=45$^\circ$ and 225$^\circ$ meridians (red areas). According to \cite{Kempf2010} these deposits grow with a rate of 0.5 mm/year in the vicinity of active sources which reduces to 10$^{-5}$ mm/year at the equator. As a consequence of the accumulation of fresh water ice grains, these two V-shaped regions appear less yellow and more neutral in color at UV-visible wavelengths on the ISS map. VIMS post-equinox temperature map (Fig. \ref{fig:figure8} - center left panel) doesn't vary but stays at a T=105 K (orange pixels) across a wide sector of the leading hemisphere, where the heavy cratered terrains are located, including the V-shaped plume deposit centered on the lon = 45$^\circ$ meridian. A similar temperature T=105 K (yellow pixels) is observed on a large part of the equatorial and mid-southern latitudes areas of the trailing hemisphere on the pre-equinox map (Fig. \ref{fig:figure8} - top left panel). In this case the plume deposit accumulates across the fractured and ridged plains.
Conversely, a small decrease of temperature (T$<$100 K) between 0$^\circ < lon < 30^\circ$ in the south hemisphere and towards the lon=180$^\circ$ meridian (antisaturnian point) are observed. Since these two areas are immediately outside the plume deposits, a similar behavior implies a change in composition, grain size and thermal inertia or a combination among these factors.   
Noteworthy, in correspondence of the V-shape deposits, VIMS has measured higher water ice band depth and larger grain sizes (up to 40 $\mu m$) with respect to the nearby equatorial regions \citep{Jaumann2008}. A similar result implies the presence of a maturation process, like annealing or sintering, in which the small ($<5\ \mu m$) plume particles coalesce in larger grains resulting in a more compact surface layer. As a consequence of these evidences, one should expect high thermal inertia across the two V-shaped plume deposits and low elsewhere. However, VIMS temperature data taken around noon local solar time show a similar trend.     

 \begin{figure}[h!]
	\centering
		\includegraphics[width=14.5cm]{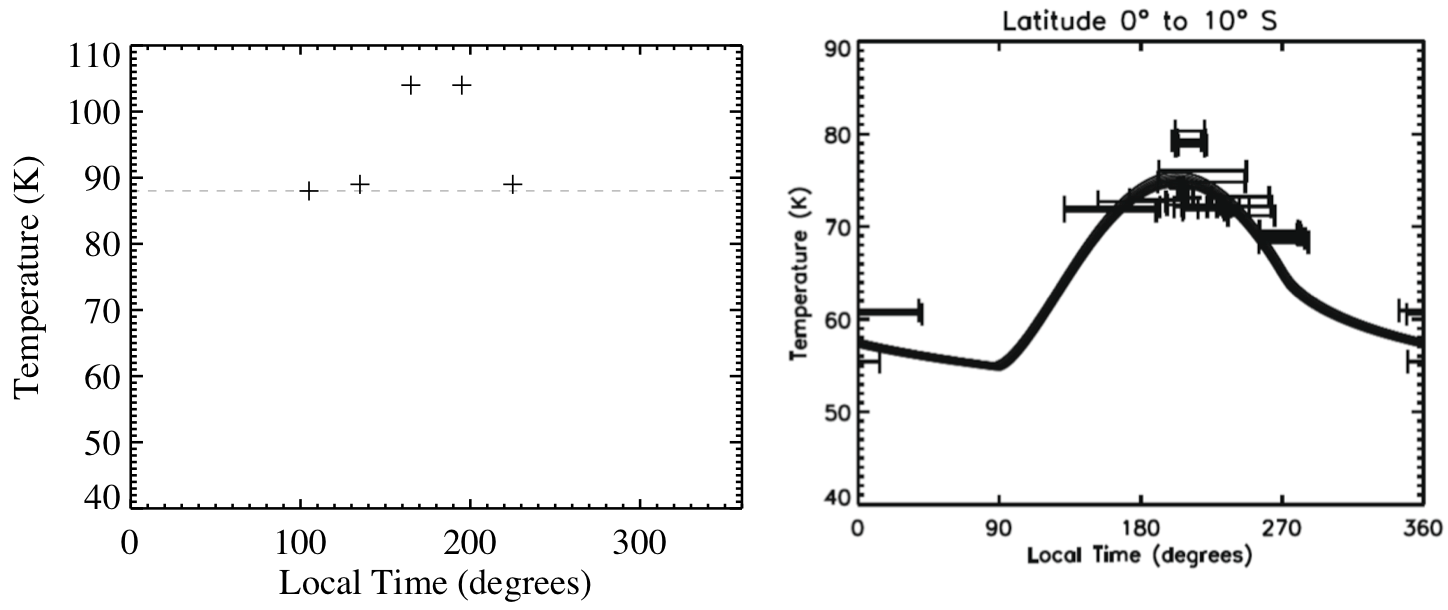}
		\caption{Enceladus daytime temperature variations. Left panel: VIMS daytime temperature variation at lon=95$^\circ$, lat=-3$^\circ$. Temperature retrieval is limited to T$\ge$88 K (horizontal dashed line). VIMS point at local time 105$^\circ$ is an upper limit. Right panel: CIRS daytime and nocturnal temperature cycle for points at $-10^\circ \le \ lat \ \le 0^\circ$ \citep{Howett2010}.}
	\label{fig:figure9}
\end{figure}

Finally, thanks to the large redundancy of the Enceladus dataset, it is possible to investigate the daytime temperature variation for a given point of the surface. In order to follow daytime temperature variations, the VIMS geometry archive has been exploited in order to search the point having the more extended coverage in local solar time. From this search results that the $1^\circ \times 1^\circ$ bin centered at lon=95$^\circ$, lat=-3$^\circ$ has the maximum temporal coverage, being observed between 6 hr $\le$ LST $\le$ 16 hr and therefore covering about all daytime times. The resulting temperature daytime curve (Fig. \ref{fig:figure9} - left panel) shows the increase of temperature from T$\le$ 88 K at LST=105$^\circ$ to T=89 K at LST=225$^\circ$. The maximum T=104 K is observed around noon at LST=165-195$^\circ$. This to demonstrate that the method based on the 3.6 $\mu m$ continuum peak wavelength has the sensitivity to follow daytime variation of the temperature. As mentioned before, temperature values retrieved by VIMS with this method appear higher than similar ones measured by CIRS at longer wavelengths. The daytime temperature curve measured by CIRS on points located at $-10^\circ \le \ lat \ \le 0^\circ$ are in fact lower (Fig. \ref{fig:figure9} - right panel, from \cite{Howett2010}). At noon CIRS measures a temperature T=80 K, about 25 K lower than VIMS. The different skindepth sensed by two instruments can be the cause of this effect.

\subsection{Tethys}
Tethys' surface is densely cratered, showing among the largest impact basins in the Saturnian system, like the 400 km-wide Odysseus crater. Cutting across the Saturn-facing meridian is located the 2500 km-long Ithaca Chasma, a 100 km-wide rift running approximately along the north-south direction with multiple parallel fractures. Tethys, like Dione and Rhea, show an evident albedo asymmetry between the leading (bright) and trailing (dark) hemispheres \citep{VerbiscerVeverka1989, Buratti1990, Schenk2011}. Tethys' equatorial lens on the leading hemisphere equator is the place where the magnetosphere focuses high energy (E$>$1 MeV) electrons resulting in alteration of surface grains \cite{Paranicas2010a, Paranicas2010b}. As a consequence of this processes an enhancement of the UV signal and a darkening in the infrared has been observed \citep{Schenk2011}. The trailing hemisphere experiences the interaction with the plasma flow particles resulting in an accumulation of non-icy material \citep{Jaumann2009}.  

\begin{figure}[h!]
	\centering
		\includegraphics[width=14.5cm]{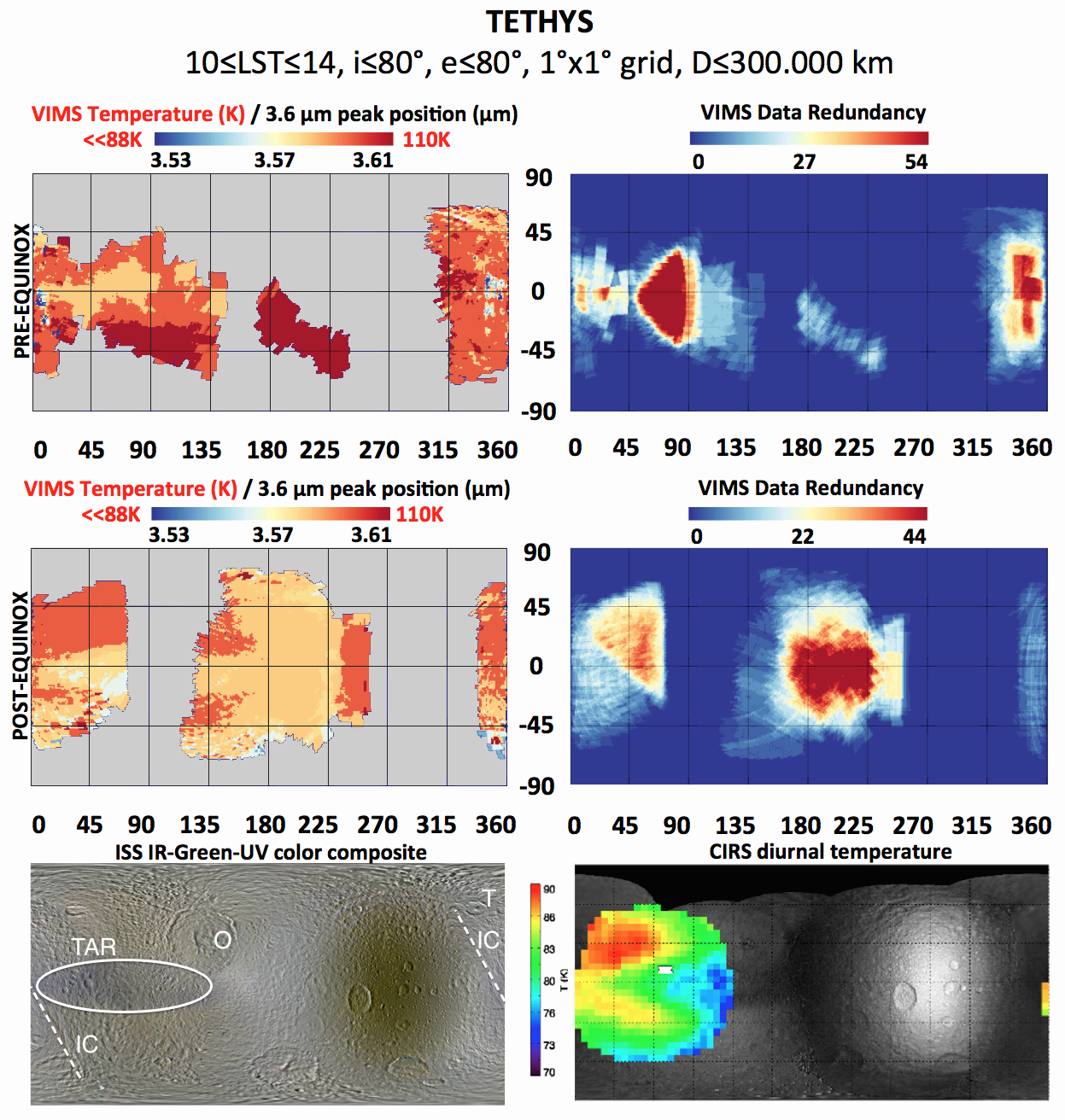}
		\caption{Tethys maps. Top left panel: Pre-equinox temperature map. Top Right panel: VIMS pre-equinox data redundancy map. Center left panel: Post-equinox temperature map. Center Right panel: VIMS pre-equinox data redundancy map. Bottom left panel: context image map based on Cassini-ISS IR-Green-UV color composite (by P. Schenk, available from http://photojournal.jpl.nasa.gov/catalog/PIA18439). Locations of Thermal Anomaly Region (TAR), Odysseus crater (O), Ithaca Chasma (IC) and Telemachus crater (T) are indicated. Bottom right panel: CIRS daytime temperature on ISS IR/UV basemap (courtesy C. Howett, personal communication).}
	\label{fig:figure10}
\end{figure}

A wide coverage of Tethys's surface is available for both pre and post-equinox datasets ( Fig. \ref{fig:figure10} - top and center right panels). Thank to this is possible to map the temperature changes across the leading hemisphere Thermal Anomaly Region (TAR) feature and to follow seasonal temperature changes occurring on the north and south hemisphere regions. The cold TAR, located in the equatorial region of the leading hemisphere, appears at T=100 K (light orange pixels) on both pre (Fig. \ref{fig:figure10} - top left panel) and post-equinox Fig. \ref{fig:figure10} - center left panel) VIMS datasets. On the post-equinox map the equatorial border of the TAR located on the antisaturnian hemisphere appears less contrasted respect to the saturnian side.  
A similar ``Pac-Man" feature \citep{Howett2012} is visible on Cassini-CIRS data (Fig. \ref{fig:figure10} - bottom right panel). On the pre-equinox map the oval-shaped boundary of the feature is noisy as a consequence of the scarcity of high spatial resolution observations available, in particular in the antisaturnian quadrant towards the Odysseus crater (Fig. \ref{fig:figure10} - top right panel). On Ithaca Chasma (IC), the 100 km-wide rift spanning from about lon=45$^\circ$, lat=-75$^\circ$ to lon=315$^\circ$, lat=75$^\circ$ and intersecting the equatorial lens border at about lon=0$^\circ$, VIMS observes a low T$\le$ 88 K (blue pixels). This decrease of temperature is a consequence of the presence of bright material and shadows occurring in the very irregular rift structure. 
Moreover, VIMS data indicate that the seasonal cycle changes the temperature distribution: the south hemisphere, warmer during the pre-equinox period, becomes colder during the post-equinox while the reverse is happening on the north hemisphere. The TAR remains the coldest region of the leading hemisphere during the entire seasonal cycle.      

Starting form the TAR, moving in the south summer hemisphere down to lat=-50$^\circ$, the temperature rises to T=110 K (red pixels). Conversely, moving northward of the TAR in the winter hemisphere the temperature remains below T=105 K (orange pixels). The center of Tethys' trailing hemisphere is redder and darken than the leading, as shown in the ISS map in Fig. \ref{fig:figure10} - bottom left panel. As a consequence of this low albedo feature, the temperature increases on the trailing hemisphere. Despite limited by the coverage, VIMS map shows that in the south trailing hemisphere, antisaturnian quadrant, between 175$^\circ$ $\le$ lon $\le$ 230$^\circ$, 0$^\circ$ $\le$ lon $\le$ -60$^\circ$ a wide region is at T=110 K (red pixels). Similar high temperature is observed in the north hemisphere around lon=315$^\circ$.  Finally, on the saturnian quadrant of the trailing hemisphere at about lon=350$^\circ$-360$^\circ$ lat=0$^\circ$-10$^\circ$ is visible the north branch of Itacha Chasma at T$\le$ 88 K (blue pixels). Lower temperatures are seen around Telemachus crater (lon=340$^\circ$, lat=55$^\circ$).

\subsection{Dione}
Dione's surface is characterized by geomorphological and albedo differences between the leading and trailing hemispheres. Old crater units are located on the bright leading hemisphere while the dark trailing shows a complex network of chasmata, probably caused by extensional tectonism, resulting in the surfacing of bright wispy lineaments \citep{Jaumann2009}.  

\begin{figure}[h!]
	\centering
		\includegraphics[width=14.5cm]{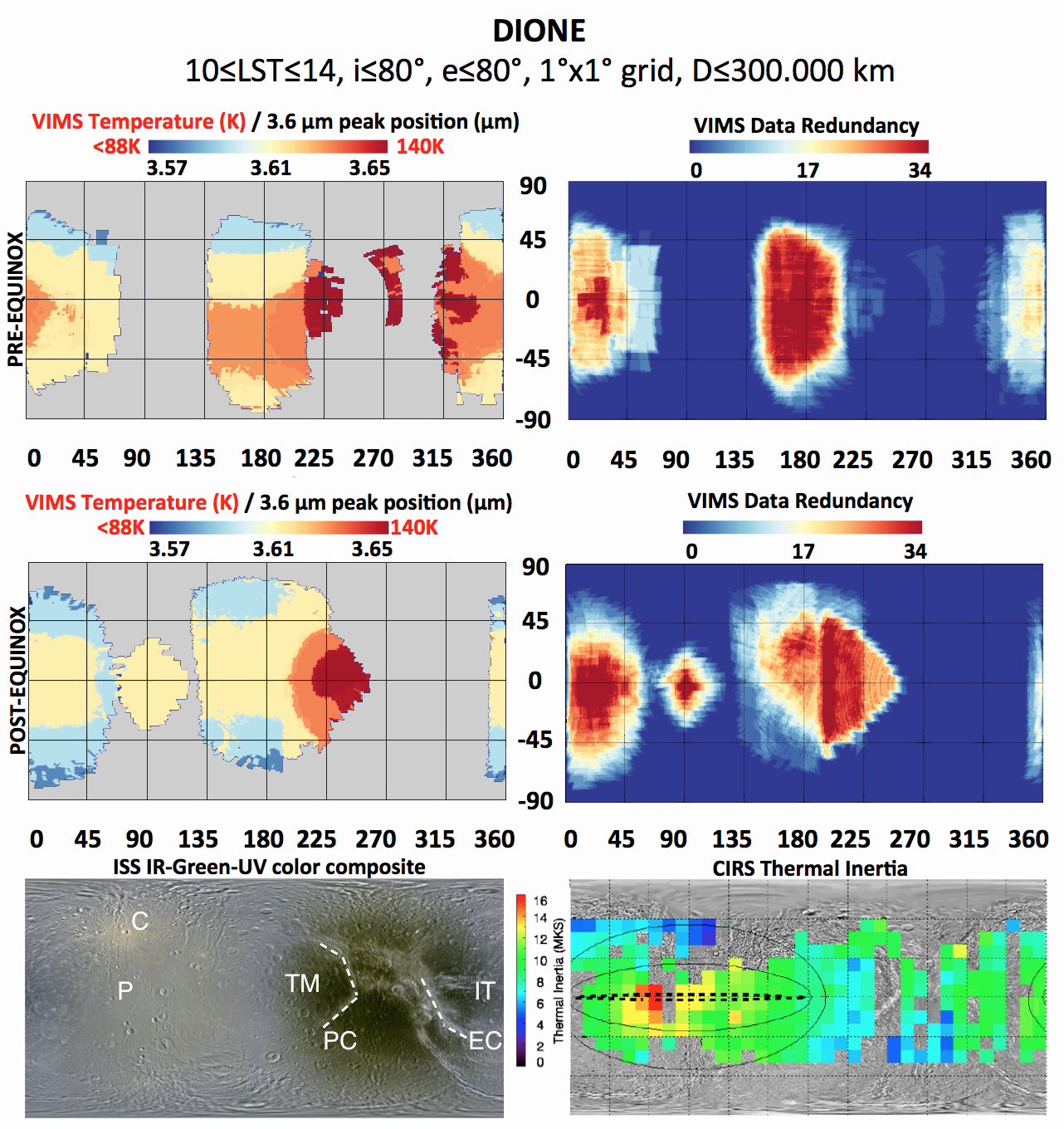}
		\caption{Dione maps. Top left panel: Pre-equinox temperature map. Top Right panel: VIMS pre-equinox data redundancy map. Center left panel: Post-equinox temperature map. Center Right panel: VIMS pre-equinox data redundancy map. Bottom left panel: context image map based on Cassini-ISS IR-Green-UV color composite (by P. Schenk, available from http://photojournal.jpl.nasa.gov/catalog/PIA18434). Locations of Ilia-Tumus craters (IT), Tiburtus-Mezentium craters (TM), Eurotas Chasmata (EC), Creusa crater (C), Padua Chasmata (PC) and Palinurus crater (P) are indicated. Bottom right panel: CIRS thermal inertia map \citep{Howett2014}.}
	\label{fig:figure11}
\end{figure}

VIMS coverage on Dione is mainly available across two wide regions centered across lon=0$^\circ$ and 180$^\circ$ meridians (see Fig. \ref{fig:figure11} - right top and center panels). The pre-equinox temperature map (Fig. \ref{fig:figure11} - top left panel) shows that the maximum temperature T=140 K is correlated with the dark material units at the center of the trailing hemisphere (see ISS IR-Green-UV color composite map in Fig. \ref{fig:figure11} - bottom left panel), corresponding to Ilia-Tumus (IT) and Tiburtus-Mezentium (TM) craters. The bright Eurotas Chasmata (EC) and the meridional part of the antisaturnian hemisphere up to lat =85$^\circ$ are cooler at T=125 (orange pixels). 

A great part of the leading hemisphere between -60$^\circ$ $\le$ lat $\le$ 40$^\circ$ is at T=115 K (yellow pixels). A temperature of 90 K (light blue pixels) is reached in the north regions, in winter season, above lat=40$^\circ$. A few pixels are at T$\le$ 88 K in the bright terrains near the recent Creusa crater (C). The low temperature measured in this area is compatible with the prominent water ice band depth measured by VIMS on the crater and on the bright ejecta rays departing from it \citep{Stephan2010, Scipioni2013}.      

The seasonal change is well-evident comparing the pre-equinox with the post-equinox map (Fig. \ref{fig:figure11} - center left panel): despite the lower coverage on the trailing hemisphere is possible to distinguish how the south hemisphere regions become cooler, down to T $\le$ 88 K for lat $\le$ 50 K (dark blue pixels). Conversely, the dark units on the trailing hemisphere remain stable at T=140 K (red pixels). In the south part of the maximum temperature spot is visible the meridional segment of Padua Chasmata (PC) where the temperature drops to T=135 K (light red pixels). The equatorial belt of the leading hemisphere is at T=115 K (yellow pixels) for lat = $\pm$ 40$^\circ$. Mid-latitude regions in both hemispheres are at T=90 K (light blue pixels).
Finally, VIMS temperature maps don't show evidence of thermal anomaly on Dione's leading hemisphere nor clear temperature changes in correspondence of the thermal inertia peak (at 16 $J \ m^{-2} \ K^{-1} \ s^{-1/2}$) above Palinurus crater (P) as reported by CIRS \citep{Howett2014} and shown in Fig. \ref{fig:figure11} - bottom right panel.

\subsection{Rhea}
\begin{figure}[h!]
	\centering
		\includegraphics[width=14.5cm]{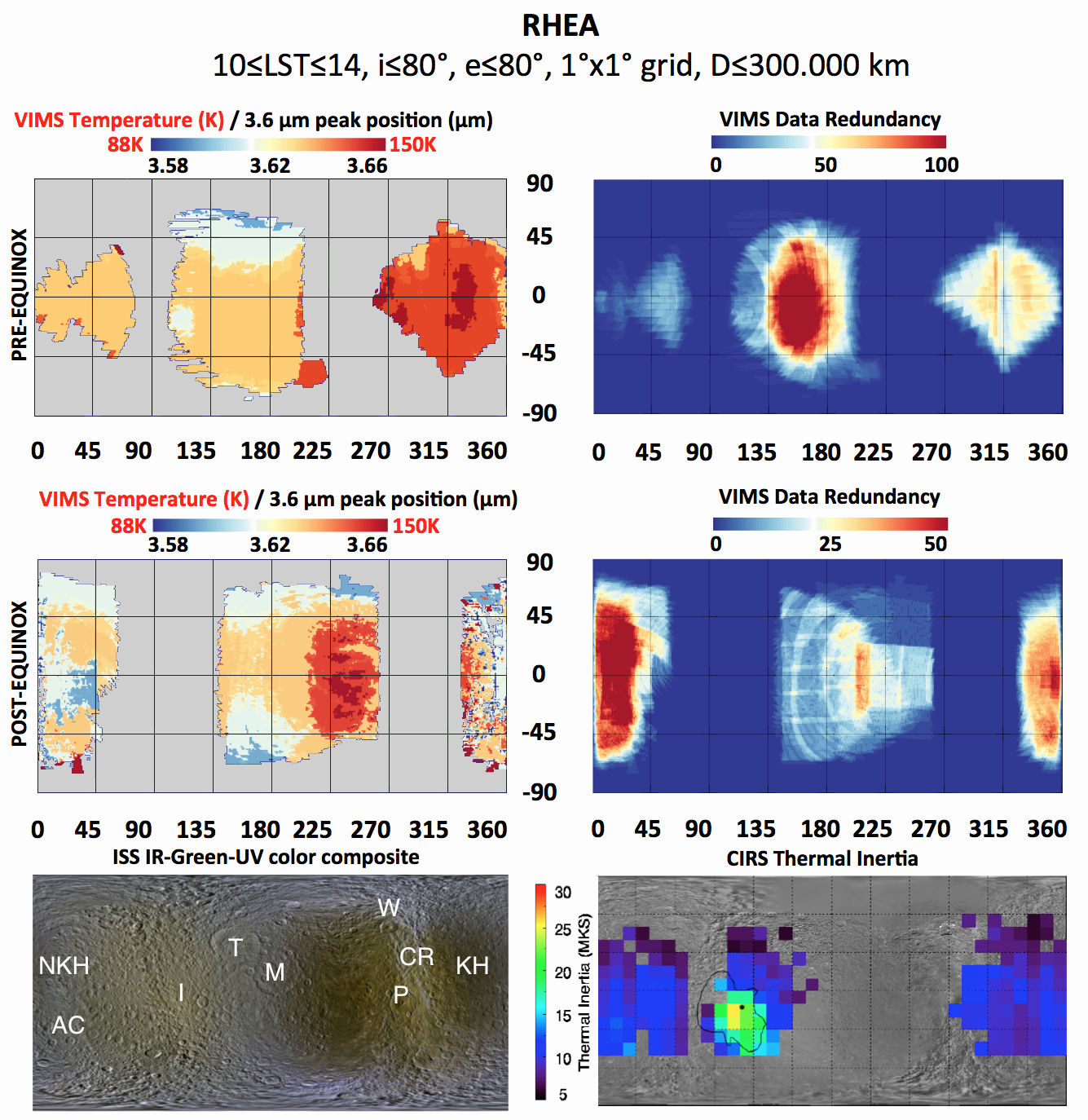}
		\caption{Rhea maps. Top left panel: Pre-equinox temperature map. Top Right panel: VIMS pre-equinox data redundancy map. Center left panel: Post-equinox temperature map. Center Right panel: VIMS pre-equinox data redundancy map. Bottom left panel: context image map based on Cassini-ISS IR-Green-UV color composite (by P. Schenk, available from http://photojournal.jpl.nasa.gov/catalog/PIA18438). Locations of Tirawa (T) and Mamaldi (M) impact basins, Kumpara-Heller craters (KH), Chasmata Region (CR), Powehiwehi crater (P), Wakonda crater (W), Inktomi crater (I), Nzame-Karora-Haoso craters (NKH) and Arunaka-Con craters (AC) are indicated. Bottom right panel: CIRS thermal inertia map \citep{Howett2014}.}
	\label{fig:figure12}
\end{figure}

Similarly to Dione, also Rhea's surface possesses geomorphological and albedo differences between the leading and trailing hemispheres. The leading hemisphere appear bright, showing a great variety of impact craters, from small ones to 400 km-wide Tirawa and Mamaldi basins. Among the small craters on the leading hemisphere is the remarkable Inktomi, the source of many extended ejecta rays. The trailing hemisphere is darker, with bright highlands crossed by a complex system of chasmata, appearing as wispy filaments, which extends along the north-south direction. A resurfacing event could be the cause of these features \citep{Jaumann2009}.
The albedo asymmetry between the two hemispheres is probably caused by the exposure of the leading surface to E ring particles while magnetosphere focuses charged particles on the trailing hemisphere \citep{Schenk2011} resulting in accumulation of non-icy material. The first process causes the layering of bright water ice grains while the second causes reddening and darkening through the implantation of charged particles.   
For reference, the ISS color context image with the position of the geomorphological features cited in the text is shown in Fig. \ref{fig:figure12} - bottom left panel.
The VIMS dataset offers a wide coverage of Rhea's surface with missing parts at about lon=90$^\circ$ and 270$^\circ$ on both pre and post-equinox maps (Fig. \ref{fig:figure12} - top and center right panels). On the pre-equinox map (Fig. \ref{fig:figure12} - top left panel) there is a clear temperature difference between the leading hemisphere (which is almost a uniform 125 K, shown by the orange pixels) and the trailing hemisphere (which has a maximum temperature of 150, shown by the red pixels, particularly around the Kumpara-Heller craters (KH) and above Powehiwehi crater (P).  Above Galunlati, Avalki, Avaiki Chasmata (or Chasmata Region, CR) running along north-south direction between lon=250$^\circ$-300$^\circ$ meridians a temperature T=140 K (light red pixels) is retrieved. The two ancient Tirawa (T) and Mamaldi (M) impact basins don't show a temperature change with respect to the surroundings. However, moving beyond the north rim of Tirawa, the temperature drops to T=110 K. 

A temperature T=135-140 K (light red pixes) is measured across the low albedo units in the trailing hemisphere, including the ones around the CR. The region around  Wakonda (W) crater appears colder at T=125 K (orange pixels).
The colder temperatures T$<$ 110 K (light and dark pixels) are measured in the north (winter) hemisphere during the pre-equinox period above lat=20$^\circ$ in the antisaturnian hemiphere. A significant temperature anomaly is observed around the Inktomi (I) crater, placed in the south (summer) leading hemisphere at lat=-12.5$^\circ$: while surrounding terrains show T=125 K the crater has a T=110 K (light blue pixels). A similar thermal anomaly feature is visible on the thermal inertia map derived by \cite{Howett2014} from Cassini-CIRS data (Fig. \ref{fig:figure12} - bottom right panel). 
VIMS and CIRS are in agreement because an increase of thermal anomaly causes a decrease of daytime temperature.
Temperature differences between leading and trailing hemisphere continue to be present in the post-equinox map (Fig. \ref{fig:figure12} - center left panel). The equatorial region of the leading hemisphere is at T=120 K (light orange pixles) with the exclusion of a wide region at T$<$110 K (light and dark pixels) spanning from equator close to Nzame-Karora-Haoso craters (NKH) to lat=-30$^\circ$, where Arunaka-Con craters (AC) are located. Another region with similar low temperature is observed in the antisaturnian meridional quadrant (around lon=150$^\circ$, lat=-50$^\circ$) and seems to be associated with the bright material surrounding the dark units in the trailing hemisphere (see ISS color context map in Fig. \ref{fig:figure12} - bottom left panel). Large variations of temperature are seen in the surroundings of CR on the trailing hemisphere. These are caused by the albedo differences between the bright rays departing from the CR features, which are at T=125 K (orange pixels) and the neighbor dark material units between T=150 K (red pixels) and T=135-140 (light red pixels).

\subsection{Hyperion}
Hyperion's irregular and porous body shows two distinct morphological units. A strong water ice signature and high albedo characterize the first unit class. Carbon dioxide and C-N, or possibly adsorbed H$_2$ \citep{Clark2012}, are linked to dark material accumulated on the bottoms of cup-like craters, forming a second unit \citep{Cruikshank2007}.

\begin{figure}[h!]
	\centering
		\includegraphics[width=14.5cm]{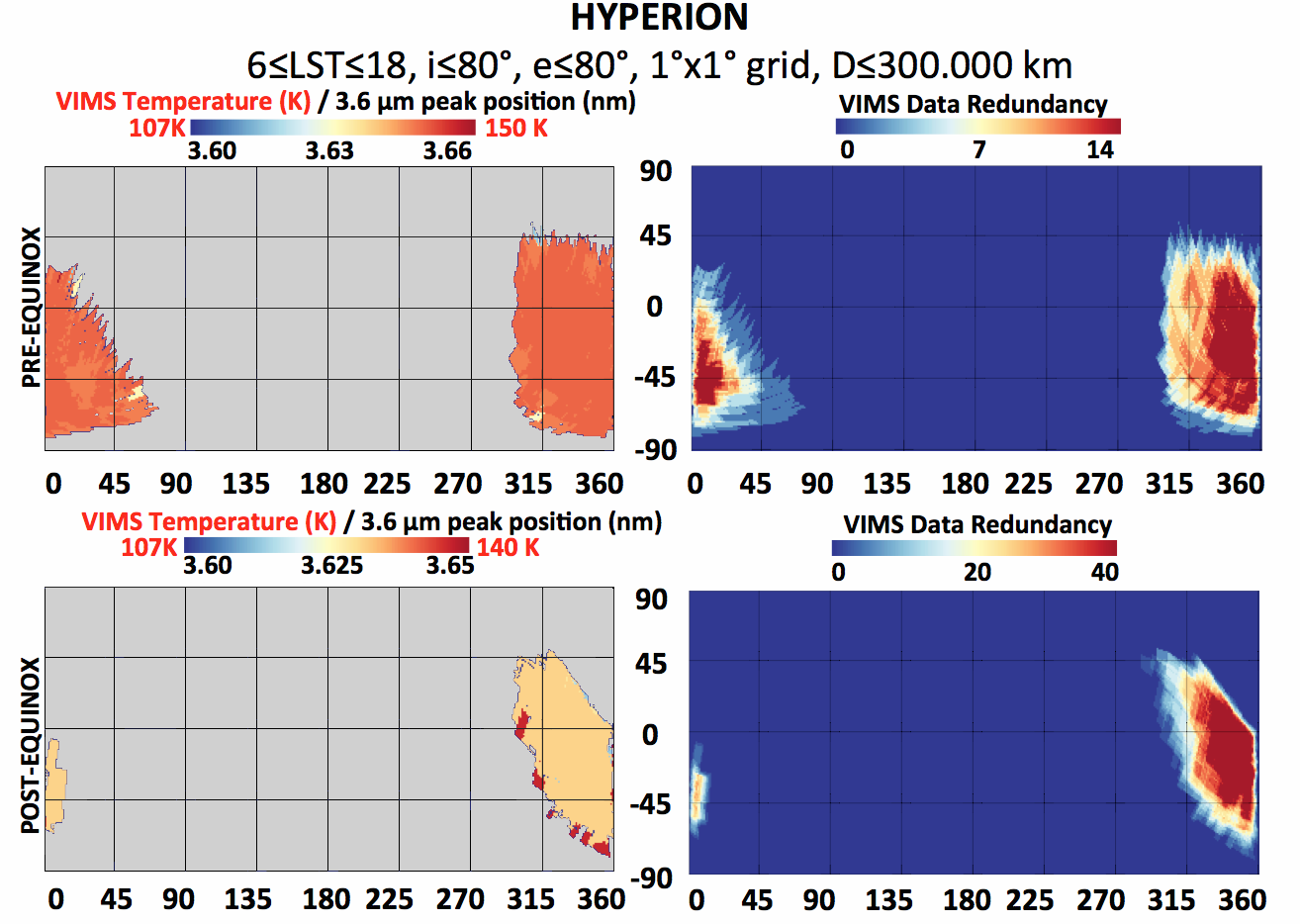}
		\caption{Hyperion maps. Top left panel: Pre-equinox temperature map for 6$\le$LST$\le$18. Top Right panel: VIMS pre-equinox data redundancy map. Center left panel: Post-equinox temperature map for 6$\le$LST$\le$18. Center Right panel: VIMS pre-equinox data redundancy map.}
	\label{fig:figure13}
\end{figure}

The limited number of available Hyperion observations fulfilling the selection rules exposed at the beginning of this section has forced us to increase the range of local solar times from 10$\le$LST$\le$14 hr to 6$\le$LST$\le$18 hr. As a consequence of this change, and differently from all the other satellites except Iapetus, for Hyperion we are reporting average daytime temperature maps in Fig. \ref{fig:figure13} - top and bottom left panels. The satellite's surface temperature is quite uniform on both pre and post-equinox datasets, equal to T=140 K and T=130 K, respectively. This difference between the pre and post-equinox periods is probably caused by the averaging of daytime temperature taken at very different local times. Certainly VIMS data indicate that the Hyperion's surface is almost isothermal.

\subsection{Iapetus}
At a mean distance of 59 Saturn radii from Saturn, Iapetus is the only regular moon in the Solar System orbiting at the largest distance from its planet. Moreover, the moon is orbiting outside Saturn's magnetosphere (which extends up to 20 Saturn's radii, \cite{Gombosi2009}) thus exposing the surface to solar weathering. Iapetus' leading hemisphere surface, corresponding to the Cassini Regio (CR), appears strongly contaminated by a layer of dark organic material and carbon dioxide \citep{Buratti2005, Filacchione2007, Cruikshank2008, Cruikshank2010, Clark2012}. The trailing hemisphere shows a more pristine water ice-rich surface across the northern Roncevaux Terra (RT) and the southern Saragossa Terra (ST) where the large Engelier crater (E) is located \citep{Denk2010}. The deposition of dark material and the high diurnal temperatures measured on the leading hemisphere have been suggested as possible causes of thermal migration of water ice \citep{SpencerDenk2010}.
   
A discussion about the origin and composition of the aromatic and aliphatic hydrocarbons seen on Iapetus' leading hemisphere surface is given by \cite{Cruikshank2014}. Different pathways have been investigated to trace the possible sources of the dark material, including the transfer of dust from Phoebe \citep{Burns1979, Tosi2010, Tamayo2011}, the dynamics of Phoebe's ring particles \citep{Verbiscer2009, Hamilton2015} and the dust resulting from collisions among the outer irregular satellites \citep{Bottke2010, Tosi2010, Tamayo2011}. Morevoer, Cassini Radar observations at 2.2 cm wavelength taken above the leading hemisphere have indicated that Iapetus' icy subsurface is more consolidated than the uppermost dark material layers \citep{Legall2014}.
   
\begin{figure}[h!]
	\centering
		\includegraphics[width=14.5cm]{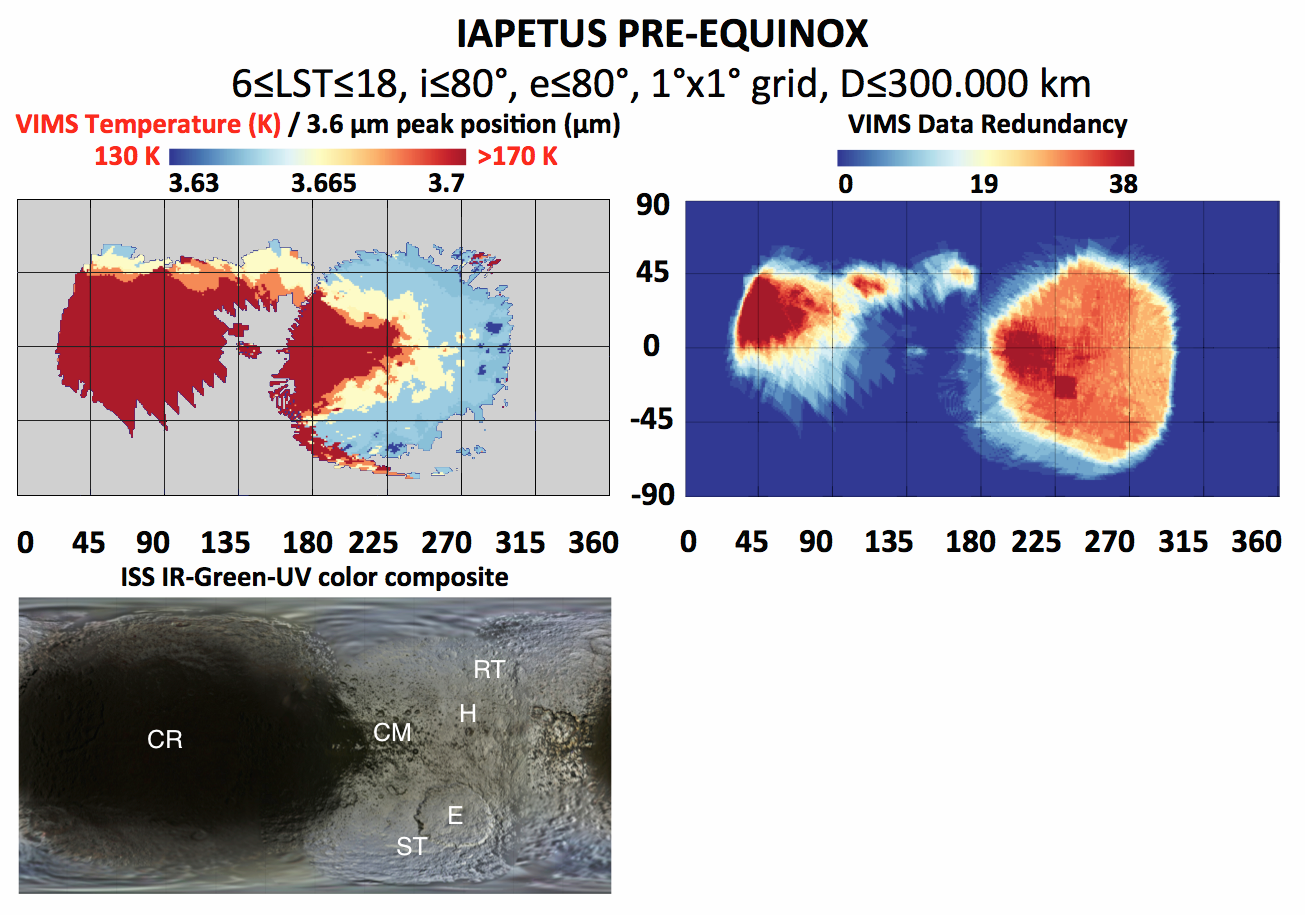}
		\caption{Iapetus maps. Top left panel: Pre-equinox temperature map. Top Right panel: VIMS pre-equinox data redundancy map. Bottom left panel: context image map based on Cassini-ISS IR-Green-UV color composite (by P. Schenk, available from http://photojournal.jpl.nasa.gov/catalog/PIA18436). Locations of Cassini Regio (CR), Carcassone Montes (CM), Roncevaux Terra (RT), Saragossa Terra (ST), Hamon crater (H) and Engelier crater (E) are indicated.}
	\label{fig:figure14}
\end{figure}

Only pre-equinox data are available for Iapetus, but they give nearly complete coverage  in the 30$^\circ$ $\le$ lon $\le$ 300$^\circ$ range (Fig. \ref{fig:figure14} - top right panel). For Iapetus, as for Hyperion, the interval of local solar times has been increased from 10$\le$LST$\le$14 hr to 6$\le$LST$\le$18 hr in order to achieve better spatial coverage. Therefore the VIMS temperature map of Iapetus shown in Fig. \ref{fig:figure14} - top left panel, is built by selecting observations taken during all daytime hours. The maximum retrieved temperature T$>$ 170 K is measured above the dark Cassini Regio (CR) in the leading hemisphere where it appears uniformly distributed and well correlated with the dark material unit. Moving towards the trailing hemisphere, VIMS has observed the Carcassone Montes (CM) transition region between the dark and bright units where temperature gradually drops to T=160 K (orange pixels) and then to T=150 K (yellow pixels). This is the same region where \cite{Denk2010} have observed color differences within the dark terrain, which shows a less reddish color than the uniformly dark CR.

The bright icy terrains on the north trailing hemisphere in Roncevaux Terra (RT) and in the south Saragossa Terra (ST) are at T=140 K (light blue). The Hamon crater (H) and other three nearby craters, whose floors are filled by dark material, are remarkably warmer than neighboring terrains: around these craters a temperature peak of T=160 K with an external corona at T=150 K are measured, resulting in a clear contrast to the nearby terrains at T=140 K (light blue).
Finally, a thermal contrast is seen between the central peak of the large Engelier crater (E) at T=150 K (yellow pixels) and the crater's floor at T=140 K (light blue).

\section{Conclusions}\label{sct:f}

\begin{figure}[h!]
	\centering
		\includegraphics[width=16cm]{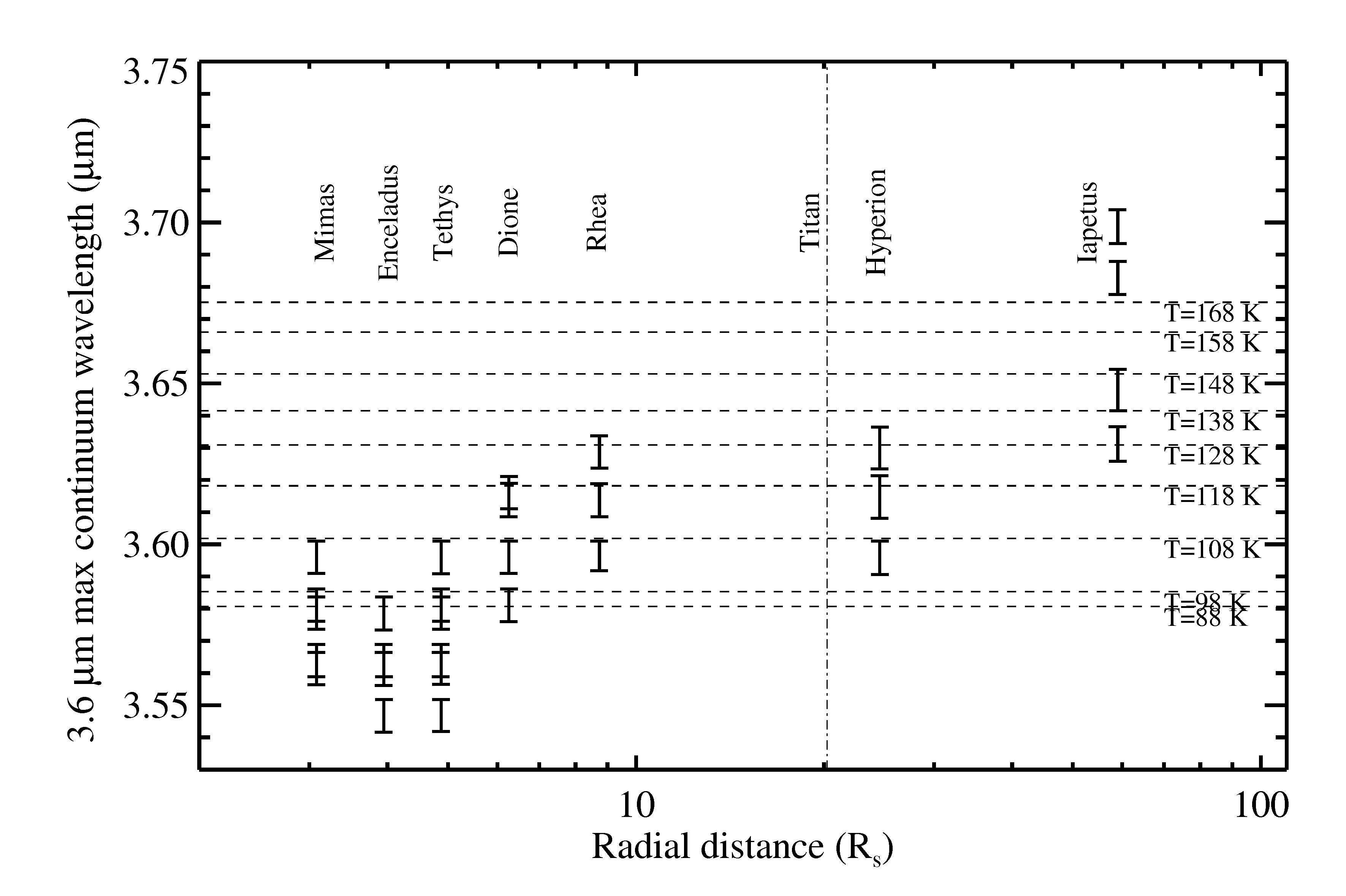}
		\caption{Temperature distribution as function of satellites orbital distance from Saturn as derived from VIMS disk-integrated observations.}
	\label{fig:figure15}
\end{figure}

Daytime temperature of Saturn's icy satellites are derived from Cassini-VIMS data by fitting the position of the continuum at 3.6 $\mu m$ on I/F spectra of water ice. This spectral feature is temperature-dependent and has been calibrated by using laboratory data at some specific conditions. However this retrieval is based on the assumption of pure water ice composition. The presence of other endmembers could alter the shape of the continuum at 3.6 $\mu m$, implying uncertainties in the temperature derivation process. However, since water ice is the dominant species on many satellites and the 3.6 $\mu m$ peak is always present in their spectra, even on Iapetus' dark terrain \citep{Filacchione2007}, we have assumed that the method can be exploited within the framework of these limitations. The availability of laboratory reflectance data allows the derivation of temperature in the range 88$\le$T$\le$168 K. The error associated with the fitting technique is of the order of 5 K.  
Further modeling is necessary to disentangle composition, including contaminant fraction, mixing modality and grain size distribution, from thermal inertia. While composition can be inferred from VIMS data, the 3.6 $\mu m$ reflectance peak method allows us to determine only diurnal temperatures. In absence of night time temperature measurements, thermal inertia cannot be properly inferred. 
The method has been applied to different VIMS datasets with different spatial resolutions, from disk-integrated observations to regional coverage data suitable to build maps through data filtering, averaging and mapping techniques.
Disk-integrated observations allow us to explore large scale thermal response between leading and trailing hemispheres of the satellites, resulting in the following properties:
\begin{enumerate} 
\item Mimas, Enceladus and Hyperion don't show noteworthy temperature differences between the two hemispheres.
\item Satellites possessing an albedo asymmetry between leading and trailing hemispheres, e.g., Tethys, Dione and Rhea, show temperature higher of about 10 K on the dark trailing than on the bright leading hemisphere;
\item Much higher temperature is observed on the dark leading hemisphere of Iapetus than on the bright trailing hemisphere.
\end{enumerate}
Since at disk-integrated spatial resolution is not possible to resolve geomorphological classes of interest on the surfaces, the temperature is correlated with albedo at hemispherical scale. Moreover, VIMS results clearly show the existence of temperature trend among satellites. In fact, temperature increases with the orbital distance from Saturn: at T$\le$ 88 K Mimas, Enceladus and Tethys are the colder satellites (see Fig. \ref{fig:figure15}). The minimum position of the 3.6 $\mu m$ peak is observed on Enceladus. Moving towards larger orbital distances the noon local time temperature progressively increases: on Dione T=98-118 K, on Rhea T=108-128 K, on Hyperion T=118-128 K. On Iapetus' trailing hemisphere T=128-148 K is measured. The maximum temperature T$>$168 K is observed on Iapetus leading hemisphere. 
This trend matches the distribution of the albedo and consequently of water ice and chromophores across the Saturn satellites' surfaces. Compositional gradients have been traced by means of the 2.0 $\mu m$ water ice band depth and the visible spectral slopes derived from VIMS reflectance spectra: while the water ice abundance is almost constant, spectral slopes show a reddening as they move from inner to outer satellites \citep{Filacchione2013}.

The correlation of temperature with geomorphological features becomes evident when VIMS observations taken at regional scale are projected in cylindrical maps. VIMS temperature maps taken around noon time (10$\le$LST$\le$14) show the following findings:   
\begin{itemize}
\item Detection of the thermal anomaly across the equatorial lens of Mimas and Tethys;
\item Measurement of the temperature T$>$115K close to Enceladus' Damascus and Alexandria sulci in the south pole region;
\item Verification of seasonal temperature changes: on Tethys, Dione and Rhea higher temperature are measured above the south hemisphere during pre-equinox and above north hemisphere during post-equinox epochs; 
\item The temperature distribution appears correlated with albedo features: the temperature increases on low albedo units located on Tethys, Dione and Rhea trailing hemispheres. 
\item The thermal anomaly region on Rhea's Inktomi crater detected by CIRS \citep{Howett2014} is confirmed by VIMS: this area appears colder with respect to surrounding terrains when observed at the same local solar time. 
\end{itemize}
Moreover, the method is able to follow daytime temperature variations: on the equatorial region of Enceladus values of 88$\le$T$\le$110 K have been measured between 6$\le$LST$\le$16 hr.

VIMS daytime temperature maps (6$\le$LST$\le$18 hr) are available for Hyperion, which appears remarkably uniform at about T=130-140 K, and for Iapetus, which shows a remarkable temperature asymmetry between the dark Cassini Regio (T$>$ 168 K) and the bright plains on the trailing hemisphere (T=140 K).

Considering the high temperatures derived by VIMS, a question arises: how is possible to maintain $CO_2$ on the warm surface units of Iapetus, where T$>$168 K? These high temperatures put some severe constraints on the occurrence of $CO_2$. VIMS has observed the 4.255 $\mu m$ absorption band of $CO_2$ adsorpted in water ice above Iapetus' dark terrain units \citep{Buratti2005, Filacchione2007, Cruikshank2010}. Since carbon dioxide is not among the pristine volatile species available at the time of formation of the Saturn's satellites, or all of the original $CO_2$ near the surface has been lost, or it must be synthesized by UV and cosmic rays irradiation of water ice and a carbon source \citep{Moore1983} in the lower subsurface layers. If all the $CO_2$ is complexed with $H_2O$, as discussed in \cite{Cruikshank2010}, and is a true clathrate, then the $H_2O$ cage that holds the $CO_2$ can maintain it even at relatively high temperatures: \cite{SandfordAllamandola1990} have observed $CO_2$ bound to $H_2O$ ice up to T=150 K. 
More recent studies have reported that the $H_2O-CO_2$ clathrate hydrates undergo to "self-preservation" mode, an anomalous slow decomposition of the cage happening below the melting point of the ice \citep{Oancea2012}. In this regime the $CO_2$ hydrated core is maintained insulated by a very thin water ice crust preventing the sublimation of the $CO_2$ \citep{FalentyKuhs2009}. This phenomenon is commonly observed at 1 bar between T=200-270 K, but it has also been observed under vacuum or at moderate pressure. A similar scenario is therefore plausible to maintain $CO_2$ bound to $H_2O$ at T$>$ 168 K on Iapetus dark terrains.

Keeping in mind its intrinsic limitations, the 3.6 $\mu m$ continuum peak wavelength technique can be employed to remotely sense surfaces' temperature of water ice-rich bodies in the outer solar system. In these cases infrared spectroscopy offers the major advantage to allow us to retrieve simultaneously the surface composition, grain size distribution and temperature from the same dataset.

\section*{Acknowledgments}
The authors acknowledge the financial support from INAF-IAPS, ASI-Italian Space Agency and NASA through the Cassini project. We gratefully thank the Cassini Icy Satellites Working group, VIMS technical team at Lunar and Planetary Lab (University of Arizona, Tucson, AZ) and the Cassini-Huygens Project at JPL (Pasadena, CA) for observations planning, sequencing and data archiving. Without all them this study would be impossible. This research has made use of NASA's Astrophysics Data System.


\section*{References}
\begin{thebibliography}{80}

\bibitem[Acton(1996)]{Acton1996}
Acton, C. H., 1996. Ancillary data services of NASA's Navigation and Ancillary Information Facility, \emph{PSS}, 44, Issue 1, 65-70, doi: 10.1016/0032-0633(95)00107-7

\bibitem[Bottke et al.(2010)]{Bottke2010}
Bottke, W. F., Nesvorny, D., Vokrouhlicky, D. Morbidelli, A., 2010. The Irregular Satellites: The Most Collisionally Evolved Populations in the Solar System, \emph{AJ}, 139, 994-1014, doi: 10.1088/0004-6256/139/3/994.


\bibitem[Brown et al.(2004)]{Brown2004}
Brown, R.H., Baines, K.H., Bellucci, G., Bibring, J.-P., Buratti, B.J., Capaccioni, F., Cerroni, P., Clark, R.N., Coradini, A., Cruikshank, D.P., Drossart, P., Formisano, V., Jaumann, R., Langevin, Y., Matson, D.L., McCord, T.B., Mennella, V., Miller, E., Nelson, R.M., Nicholson, P.D., Sicardy, B., Sotin, C., 2004. The Cassini Visual and Infrared Mapping Spectrometer (VIMS) investigation. \emph{Space Sci. Rev.} 115 (1-4), 111-168.

\bibitem[Brown et al.(2006)]{Brown2006}
Brown, R.~H., Clark, R.~N., Buratti, B.~J., Cruikshank, D.~P., Barnes, J.~W., Mastrapa, R.~M.~E., Bauer, J. Newman, S., Momary, T., Baines, K.~H., Bellucci, G., Capaccioni, F., Cerroni, P., Combes, M., Coradini, A., Drossart, P., Formisano, V., Jaumann, R., Langevin, Y., Matson, D.~L., McCord, T.~B., Nelson, R.~M., Nicholson, P.~D., Sicardy, B., Sotin, C., 2006. Composition and Physical Properties of Enceladus' Surface, \emph{Science}, 311, 1425-1428, doi: 10.1126/science.1121031.

\bibitem[Buratti et al.(1990)]{Buratti1990}
Buratti, B.J., Mosher, J.A., Johnson, T.V., 1990. Albedo and color maps of the saturnian satellites, \emph{Icarus}, 87, 339-357.

\bibitem[Buratti et al.(2005)]{Buratti2005}
Buratti, B. J., Cruikshank, D. P., Brown, R. H., Clark, R. N., Bauer, J. M., Jaumann, R., McCord, T. B., Simonelli, D. P., Hibbitts, C. A., Hansen, G. B., Owen, T. C., Baines, K. H., Bellucci, G., Bibring, J.-P., Capaccioni, F., Cerroni, P., Coradini, A., Drossart, P., Formisano, V., Langevin, Y., Matson, D. L., Mennella, V., Nelson, R. M., Nicholson, P. D., Sicardy, B., Sotin, C., Roush, T. L., Soderlund, K., Muradyan, A., 2005. Cassini Visual and Infrared Mapping Spectrometer Observations of Iapetus: Detection of $CO_2$, \emph{ApJ}, 622, L149-L152, doi: 10.1086/429800. 


\bibitem[Buratti et al.(2008)]{Buratti2008}
Buratti, B. J., Soderlund, K., Bauer, J., Mosher, J. A., Hicks, M. D., Simonelli, D. P., Jaumann, R., Clark, R. N., Brown, R. H., Cruikshank, D. P., Momary, T., 2008. Infrared (0.83-5.1 $\mu$ m) photometry of Phoebe from the Cassini Visual Infrared Mapping Spectrometer, \emph{Icarus}, 193, 309-322, doi: 10.1016/j.icarus.2007.09.014.

\bibitem[Burns et al.(1979)]{Burns1979}
Burns, J. A., Lamy, P. L., and Soter, S., 1979. Radiation forces on small particles in the solar system, \emph{Icarus}, 40, 1-48.

\bibitem[Carlson et al.(1996)]{Carlson1996}
Carlson, R., Smythe, W., Baines, K., Barbinis, E., Becker, K., Burns, R., Calcutt, S., Calvin, W., Clark, R., Danielson, G., Davies, A., Drossart, P., Encrenaz, T., Fanale, F., Granahan, J., Hansen, G., Herrera, P., Hibbitts, C., Hui, J., Irwin, P., Johnson, T., Kamp, L., Kieffer, H., Leader, F., Lellouch, E., Lopes-Gautier, R., Matson, D., McCord, T., Mehlman, R., Ocampo, A., Orton, G., Roos-Serote, M., Segura, M., Shirley, J., Soderblom, L., Stevenson, A., Taylor, F., Torson, J., Weir, A., Weissman, P., 1996. Near-Infrared Spectroscopy and Spectral Mapping of Jupiter and the Galilean Satellites: Results from Galileo's Initial Orbit, \emph{Science}, 274, 385-388, doi: 10.1126/science.274.5286.385

\bibitem[Ciarniello et al.(2011)]{Ciarniello2011}
Ciarniello, M., Capaccioni, F., Filacchione, G., Clark, R. N., Cruikshank, D. P., Cerroni, P., Coradini, A., Brown, R. H., Buratti, B. J., Tosi, F., Stephan, K., 2011. Hapke modeling of Rhea surface properties through Cassini-VIMS spectra, \emph{Icarus}, 214, 541-555, DOI:10.1016/j.icarus.2011.05.010.

\bibitem[Ciarniello et al.(2014)]{Ciarniello2014}
Ciarniello, M., Capaccioni, F., Filacchione, G., 2014. A test of Hapke's model by means of Monte Carlo ray-tracing, \emph{Icarus}, 237, 293-305, DOI: 10.1016/j.icarus.2014.04.045.


\bibitem[Clark et al.(2005)]{Clark2005}
Clark, R.~N., Brown, R.~H., Jaumann, R., Cruikshank, D.~P., Nelson, R.~M., Buratti, B.~J., McCord, T.~B., Lunine, J., Baines, K.~H., Bellucci, G., Bibring, J.-P., Capaccioni, F., Cerroni, P., Coradini, A., Formisano, V., Langevin, Y., Matson, D.~L., Mennella, V., Nicholson, P.~D., Sicardy, B., Sotin, C., Hoefen, T.~M., Curchin, J.~M., Hansen, G., Hibbitts, K., Matz, K.-D., 2005. Compositional maps of Saturn's moon Phoebe from imaging spectroscopy, \emph{Nature}, 435, 66-69, doi: 10.1038/nature03558.

\bibitem[Clark et al.(2008)]{Clark2008}
Clark, R.~N., Curchin, J.~M., Jaumann, R., Cruikshank, D.~P., Brown, R.~H., Hoefen, T.~M., Stephan, K., Moore, J.~M., Buratti, B.~J., Baines, K.~H., Nicholson, P.~D., Nelson, R.~M., 2008. Compositional mapping of Saturn's satellite Dione with Cassini VIMS and implications of dark material in the Saturn system, \emph{Icarus}, 193, 372-386, doi: 10.1016/j.icarus.2007.08.035.
	
\bibitem[Clark et al.(2012)]{Clark2012}
Clark, R. N., Cruikshank, D. P., Jaumann, R., Brown, R. H., Curchin, J. M., Hoefen, T. D., Stephan, K., Buratti, B. J., Filacchione, G., Baines, K. H., Nicholson, P. D., 2012. The Composition of Iapetus: Mapping Results from Cassini VIMS, 2012. \emph{Icarus}, 218, 831-860, doi:10.1016/j.icarus.2012.01.008

\bibitem[Coradini et al.(2008)]{Coradini2008}
Coradini, A., Tosi, F., Gavrishin, A.~I., Capaccioni, F., Cerroni, P., Filacchione, G., Adriani, A., Brown, R.~H., 	Bellucci, G., Formisano, V., D'Aversa, E., Lunine, J.~I., Baines, K.~H., Bibring, J.-P., Buratti, B.~J., Clark, R.~N., Cruikshank, D.~P., Combes, M., Drossart, P., Jaumann, R., Langevin, Y., Matson, D.~L., McCord, T.~B., Mennella, V., 	Nelson, R.~M., Nicholson, P.~D., Sicardy, B., Sotin, C., Hedman, M.~M., Hansen, G.~B., Hibbitts, C.~A., Showalter, M., Griffith, C., Strazzulla, G., 2008. Identification of spectral units on Phoebe, \emph{Icarus}, 193, 233-251, doi: 10.1016/j.icarus.2007.07.023.

\bibitem[Coradini et al.(2011)]{Coradini2011}
Coradini, A. and 48 colleagues, 2011. The Surface Composition and Temperature of Asteroid 21 Lutetia As Observed by Rosetta/VIRTIS. \emph{Science}, 334, 492-494, doi: 10.1126/science.1204062

\bibitem[Cruikshank et al.(2007)]{Cruikshank2007}
Cruikshank, D.~P., Dalton, J.~B., Ore, C.~M.~D., Bauer, J., Stephan, K., Filacchione, G., Hendrix, A.~R., Hansen, C.~J., Coradini, A., Cerroni, P., Tosi, F., Capaccioni, F., Jaumann, R., Buratti, B.~J., Clark, R.~N., Brown, R.~H., Nelson, R.~M., McCord, T.~B., 	Baines, K.~H., Nicholson, P.~D., Sotin, C., Meyer, A.~W., Bellucci, G., Combes, M., Bibring, J.-P., Langevin, Y., Sicardy, B., Matson, D.~L., Formisano, V., Drossart, P., Mennella, V., 2007. Surface composition of Hyperion, \emph{Nature}, 448, 54-56, doi: 10.1038/nature05948.
	
\bibitem[Cruikshank et al.(2008)]{Cruikshank2008}	
Cruikshank, D.~P., Wegryn, E., Dalle Ore, C.~M., Brown, R.~H., Bibring, J.-P., Buratti, B.~J., Clark, R.~N., McCord, T.~B., Nicholson, P.~D., Pendleton, Y.~J., Owen, T.~C., Filacchione, G., Coradini, A., Cerroni, P., Capaccioni, F., Jaumann, R., Nelson, R.~M., Baines, K.~H., Sotin, C., Bellucci, G.,Combes, M., Langevin, Y., Sicardy, B., Matson, D.~L., Formisano, V., Drossart, P., Mennella, V., 2008. Hydrocarbons on Saturn's satellites Iapetus and Phoebe, \emph{Icarus}, 193, 334-343, doi: 10.1016/j.icarus.2007.04.036.	

\bibitem[Cruikshank et al.(2010)]{Cruikshank2010}	
Cruikshank, D.~P., Meyer, A.~W., Brown, R.~H., Clark, R.~N., Jaumann, R., Stephan, K., Hibbitts, C.~A., Sandford, S.~A., Mastrapa, R.~M.~E., Filacchione, G., Ore, C.~M.~D., Nicholson, P.~D., Buratti, B.~J., McCord, T.~B., Nelson, R.~M., Dalton, J.~B., Baines, K.~H., Matson, D.~L., 2010. Carbon dioxide on the satellites of Saturn: Results from the Cassini VIMS investigation and revisions to the VIMS wavelength scale, \emph{Icarus}, 206, 561-572, doi: 10.1016/j.icarus.2009.07.012.

\bibitem[Cruikshank et al.(2014)]{Cruikshank2014}	
Cruikshank, D. P., Dalle Ore, C. M., Clark, R. N., Pendleton, Y. J., 2014. Aromatic and aliphatic organic materials on Iapetus: Analysis of Cassini VIMS data, \emph{Icarus}, 233, 306-315, doi: 10.1016/j.icarus.2014.02.011.

\bibitem[Denk et al.(2010)]{Denk2010}
Denk, T., Neukum, G., Roatsch, T., Porco, C. C., Burns, J. A., Galuba, G. G., Schmedemann, N., Helfenstein, P., Thomas, P. C., Wagner, R. J., West, R. A., 2010. Iapetus: Unique Surface Properties and a Global Color Dichotomy from Cassini Imaging, \emph{Science}, 327, 435-439, doi: 10.1126/science.1177088


\bibitem[Falenty and Kuhs(2009)]{FalentyKuhs2009}
Falenty, A, Kuhs, W. F., 2009. Self-Preservation of CO2 Gas Hydrates Surface Microstructure and Ice Perfection, \emph{J. Phys. Chem. B}, 113 (49), 15975-15988, doi: 10.1021/jp906859a.

\bibitem[Filacchione et al.(2007)]{Filacchione2007}
Filacchione, G., Capaccioni, F., McCord, T.~B. Coradini, A., Cerroni, P., Bellucci, G., Tosi, F., D'Aversa, E., Formisano, V., Brown, R.~H., Baines, K.~H., Bibring, J.~P., Buratti, B.~J., Clark, R.~N., Combes, M., Cruikshank, D.~P., Drossart, P., Jaumann, R., Langevin, Y., Matson, D.~L., Mennella, V., Nelson, R.~M., Nicholson, P.~D., Sicardy, B., Sotin, C., Hansen, G., Hibbitts, K., Showalter, M., Newman, S., 2007. Saturn's icy satellites investigated by Cassini-VIMS. I. Full-disk properties: 350-5100 nm reflectance spectra and phase curves, \emph{Icarus}, 186, 259-290, doi: 10.1016/j.icarus.2006.08.001.

\bibitem[Filacchione et al.(2010)]{Filacchione2010}
Filacchione, G., Capaccioni, F., Clark, R.~N., Cuzzi, J.~N., Cruikshank, D.~P., Coradini, A., Cerroni, P., Nicholson, P.~D., McCord, T.~B., Brown, R.~H., Buratti, B.~J., Tosi, F., Nelson, R.~M., Jaumann, R., Stephan, K., 2010. Saturn's icy satellites investigated by Cassini-VIMS. II. Results at the end of nominal mission, \emph{Icarus}, 206, 507-523, doi: 10.1016/j.icarus.2009.11.006.

\bibitem[Filacchione et al.(2012)]{Filacchione2012}
Filacchione, G., Capaccioni, F., Ciarniello, M., Clark, R. N., Cuzzi, J. N., Nicholson, P. D., Cruikshank, D. P., Hedman, M. M., Buratti, B. J., Lunine, J. I., Soderblom, L. A., Tosi, F., Cerroni, P., Brown, R. H., McCord, T. B., Jaumann, R., Stephan, K., Baines, K. H., Flamini, E., 2012. Saturn's icy satellites and rings investigated by Cassini-VIMS: III - Radial compositional variability, \emph{Icarus}, 220, 1064-1096, doi: 10.1016/j.icarus.2012.06.040.

\bibitem[Filacchione et al.(2013)]{Filacchione2013}
Filacchione, G., Capaccioni, F., Clark, R. N., Nicholson, P. D., Cruikshank, D. P., Cuzzi, J. N., Lunine, J. I., Brown, R. H., Cerroni, P., Tosi, F., Ciarniello, M., Buratti, B. J., Hedman, M. M., Flamini, E., 2013. The radial distribution of water ice and chromophores across Saturn's system, \emph{ApJ}, 766, issue 2, article id. 76, doi: 10.1088/0004-637X/766/2/76.

\bibitem[Filacchione et al.(2014)]{Filacchione2014}
Filacchione, G., Ciarniello, M., Capaccioni, F., Clark, R. N., Nicholson, P. D., Hedman, M. M., Cuzzi, J. N., Cruikshank, D. P., Dalle Ore, C. M., Brown, R. H., Cerroni, P., Altobelli, N., Spilker, L. J., 2014. Cassini-VIMS observations of Saturn's main rings: I. Spectral properties and temperature radial profiles variability with phase angle and elevation, \emph{Icarus}, 241, 45-65, doi: 10.1016/j.icarus.2014.06.001.

\bibitem[Fink and Larson(1975)]{FinkLarson1975}
Fink, U., Larson, H. P., 1975. Temperature dependence of the water ice spectrum between 1 and 4microns: application to Europa, Ganymede and Saturn's rings. \emph{Icarus}, 24, 411-420.

\bibitem[Flasar et al.(2004)]{Flasar2014}
Flasar, F. M., Kunde, V. G., Abbas, M. M., Achterberg, R. K., Ade, P., Barucci, A., Bézard, B., Bjoraker, G. L., Brasunas, J. C., Calcutt, S., Carlson, R., Césarsky, C. J., Conrath, B. J., Coradini, A., Courtin, R., Coustenis, A., Edberg, S., Edgington, S., Ferrari, C., Fouchet, T., Gautier, D., Gierasch, P. J., Grossman, K., Irwin, P., Jennings, D. E., Lellouch, E., Mamoutkine, A. A., Marten, A., Meyer, J. P., Nixon, C. A., Orton, G. S., Owen, T. C., Pearl, J. C., Prangé, R., Raulin, F., Read, P. L., Romani, P. N., Samuelson, R. E., Segura, M. E., Showalter, M. R., Simon-Miller, A. A., Smith, M. D., Spencer, J. R., Spilker, L. J., Taylor, F. W., 2004. Exploring The Saturn System In The Thermal Infrared: The Composite Infrared Spectrometer, \emph{Space Science Reviews},15, 1-4, 169-297. doi: 10.1007/s11214-004-1454-9. 

\bibitem[Fletcher(1970)]{Fletcher1970}
Fletcher, N. H., 1970. The Chemical Physics of Ice,  Cambridge University Press.

\bibitem[Goguen et al.(2013)]{Goguen2013}
Goguen, J. D., Buratti, B. J., Brown, R. H., Clark, R. N., Nicholson, P. D., Hedman, M. M., Howell, R. R., Sotin, C., Cruikshank, D. P., Baines, K. H., Lawrence, K. J., Spencer, J. R., Blackburn, D. G., 2013. The temperature and width of an active fissure on Enceladus measured with Cassini VIMS during the 14 April 2012 South Pole flyover, \emph{Icarus}, 226, 1128-1137, doi: 10.1016/j.icarus.2013.07.012.

\bibitem[Gombosi et al., 2009]{Gombosi2009}
Gombosi, T. I., Armstrong, T. P., Arridge, C. S., Khurana, K. K., Krimigis, S. M., Krupp, N., Persoon, A. N., Thomsen, M. F., 2009. Saturn's magnetospheric configuration, \emph{Saturn after Cassini/Huygens}, 203-256, Springer, doi:10.1007/978-1-4020-9215-2.


\bibitem[Grundy et al.(1999)]{Grundy1999}
Grundy, W. M., Buie, M. W., Stansberry, J. A., Spencer, J. R., Schmitt, B., 1999. Near infrared spectra of icy outer solar system surfaces: remote determination of $H_2O$ ice temperatures. \emph{Icarus} 142, 536-549.

\bibitem[Grundy and Schmitt(1998)]{GrundySchmitt1998}
Grundy, W. M., Schmitt, B., 1998. The temperature dependent near infrared absorption spectrum of hexagonal $H_2O$ ice. \emph{Journal of Geophysical Research}, 103, 25809-25822.

\bibitem[Hamilton et al.(2015)]{Hamilton2015}
Hamilton, D. P., Skrutskie, M. F., Verbiscer, A. J., Masci, F. J. 2015.  Small particles dominate Saturn?s Phoebe ring to surprisingly large distances, \emph{Nature}, 522, 185-187, doi: 10.1038/nature14476.


\bibitem[Hansen et al.(2006)]{Hansen2006}
Hansen, C. J., Esposito, L., Stewart, A. I. F., Colwell, J., Hendrix, A. Pryor, W., Shemansky, D., West, R., 2006. Enceladus' Water Vapor Plume, \emph{Science}, 311, 1422-1425, doi:10.1126/science.1121254

\bibitem[Hobbs(1974)]{Hobbs1974}
Hobbs, P. V., 1974. Ice Physics, Clarendon Press. Oxford.

\bibitem[Hedman et al.(2013)]{Hedman2013}
Hedman, M. M., Gosmeyer, C. M., Nicholson, P. D., Sotin, C., Brown, R. H., Clark, R. N., Baines, K. H., Buratti, B. J., Showalter, M. R., 2013. An observed correlation between plume activity and tidal stresses on Enceladus. \emph{Nature}, 500, 182-184, doi: 10.1038/nature12371.

\bibitem[Howett et al.(2010)]{Howett2010}
Howett, C. J. A., Spencer, J. R., Pearl, J., Segura, M., 2010. Thermal inertia and bolometric Bond albedo values for Mimas, Enceladus, Tethys, Dione, Rhea and Iapetus as derived from Cassini/CIRS measurements. \emph{Icarus}, 206, 573-593, doi: 10.1016/j.icarus.2009.07.016.

\bibitem[Howett et al.(2011a)]{Howett2011a}
Howett, C. J. A., Spencer, J. R., Schenk, P., Johnson, R. E., Paranicas, C., Hurford, T. A., Verbiscer, A., Segura, M., 2011. A high-amplitude thermal inertia anomaly of probable magnetospheric origin on Saturn's moon Mimas. \emph{Icarus}, 216, 221-226, doi: 10.1016/j.icarus.2011.09.007.

\bibitem[Howett et al.(2011b)]{Howett2011b}
Howett, C. J. A., Spencer, J. R., Pearl, J., Segura, M., 2011. High heat flow from Enceladus' south polar region measured using 10-600 $cm^{-1}$ Cassini/CIRS data, \emph{JGR}, 116, E03003, doi:10.1029/2010JE003718. 

\bibitem[Howett et al.(2012)]{Howett2012}
Howett, C. J. A., Spencer, J. R., Hurford, T., Verbiscer, A., Segura, M., 2012. PacMan returns: An electron-generated thermal anomaly on Tethys, \emph{Icarus}, 221, 1084-1088, doi:10.1016/j.icarus.2012.10.013.

\bibitem[Howett et al.(2014)]{Howett2014}
Howett, C. J. A., Spencer, J. R., Hurford, T., Verbiscer, A., Segura, M., 2014. Thermophysical property variations across Dione and Rhea. \emph{Icarus}, 241, 239-247, doi: 10.1016/j.icarus.2014.05.047.

\bibitem[Imanaka et al.(2012)]{Imanaka2012}
Imanaka, H., Cruikshank, D.P., Khare, B.N., McKay, C.P., 2012. Optical constants of Titan tholins at mid-infrared wavelengths (2.5-25 $\mu m$) and the possible chemical nature of Titan's haze particles, \emph{Icarus}, 218, 247-261.

\bibitem[Jaumann et al.(2006)]{Jaumann2006}
Jaumann, R., Stephan, K., Brown, R. H., Buratti, B. J., Clark, R. N., McCord, T. B., Coradini, A., Capaccioni, F., Filacchione, G., Cerroni, P., Baines, K. H., Bellucci, G., Bibring, J.-P., Combes, M., Cruikshank, D. P., Drossart, P., Formisano, V., Langevin, Y., Matson, D. L., Nelson, R. M., Nicholson, P. D., Sicardy, B., Sotin, C., Soderbloom, L. A., Griffith, C., Matz, K.-D., Roatsch, Th., Scholten, F., Porco, C. C., 2006. High-resolution CASSINI-VIMS mosaics of Titan and the icy Saturnian satellites, \emph{Planetary and Space Science}, 54, 1146-1155, doi: 10.1016/j.pss.2006.05.034.

\bibitem[Jaumann et al.(2008)]{Jaumann2008}
Jaumann, R., Stephan, K., Hansen, G.~B., Clark, R.~N., Buratti, B.~J., Brown, R.~H., Baines, K.~H., Newman, S.~F., Bellucci, G., Filacchione, G., Coradini, A., Cruikshank, D.~P., Griffith, C.~A., Hibbitts, C.~A., McCord, T.~B., Nelson, R.~M., Nicholson, P.~D., Sotin, C., Wagner, R., 2008. Distribution of icy particles across Enceladus' surface as derived from Cassini-VIMS measurements. \emph{Icarus}, 193, 407-419, doi: 10.1016/j.icarus.2007.09.013.

\bibitem[Jaumann et al.(2009)]{Jaumann2009}
Jaumann, R., Clark, R. N., Nimmo, F., Hendrix, A. R., Buratti, B. J., Denk, T., Moore, J. M., Schenk, P. M., Ostro, S. J., Srama, R., 2009. Icy Satellites: Geological Evolution and Surface Processes, in \emph{Saturn from Cassini-Huygens}, edited by M.K. Dougherty, L.W. Esposito, and S.M. Krimigis, Springer, Berlin, 459-509, doi: 10.1007/978-1-4020-9217-6$\_$20.


\bibitem[Kempf et al.(2010)]{Kempf2010}
Kempf, S., Beckmann, U, Schmidt, J., (2010). How the Enceladus dust plume feeds Saturn?s E ring, \emph{Icarus}, 206, 446-457, doi: 10.1016/j.icarus.2009.09.016.

\bibitem[Kuiper(1957)]{Kuiper1957}
Kuiper, G.P., 1957. Infrared observations of planets and satellites. \emph{Astronomical Journal}, 62, 295.

\bibitem[Le Gall et al.(2014)]{Legall2014}
Le Gall, A., Leyrat, C., Janssen, M. A., Keihm, S., Wye, L. C., West, R., Lorenz, R. D., 2014. Iapetus' near surface thermal emission modeled and constrained using Cassini RADAR Radiometer microwave observations, \emph{Icarus}, 241, 221-238, doi: 10.1016/j.icarus.2014.06.011. 

\bibitem[Leto et al.(2005)]{Leto2005}
Leto, G., Gomis,O., Strazzulla,G., 2005. The reflectance spectrum of water ice: is the 1.65 $\mu m$ peak a good temperature probe? \emph{Memorie della Società Astronomica Italiana Supplement}, 6, 57.


\bibitem[Mastrapa et al.(2008)]{Mastrapa2008}
Mastrapa, R., Bernstein, M., Sandford, S., Roush, T., Cruikshank, D., Dalle Ore, C., 2008. Optical constants of amorphous and crystalline H$_2$O-ice in the near infrared from 1.1 to 2.6 $\mu m$. \emph{Icarus}, 197, 307-320, doi: 10.1016/j.icarus.2008.04.008.

\bibitem[Mastrapa et al.(2009)]{Mastrapa2009}
Mastrapa, R. M., Sandford, S. A., Roush, T L., Cruikshank, D. P., Dalle Ore, C M., 2009. Optical constants of amorphous and crystalline H$_2$O-ice: 2.5 - 22 $\mu m$ (4000 - 455 cm$^{-1}$) optical constants of H$_2$O-ice. \emph{Astrophysical Journal}, 701, 1347-1356, doi: 10.1088/0004-637X/701/2/1347.

\bibitem[Moore et al.(1983)]{Moore1983}
Moore, M. H., Donn, B., Khanna, R., A'Hearn, M. F., 1983. Studies of proton-irradiated cometary-type ice mixtures, \emph{Icarus}, 54, 388-505, doi: 10.1016/0019-1035(83)90236-1. 

\bibitem[Nimmo et al.(2007)]{Nimmo2007}
Nimmo, F., Spencer, J. R., Pappalardo, R. T., Mullen, M. E., 2007. Shear heating as the origin of the plumes and heat flux on Enceladus. \emph{Nature}. 447, 289?291, doi:10.1038/nature05783 

\bibitem[Oancea et al.(2012)]{Oancea2012}
Oancea, A., Grasset, O., Le Menn, E., Bollengier, O., Bezacier, L., Le Mouelic, S., Tobie, G., 2012.
Laboratory infrared reflection spectrum of carbon dioxide clathrate hydrates
for astrophysical remote sensing applications, \emph{Icarus}, 221, 900-910, doi: 10.1016/j.icarus.2012.09.020.

\bibitem[Paranicas et al.(2010a)]{Paranicas2010a}
Paranicas, C., Mitchell, D. G., Krimigis, S. M., Carbary, J. F., Brandt, P. C., Turner, F. S., Roussos, E., Krupp, N., Kivelson, M. G., Khurana, K. K., Cooper, J. F., Armstrong, T. P., Burton, M., 2010a. Asymmetries in Saturn's radiation belts, \emph{JGR}, 115, Issue A7, CiteID A07216, doi: 10.1029/2009JA014971.

\bibitem[Paranicas et al.(2010b)]{Paranicas2010b}
Paranicas, C., Mitchell, D. G., Roussos, E., Kollmann, P., Krupp, N., M\"uller, A. L., Krimigis, S. M., Turner, F. S., Brandt, P. C., Rymer, A. M., Johnson, R. E., (2010b). Transport of energetic electrons into Saturn's inner magnetosphere, \emph{JGR}, 115, Issue A9, CiteID A09214, doi: 10.1029/2010JA015853.

\bibitem[Pinilla-Alonso et al.(2011)]{Pinilla2011}
Pinilla-Alonso, N., Roush, T. L., Marzo, G., Cruikshank, D. P., Dalle Ore, C. M., 2011. Iapetus surface variability revealed from statistical clustering of a VIMS mosaic: the distribution of CO$_{2}$. \emph{Icarus}, 215, 75-82, doi: 10.1016/j.icarus.2011.07.004.

\bibitem[Pitman et al.(2010)]{Pitman2010}
Pitman, K. M.,  Buratti, B. J., Mosher, J. A., 2010. Disk-integrated bolometric Bond albedos and rotational light curves of saturnian satellites from Cassini Visual and Infrared Mapping Spectrometer. \emph{Icarus}, 206, 537-560, doi: 10.1016/j.icarus.2009.12.001.

\bibitem[Porco et al.(2006)]{Porco2006}
Porco, C. C., Helfenstein, P., Thomas, P. C., Ingersoll, A. P., Wisdom, J., West, R., Neukum, G.,
Denk, T., Wagner, R., Roatsch, T., Kieffer, S., Turtle, E., McEwen, A., Johnson, T. V.,
Rathbun, J., Veverka, J., Wilson, D., Perry, J., Spitale, J., Brahic, A., Burns, J. A.,
DelGenio, A. D., Dones, L., Murray, C. D., Squyres, S., 2006. Cassini Observes the Active
South Pole of Enceladus, \emph{Science}, 311, 1393-1401, doi:10.1126/science.1123013

\bibitem[Porco et al.(2014)]{Porco2014}
Porco, C., DiNino, D., Nimmo, F., 2014. How the Geysers, Tidal Stresses, and Thermal Emission across the South Polar Terrain of Enceladus are Related, \emph{AJ}, 148, 45, doi:10.1088/0004-6256/148/3/45.

\bibitem[Postberg et al.(2011)]{Postberg2011}
Postberg, F., Schmidt, J., Hillier, J., Kempf, S., Srama, R., 2011. A salt-water reservoir as the source of a compositionally stratified plume on Enceladus, \emph{Nature}, 474, 620-622, doi:10.1038/nature10175.

\bibitem[Sandford and Allamandola(1990)]{SandfordAllamandola1990}
Sandford, S. A.; Allamandola, L. J., 1990. The physical and infrared spectral properties of $CO_2$ in astrophysical ice analogs, \emph{ApJ}, 355, 357-372, doi: 10.1086/168770.


\bibitem[Schenk et al.(2011)]{Schenk2011}
Schenk, P., Hamilton, D. P., Johnson, R. E., McKinnon, W. B., Paranicas, C., Schmidt, J., Showalter, M. R., 2011. Plasma, plumes and rings: Saturn system dynamics as recorded in global color patterns on its midsize icy satellites. \emph{Icarus}, 211, 740-757, doi:10.1016/j.icarus.2010.08.016.

\bibitem[Scipioni et al.(2013)]{Scipioni2013}
Scipioni, F., Tosi, F., Stephan, K., Filacchione, G., Ciarniello, M., Capaccioni, F., Cerroni, P. and the VIMS
Team, 2013. Spectroscopic classification of icy satellites of Saturn I: Identification of terrain units on Dione, \emph{Icarus}, 226, 1331-1349, doi: 10.1016/j.icarus.2013.08.008.

\bibitem[Scipioni et al.(2014)]{Scipioni2014}
Scipioni, F., Tosi, F., Stephan, K., Filacchione, G., Ciarniello, M., Capaccioni, F., Cerroni, P. and the VIMS
Team, 2014. Spectroscopic classification of icy satellites of Saturn II: Identification of terrain units on Rhea, \emph{Icarus}, 234, 1-16, doi: 10.1016/j.icarus.2014.02.010.

\bibitem[Spencer et al.(2006)]{Spencer2006}
Spencer, J. R., Pearl, J. C., Segura, M., Flasar, F. M., Mamoutkine, A., Romani, P., Buratti, B. J., Hendrix, A. R., Spilker, L. J., Lopes, R. M. C., 2006. Cassini Encounters Enceladus: Background and the Discovery of a South Polar Hot Spot, \emph{Science}, 311, 1401-1405, doi:10.1126/science.1121661.

\bibitem[Spencer et al.(2009)]{Spencer2009}
Spencer, J. R., Barr, A. C., Esposito, L. W., Helfenstein, P., Ingersoll, A. P., Jaumann, R., McKay, C. P., Nimmo, F., Hunter Waite, J., 2009. Enceladus: an active cryovolcanic satellite, in \emph{Saturn from Cassini-Huygens} , edited by M.K. Dougherty, L.W. Esposito, and S.M. Krimigis, Springer, Berlin, 683-724. ISBN 978-1-4020-9216-9.


\bibitem[Spencer and Denk (2010)]{SpencerDenk2010}
Spencer, J. R., Denk, T., 2010. Formation of Iapetus? Extreme Albedo Dichotomy by Exogenically Triggered Thermal Ice Migration, \emph{Science}, 327, 432.

\bibitem[Stephan et al.(2010)]{Stephan2010}
Stephan, K., Jaumann, R., Wagner, R., Clark, R.~N., Cruikshank, D.~P., Hibbitts, C.~A. Roatsch, T., Hoffmann, H., Brown, R.~H., Filacchione, G., Buratti, B.~J., Hansen, G.~B.,  McCord, T.~B., Nicholson, P.~D., Baines, K.~H., 2010. Dione' s spectral and geological properties, \emph{Icarus}, 206, 631-652, DOI: 10.1016/j.icarus.2009.07.036.


\bibitem[Stephan et al.(2012)]{Stephan2012}
Stephan, K., Jaumann, R., Wagner, R., Clark, R. N., Cruikshank, D. P., Giese, B.,  Hibbitts, C. A., Roatsch, T., Matz, K.-D., Brown, R. H., Filacchione, G., Capaccioni, F., Scholten, F., Buratti, B. J., Hansen, G. B., Nicholson, P. D., Baines, K. H., Nelson, R. M., Matson, D. L., 2012. The saturnian satellite Rhea as seen by Cassini VIMS. \emph{Planetary and Space Science}, 61, 142-160, DOI: 10.1016/j.pss.2011.07.019.

\bibitem[Taffin et al.(2012)]{Taffin2012}
Taffin, C., Grasset, O., LeMenna, E., Bollengier, O., Giraud, M., LeMouelic, S., 2012. Temperature and grain size dependence of near-IR spectral signature of crystalline water ice: From lab experiments to Enceladus' south pole, \emph{Planetary and Space Science}, 61, 124-134, doi: 10.1016/j.pss.2011.08.015  

\bibitem[Tamayo et al.(2011)]{Tamayo2011}
Tamayo, D., Burns, J. A., Hamilton, D. P., Hedman, M. M., 2011. Finding the trigger to Iapetus' odd global albedo pattern: Dynamics of dust from Saturn's irregular satellites, \emph{Icarus}, 215, 260-278, doi: 10.1016/j.icarus.2011.06.027.


\bibitem[Tosi et al.(2010)]{Tosi2010}
Tosi, F., Turrini, D., Coradini, A., Filacchione, G. and the VIMS Team, 2010. Probing the origin of the dark material on Iapetus, \emph{MNRAS}, 403, 1113-1130, doi: 10.1111/j.1365-2966.2010.16044.x.

\bibitem[Verbiscer and Veverka(1989)]{VerbiscerVeverka1989}
Verbiscer, A.J., Veverka, J., 1989. Albedo dichotomy of Rhea: Hapke analysis of Voyager photometry, \emph{Icarus}, 82, 336-353.

\bibitem[Verbiscer et al.(2009)]{Verbiscer2009}
Verbiscer, A. J., Skrutskie, M. F., Hamilton, D. P., 2009. Saturn's largest ring, \emph{Nature}, 461, 1098-1100, doi: 10.1038/nature08515.

\bibitem[Waite et al.(2012)]{Waite2006}
Waite, J. H., Combi, M. R., Ip, W.-H., Cravens, T. E., McNutt, R. L., Kasprzak, W., Yelle, R., Luhmann, J., Niemann, H., Gell, D. Magee, B., Fletcher, G., Lunine, J., Tseng, W.-L., 2006. Cassini Ion and Neutral Mass Spectrometer: Enceladus Plume
Composition and Structure, \emph{Science}, 311, 1419-1422, doi:10.1126/science.1121290.

\bibitem[Warren(1984)]{Warren1984}
Warren, S. G.,1984. Optical constants of ice from the ultraviolet to the microwave, \emph{Applied Optics}, 23,1206-1225.

\bibitem[Zubko et al.(1996)]{Zubko1996}
Zubko, V. G., Mennella, V., Colangeli, L., Bussoletti, E., 1996. Optical constants of cosmic carbon analogue grains
I. Simulation of clustering by a modified continuous distribution of ellipsoids. \emph{Mon. Not. R. Astron. Soc.}, 282, 1321-1329.


\end{thebibliography}
\end{document}